\documentclass[12pt,showpacs,nofootinbib]{revtex4}

\usepackage{dcolumn,amsmath}
\usepackage{amsfonts}
\usepackage{amsthm}
\usepackage{mathrsfs}
\usepackage{amssymb}
\usepackage{epsfig}
\usepackage{psfrag}
\usepackage{array}
\usepackage{graphics,graphicx,float,mfpic}
\usepackage{color}

\newcommand{\be}{\begin{equation}}
\newcommand{\ee}{\end{equation}}
\newcommand{\bea}{\begin{eqnarray}}
\newcommand{\eea}{\end{eqnarray}}

\newcommand{\ssl}{\sl}

\newcommand{\bbf}{\bf}
\newcommand{\ccal}{\mathcal}
\newcommand{\bx}{{\bf x}}
\newcommand{\Q}{N}
\newcommand{\XPEH}{\it}
\newcommand{\mcal}{\mathcal}
\newcommand{\ssigma}{{\mcal N}}
\newcommand{\RRR}{{R}}
\newcommand{\ccc}{{a}}
\newcommand{\n}{{n}}
\newcommand{\nn}{{\bf \n}}

\def\be{\begin{equation}}
\def\ee{\end{equation}}
\def\bea{\begin{eqnarray}}
\def\eea{\end{eqnarray}}

\begin{document}


\title{Stationary 
ring solitons  in field theory -- \\
knots and vortons
\vspace{1 cm}
}

\author{Eugen Radu%
\footnote{{{\tt Eugen.Radu@lmpt.univ-tours.fr}}}}


\author{Mikhail S. Volkov%
\footnote{{\tt Michael.Volkov@lmpt.univ-tours.fr}}}

 \affiliation{ \vspace{0.1 cm}
{Laboratoire de Math\'{e}matiques et Physique Th\'{e}orique
CNRS-UMR 6083, \\ Universit\'{e} de Tours,
Parc de Grandmont, 37200 Tours, FRANCE\\
}
\vspace{1 cm}
}

\begin{abstract}
We review the current status of the problem of constructing 
classical field theory solutions describing {stationary}  
vortex rings in Minkowski space in $3+1$
dimensions. 
We describe the known up to date solutions of this type, such as the  
static knot solitons stabilized by the topological Hopf charge, 
the  attempts to gauge them, the anomalous solitons
stabilized by the Chern-Simons number, as well as  
the non-Abelian monopole and sphaleron rings. Passing to the 
rotating solutions, we first discuss the conditions insuring that they
do not radiate, and then 
describe the spinning $Q$-balls, their twisted 
and gauged generalizations reported here for the first time, 
spinning skyrmions, and rotating monopole-antimonopole pairs. 
We then present the first explicit construction of global vortons as solutions
of the elliptic boundary value problem, which demonstrates their non-radiating 
character. 
Finally, we describe  the analogs of vortons  
in the Bose-Einstein condensates, analogs of spinning $Q$-balls
in the non-linear optics, and also moving vortex rings
in superfluid helium and in ferromagnetics.

\end{abstract}



\maketitle
\hspace{0.7 cm}
to appear in {\it Physics Reports}

\vspace{0.1 cm}
\newpage 

\tableofcontents


\section{Introduction}

Stationary vortex loops are often discussed in the literature in various contexts
ranging from the models of condensed matter physics 
\cite{Babaev:2001jt},
\cite{Babaev:2001zy},
\cite{Battye:2001ec},
\cite{Cooper},
\cite{Metlitski:2003gj},
\cite{Protogenov:2002bt}, 
\cite{Ruostekoski:2001fc},
\cite{Ruostekoski:2003qx},
\cite{Savage:2003hh}, 
 to high energy physics and cosmology
\cite{Brandenberger:1996zp},
\cite{Davis:1988ip},
\cite{Davis:1988jp},
\cite{Davis:1988ij},
\cite{Vilenkin},
\cite{Witten:1984eb}.
Such objects could be quite interesting physically and might be  
responsible for a plenty of important physical phenomena,
starting from the structure of quantum superfluids \cite{Donnelly} and 
Bose-Einstein condensed alkali gases \cite{Leggett} 
to the baryon asymmetry of the Universe, dark matter,
and the galaxy formation \cite{Vilenkin}. These intriguing 
physical aspects of vortex loops occupy therefore most of the discussions 
in the literature, while much less attention is usually payed 
to their mathematical existence. 
In most considerations the existence of such loops is discussed 
only qualitatively, using plausibility arguments, and the problem of constructing
the corresponding field theory solutions is rarely addressed, such that almost 
no solutions of this type are explicitly known.

The physical arguments usually invoked to justify the loop existence 
arise within  the effective macroscopic description 
of vortices \cite{Abrikosov}, \cite{NO}. Viewed from large distances,
their internal structure can be neglected and the vortices can be effectively 
described as elastic thin ropes or, if they  carry currents,
as thin wires 
\cite{Carter:1989dp},
\cite{Carter:1989xk},
\cite{Carter:1990sm},
\cite{Carter:1994hn},
\cite{Martins:1998gb}. 
This suggests the following  engineering 
procedure: to `cut out' a finite piece of the vortex, then `twist' it several times
and finally `glue' its ends. 
The resulting loop will be stabilized against contraction by the potential 
energy of the twist deformation, that is by the positive pressure contributed by the 
current and stresses 
\cite{Copeland:1987th},
\cite{Copeland:1987yv},
\cite{Haws:1988ax},
\cite{Niemi:2000ny}. 
Equally, instead of twisting the rope,
one can first `make' a loop and then  
`spin it up' thus giving to it an angular momentum to 
stabilize it against shrinking 
\cite{Davis:1988jq},
\cite{Davis:1988ij}. 
Depending on whether they are spinning or not, 
such loops are often called in the literature vortons and knots (also springs), 
respectively.

These macroscopic arguments are suggestive. However, they
cannot guarantee the existence of loops as stationary field theory
objects,  since they do not take into account all the field degrees of freedom that 
could be essential. Speaking more rigorously, one can prepare twisted or spinning
loop configurations as the initial data for the fields. 
However, nothing guarantees that these data will evolve to 
non-trivial equilibrium field theory objects, since, up to few exceptions, the 
typical field theory
models under consideration do not have a topological energy bound for static loop 
solitons. For spinning loops the situation can be better, 
since the angular momentum 
can provide an additional stabilization of the system. 
However, nothing excludes the possibility 
of radiative  energy leakages from
spinning loops. Since there is typically a current
circulating along them, which means an accelerated motion of charges, 
it seems plausible that spinning loops should radiate, in which case
they would not 
be {\it stationary} but at best only quasistationary. 
In fact, loop formation is generically observed in 
dynamical vortex network simulations (although for currentless vortices),
and  these loops indeed 
radiate rapidly away all their energy 
(see $e.g.$ \cite{Bevis:2006mj}).

The lack of explicit solutions renders the situation even more controversial.
In fact, although there are explicit examples of loop solitons 
in {\it global} field theory, which means containing only scalar fields with global 
internal symmetries, almost nothing is known about such solutions in 
{\it gauge} field theory. Vortons, for example, were initially proposed 
more than 20 years ago in the context of the local U(1)$\times$U(1) 
theory of superconducting cosmic strings of  Witten 
 \cite{Witten:1984eb}. However, not a single 
field theory solution of this type has been obtained up to now.
Search for static knot solitons in the Ginzburg-Landau type gauge field theory models
has already 30 years of history, the result being always negative. 
All this  casts some doubts on 
the existence of {stationary}, non-radiating vortex loops in physically interesting 
gauge field theory models. 
However, rigorous no-go arguments are not known either. 
As a result, there are no solid arguments neither 
for nor against the existence of such solutions.

Interestingly, {\it exact} solutions describing ring-type objects are known
in curved space. These are black rings in the multidimensional generalizations 
of  General Relativity -- spinning toroidal black holes 
(see \cite{Emparan} for a review). It seems therefore that Einstein's
field equations, notoriously known for their complexity,
are easier to solve than the non-linear equations describing
loops made of interacting gauge and scalar fields in Minkowski space.

It should be emphasized that the problem here is not related to the dynamical 
stability of these loops. In order to be stable or unstable they should first of 
all exist as stationary field theory solutions. The problem is related to the very 
existence of such solutions. One should be able to decide whether they exist or not,
which is a matter of principle, and this is the main issue that we address  in this paper.
Therefore, when talking about loops stabilized by some forces, 
we shall mean the force balance that
makes possible their existence,
and not the stability of these loops with respect to all 
dynamical perturbations. If they exist, their dynamical stability should be 
analyzed separately, but we shall call {\it solitons} all  localized, globally regular, 
finite energy field theory solutions, irrespectively of their stability properties. 
It should also be stressed that we insist on the 
{\it stationarity} condition for the solitons, which 
implies the absence of radiation.  
Quasistationary loops which radiate slowly and 
live long enough could also be physically
interesting, but we are only interested in loops which could live infinitely
longtime, at least in classical theory.

Trying to clarify the situation, we review in what follows 
the known field theory solutions  in Minkowski space
in $3+1$ dimensions describing stationary ring objects. 
We shall divide them in two groups, depending on whether they do have 
or do not have an angular momentum.  Solutions without angular momentum are 
typically static, that is they do not depend on time and also do not typically 
have an electric field. 
We start by describing  the static knot solitons stabilized by the Hopf charge
in a global field theory model. Then we 
review the attempts to generalize these solutions within 
gauge field theory, with the conclusion that some additional constraints on the
gauge field are necessary, since fixing only the Hopf charge does not seem to be
sufficient  to
stabilize the system in this case. We then consider two known examples of 
static ring solitons in gauge field theory: the anomalous solitons stabilized 
by the Chern-Simons number, and the non-Abelian monopole and sphaleron rings.

Passing to the rotating solutions, we first of all discuss 
the issue  of how the presence 
of an angular momentum, associated to some internal 
motions in the system, can be 
reconciled with the absence of radiation which is normally 
generated by these motions. 
One possibility for this is to consider
time-independent solitons in theories with local internal symmetries.
They will not radiate, but it can be shown  
for a number of important field theory models that the angular momentum
vanishes in this case.  Another possibility 
arises in systems with global symmetries,
where one can consider   {\it non-manifestly} stationary 
and axisymmetric fields containing 
spinning phases. Such fields could have a non-zero 
angular momentum, but the absence of radiation is not automatically
guaranteed in this case. The general conclusion is that the 
existence of non-radiating spinning solitons, although not impossible,  
seems to be rather restricted. Such solutions seem  to exist only in some quite special
field theory models, while generic spinning field systems should radiate. 

Nevertheless, non-radiating spinning 
solitons in Minkowski space
in $3+1$ dimensions exist, and we review below all known examples: these are  the 
spinning $Q$-balls, their twisted and gauged generalizations, 
spinning Skyrmions and rotating monopole-antimonopole pairs.
In addition, there are also vortons, and below we present for the first time  
numerical solutions of elliptic equations describing vortons in the 
global field theory limit. 
Although global vortons have 
been studied before by different methods,
our construction shows that they indeed exist as stationary, 
non-radiating field theory objects. 

We also discuss  stationary ring solitons in non-relativistic physics. 
Surprisingly, it turns out that the relativistic vortons can be mapped
to the `skyrmion' solutions of the Gross-Pitaevskii equation in the theory 
of Bose-Einstein condensation.  In addition, it turns out that $Q$-balls 
can describe light pulses in media with non-linear refraction -- `light bullets'.
Finally, for the sake of completeness, we discuss 
also the moving vortex rings stabilized by the Magnus force
in continuous media theories,  such as in the 
superfluid helium and in  ferromagnetics.   

In our numerical calculations  we use an elliptic PDE solver 
with which we have managed to 
reproduce most of the solutions we describe, as well as obtain a number of 
new results presented below for the first time.  
The latter include the explicit vorton solutions, 
spinning twisted $Q$-balls, spinning 
gauged $Q$-balls, 
as well as the `Saturn', `hoop' and bi-ring solutions for the interacting $Q$-balls.   
We put the main emphasise on describing how the solutions are constructed
and not to  their physical applications, so that 
our approach is just the opposite and therefore  complementary
to the one generally adopted 
in the existing literature. 
As a result, we outline the current status of the ring soliton existence problem --
within the numerical approach. Giving mathematically rigorous existence proofs 
is an issue  that should be analyzed separately.
It seems that our global vorton solutions could be generalized 
in the context of gauge field theory. 
The natural problem to attack 
would then be to obtain vortons in the electroweak sector 
of Standard Model.

In this text the signature of the Minkowski spacetime metric $g_{\mu\nu}$ 
is chosen to be $(+,-,-,-)$, 
the spacetime coordinates are denoted by $x^\mu=(x^0,x^k)\equiv (t,\bx)$ 
with $k=1,2,3$. All physical quantities discussed below, including 
fields, coordinates, coupling constants and conserved quantities are
dimensionless.  

\section{Knot solitons}

Let us first consider solutions with zero angular momentum stabilized by their 
intrinsic deformations. 
We shall start by discussing  the famous example of 
static knotted solitons in a
non-linear sigma model. Since this  is the best known and also in some 
sense canonical example of knot solitons, we shall describe it in some detail.  
We shall then review the status of  gauge field theory 
generalizations of these solutions.  

\subsection{Faddeev-Skyrme model}

More than 30 years ago Faddeev introduced a field theory consisting of 
a non-linear O(3) sigma model augmented by adding a Skyrme-type term
\cite{Faddeev:1975},
\cite{Faddeev:1976pg}. 
This theory can also be obtained by a consistent truncation of the O(4) Skyrme model
(see Sec.\ref{versus} below). 
Its dynamical variables are three
scalar fields $\nn\equiv \n^a=(\n^1,\n^2,\n^3)$
 constraint by the condition 
$$
\nn\cdot\nn=\sum_{a=1}^3 \n^a \n^a=1,
$$
so that they span a two-sphere $S^2$. 
The Lagrangian density of the theory is 
\be                                                              \label{action0} 
{\mcal L}[\nn]=\frac{1}{32\pi^2}\left(\partial_\mu \nn\cdot\partial^\mu \nn
-{\ccal F}_{\mu\nu}{\ccal F}^{\mu\nu}\right)
\ee
where 
\be                                              \label{F}
{\ccal F}_{\mu\nu}=\frac12\,\epsilon_{abc}\n^a\partial_\mu \n^b\partial_\nu \n^c\equiv
\frac12\,\nn\cdot(\partial_\mu\nn\times\partial_\nu\nn). 
\ee
The Lagrangian field equations read 
\be                              \label{eqs-Hopf}
\partial_\mu\partial^\mu\nn+\partial_\mu{\ccal F}^{\mu\nu}(\nn\times\partial_\nu\nn)
=
(\nn \cdot\partial_\mu\partial^\mu\nn 
)\,\nn\,.
\ee  
In the static limit the 
energy of the system is  
\be                                                               \label{Faddeev}
E[\nn]=\left.\left.\frac{1}{32\pi^2}\int_{\mathbb{R}^3}\right((\partial_k \nn)^2
+({\ccal F}_{ik})^2\right)
d^3 \bx\,\equiv E_2+E_4. 
\ee
Under scale transformations, $\bx\to\Lambda\bx$,
one has $E_2\to\Lambda E_2$ and $E_4\to E_4/\Lambda$.
The energy will therefore be stationary for $\Lambda=1$ if only the 
virial relation holds,
\be
E_2=E_4.
\ee
This shows that the four derivative term $E_4$ is necessary, since otherwise  
the virial relation would require
that $E_2=0$, thus ruling out all non-trivial static
solutions -- in agreement with the  Hobart-Derrick theorem
\cite{Hobart}, 
\cite{Derrick:1964ww}. 

Any static field $\nn(\bx)$ defines a map $\mathbb{R}^3\to {S}^2$.
Since for finite energy configurations $\nn$ should approach
a constant value for $|\bx|\to\infty$, all points at infinity of $\mathbb{R}^3$ 
map to one point on $S^2$. Using the global O(3)-symmetry of the theory one can choose
this point to be the north pole of the $S^2$, 
\be                        \label{n-inf}
\lim_{|\bx|\to\infty}\nn(\bx)=\nn_\infty=(0,0,1).
\ee
Notice that this condition leaves a residual O(2) symmetry of global rotations around 
the third axis in the internal space,
\be                                       \label{residual}
n^1+in^2\to(n^1+in^2)e^{i\alpha},~~~~~~~~~~ n^3\to n^3.
\ee 
The position of the field configuration described by $\nn(\bx)$
 can now be defined as the set of points where the 
field is as far as possible from the vacuum value, that is the preimage
of the point $-\nn_\infty$ antipodal to the vacuum $+\nn_\infty$. This preimage 
forms a closed
loop (or collection of loops) called position curve. 
Solitons in the theory can therefore be viewed as string-like objects,
stabilized by their topological charge to be defined below.

At the intuitive level, these solitons can be viewed as closed loops made of 
twisted vortices.  
 Specifically, the theory 
admits solutions describing straight vortices that can be 
parametrized in cylindrical coordinates 
$\{\rho,z,\varphi\}$ as $n^3=\cos\Theta(\rho)$, 
$\n^1+i\n^2=\sin\Theta(\rho)\,e^{ipz+in\varphi}$,  where $n\in\mathbb{Z}$
is the vortex winding number \cite{Kundu}. 
The phase $pz+n\varphi$ thus changes both along and around the vortex and 
the vector $\nn$  rotates   around the third internal direction as one moves
along the vortex, which can be interpreted as {\it twisting}
of the vortex. 
It seems plausible that a loop made of a piece of length $L$ of such a twisted 
vortex, where $pL=2\pi m$, 
could be stabilized by the potential energy of the twist deformation. 
For such a twisted loop the phase increases by $2\pi n$ after a revolution around
the vortex core and by $2\pi m$ as one travels along the loop, where $m\in\mathbb{Z}$
is the number of twists. If the loop is homeomorphic to a circle, then the product
$nm$ gives the value of its topological invariant: the Hopf charge.

\subsection{Hopf charge}

\begin{figure}[h]
\hbox to\linewidth{\hss%
  \resizebox{8cm}{3cm}{\includegraphics{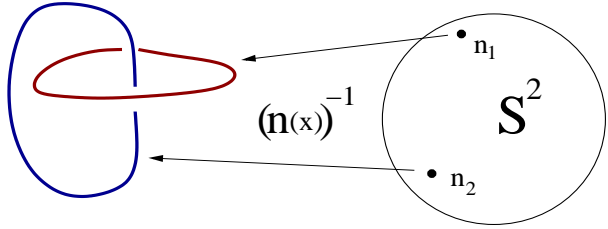}}%
\hss}
\caption{\small 
Preimages 
of any two points $\nn_1$ and $\nn_2$ on the target space $S^2$ 
are two loops. The number of mutual linking of these two loops
is the Hopf charge $\Q[\nn]$ of the map $\nn(\bx)$. Here the case of 
$\Q=1$ is schematically shown. 
}
\label{FigHopf}
\end{figure}

The condition \eqref{n-inf} allows one to view the infinity of $\mathbb{R}^3$ 
 as one point, thus effectively replacing 
$\mathbb{R}^3$ by its one-point compactification
${S}^3$. Any smooth field configuration 
can therefore be viewed as a map
\be
\nn(\bx): {S}^3\to {S}^2.
\ee 
Any such map can be characterized by the topological
charge $\Q[\nn]\in \pi_3(S^2)=\mathbb{Z}$ known as the Hopf invariant.
 This invariant has a simple
interpretation. The preimage of a generic point on the target  $S^2$ is a 
closed loop. If a field has Hopf number $\Q$ then 
the two loops consisting of preimages of two generic distinct points on  $S^2$ 
will be linked exactly $\Q$ times (see Fig.\ref{FigHopf}). 

Although there is no local formula for $\Q[\nn]$ in terms of $\nn$, one can give
a non-local expression as follows. 
The 2-form ${\ccal F}=\frac12\,{\ccal F}_{ik}dx^i\wedge dx^k$ defined by Eq.\eqref{F} is closed, 
$d{\ccal F}=0$, 
and since the second cohomology group of $S^3$ is trivial, $H(S^3)=0$, 
there globally exists 
 a vector potential ${\ccal A}={\ccal A}_k dx^k$ such that 
${\ccal F}=d{\ccal A}$. 
The Hopf index  
can then be expressed as
\be                                                              \label{Q} 
\Q[\nn]=\frac{1}{8\pi^2}\int \epsilon_{ijk}{\ccal A}_i{\ccal F}_{jk}\, d^3\bx.
\ee
For any smooth, finite energy field configuration $\nn(\bx)$ this integral is 
integer-valued \cite{Lin}. 

\begin{figure}[ht]
\hbox to\linewidth{\hss%
	\resizebox{8cm}{7cm}{\includegraphics{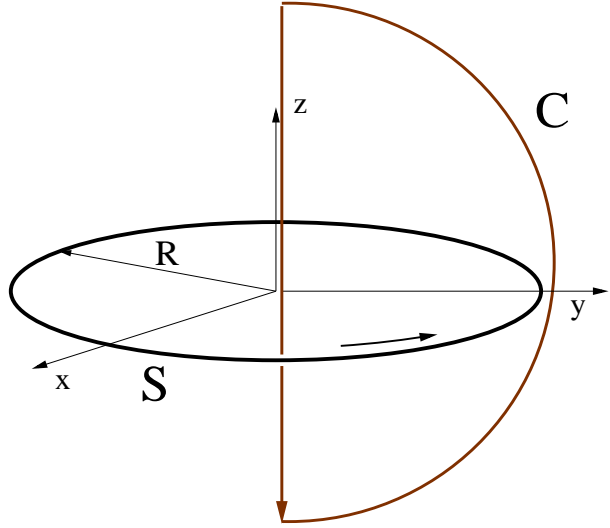}}
\hspace{5mm}%
\hss}
	\caption{Knot topology: the complex phase of the field in Eq.\eqref{Hopf:axial}
winds along two orthogonal directions: along $S$  and along the
 contour $C$ consisting of the $z$-axis and a semi-circle 
whose radius expands to infinity. A similar winding of phases 
is found for other systems to be
discussed below: for vortons, skyrmions, and twisted $Q$-balls.}
\label{FigS}
\end{figure}

It can be shown that the maximal symmetry of $\nn(\bx)$  compatible with 
a non-vanishing Hopf charge is O(2) \cite{Kundu}. 
It follows that spherically symmetric fields are topologically trivial.  
However, axially symmetric fields can have any value of the Hopf charge.  
Using cylindrical coordinates 
such fields can be parametrized as \cite{Kundu} 
\be                                              \label{Hopf:axial}
\n^1+i\n^2= e^{i(m\varphi-n\psi)}\sin\Theta,~~~~~n^3=\cos\Theta
\ee
where $n,m\in\mathbb{Z}$ and $\Theta,\psi$ are functions of $\rho,z$. 
Since  $\nn\to\nn_\infty$ asymptotically, 
$\Theta$ should vanish for $r=\sqrt{\rho^2+z^2}\to\infty$. 
The regularity at the $z$-axis requires for $m\neq 0$ that $\Theta$ 
should vanish also there. 
As a result, one has $\nn=\nn_\infty$ both at the $z$-axis and at infinity, 
that is at the contour  
$C$ shown in Fig.\ref{FigS}.
Next,  one assumes that $\nn=-\nn_\infty$ on a circle $S$ 
around the $z$-axis which is linked to $C$ as shown in Fig.\ref{FigS} 
(more generally, one can have $\nn=-\nn_\infty$ on several circles
around the $z$-axis).   
The phase function $\psi$ 
is supposed to increase by $2\pi$ after one revolution along $C$.
Since  $\cos\Theta$ interpolates
between $-1$ and $1$ 
on every trajectory from $S$ to $C$, it follows that  
surfaces of constant $\Theta$ are homeomorphic to tori. 
 
The preimage of the point $-\nn_\infty$ consists of $m$ copies of the circle $S$.  
The preimage of $\nn_\infty$ consists of $n$ copies of the contour $C$.
These two preimages   
 are therefore linked $mn$ times.  

One can also compute 
${\ccal F}=d{\ccal A}$ according to \eqref{F}, from where one finds 
\be                                        \label{Hopf:axial1}
{\ccal A}=n\cos^2\frac{\Theta}{2}\,d\psi+
m\sin^2\frac{\Theta}{2}\,d\varphi
\ee
 so that  
 ${\ccal A}\wedge{\ccal F}=nm\cos^2\frac{\Theta}{2}\sin\Theta\, 
d\psi\wedge d\Theta\wedge d\varphi$.
Inserting this to \eqref{Q} gives the Hopf charge
\be						\label{Hopf-axial}
\Q[\nn]=mn.
\ee
Following Ref.\cite{Sutcliffe:2007ui}, we shall call the fields given by the ansatz 
\eqref{Hopf:axial}  ${\bf A}_{mn}$. 


Let us consider two explicit examples of the ${\bf A}_{mn}$ field. Let us introduce 
toroidal coordinates  $\{u,v,\varphi\}$ such that 
\be                                   \label{toroidal}
\rho=(R/\tau)\,{\sinh u},~~~~
z=(R/\tau)\,{\sin v}, 
\ee
where $\tau=\cosh u-\cos v$ with 
$u\in[0,\infty)$, $v=[0,2\pi)$.
The correct boundary conditions
for the field 
will then be achieved by choosing in the ansatz \eqref{Hopf:axial} 
\be                                   \label{toroidal1}
\Theta=\Theta(u),~~~~\psi=v,
\ee
where $\Theta(0)=0$, $\Theta(\infty)=\pi$.  

Another useful parametrization is achieved 
by expressing $\n^a$ in terms of its complex stereographic projection  coordinate
\be                                   \label{W-Hopf}
W=\frac{\n^1+i\n^2}{1+\n^3}.
\ee 
The values $W=0,\infty$ correspond to the vacuum and to the position curve of the soliton, 
respectively.  Passing to spherical coordinates $\{r,\vartheta,\varphi\}$ one introduces 
\be                                 \label{W-Hopf1}
\phi=\cos \chi(r)+i\sin\chi(r)\cos\vartheta,~~~~\sigma=\sin\chi(r)\sin\vartheta e^{i\varphi},
\ee
 such that $|\phi|^2+|\chi|^2=1$, where $\chi(0)=\pi$, $\chi(\infty)=0$. 
A particular case of the field \eqref{Hopf:axial}
is then obtained by setting
\be                                    \label{WW-Hopf}
W=\frac{(\sigma)^m}{(\phi)^n}.
\ee

\subsection{Topological bound and the $\Q=1,2$ hopfions}

The following inequality for 
the energy \eqref{Faddeev} and Hopf charge \eqref{Q} has been 
established by Vakulenko and Kapitanski 
\cite{Vakulenko:1979uw}, 
\be                                         \label{Hopf} 
{E[\nn]\geq c|\Q[\nn]|^{3/4}},
\ee
where $c=(3/16)^{3/8}$ \cite{Kundu}. Its derivation is non-trivial and proceeds
via considering a sequence of Sobolev inequalities. It is worth noting that a fractional
power of the topological charge occurs in this topological bound, whose value is 
optimal \cite{Lin}. On the other hand, it seems that the value of $c$ can be improved,
that is increased. Ward conjectures \cite{Ward:1998pj} that the bound holds for $c=1$,
which has not been proven but is compatible with all the data available. 
The existence of this bound shows that smooth fields
attaining it, if exist,  describe topologically stable solitons.
Constructing them implies  minimizing the energy \eqref{Faddeev} with 
fixed Hopf charge \eqref{Q}. Such minimum energy configurations 
are sometimes called in the literature Hopf solitons or hopfions, 
and we shall call them fundamental or ground state
hopfions if they have the least possible energy for a given $\Q$. 

Hopfions have been constructed for the first time 
for the lowest two values of the Hopf charge, $\Q=1,2$
by Gladikowski and Hellmund \cite{Gladikowski:1996mb} 
and almost simultaneously (although somewhat more qualitatively) 
by Faddeev and Niemi \cite{Faddeev:1996zj}.  
Gladikowski and Hellmund  
used the axial ansatz \eqref{Hopf:axial} expressed in toroidal 
coordinates, with $\Theta=\Theta(u,v)$ and $\psi=v+\psi_0(u,v)$,
where $\Theta(0,v)=0$, $\Theta(\infty,v)=\pi$. 
Assuming the functions $\Theta(u,v)$ and $\psi_0(u,v)$
to be periodic in $v$, they discretized the variables $u,v$ 
and numerically minimized the discretized expression for the energy with respect 
to the lattice cite values of $\Theta,\psi_0$. They found a smooth minimum energy  
configuration  of the ${\bf A}_{11}$ type for $\Q=1$, 
while for ${\Q}=2$ they obtained two solutions,  ${\bf A}_{21}$
and ${\bf A}_{12}$, the latter being more energetic than the former. 

\begin{figure}[h]
\hbox to\linewidth{\hss%
\resizebox{5cm}{4cm}{\includegraphics{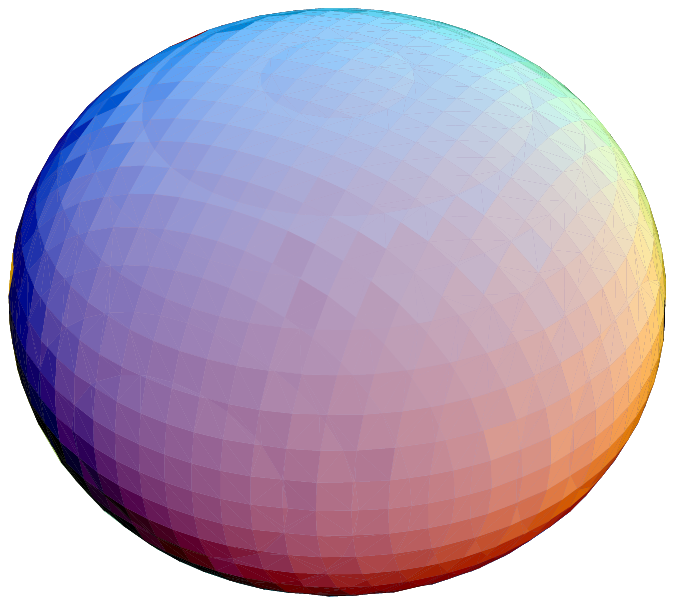}}%
\hspace{5mm}%
\resizebox{5cm}{4cm}{\includegraphics{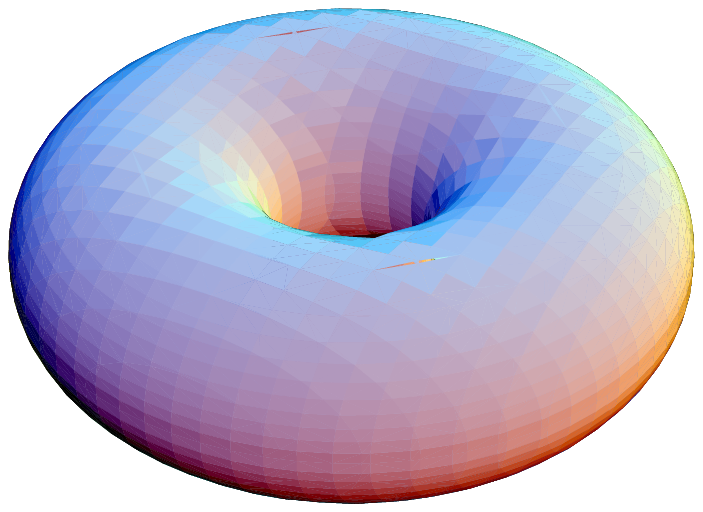}}%
\hss}
\caption{\small The energy isosurfaces for the $\Q=1$ (left) and $\Q=2$ (right)
fundamental hopfions. 
}
\label{FN-en}
\end{figure}

We have verified the results of Gladikowski and Hellmund by integrating the field
equations \eqref{eqs-Hopf} in the static, axially symmetric sector. 
Using the axial ansatz
\eqref{Hopf:axial}, the azimuthal variable $\varphi$ decouples, and the equations reduce to 
two coupled PDE's for $\Theta(\rho,z)$, $\psi(\rho,z)$. Unfortunately, these equations 
are rather complicated and it is not possible to reduce them to ODE's by further separating
variables, via passing to toroidal coordinates, say. 
In fact, we are unaware of any attempts to solve these differential equations. 
We applied therefore our numerical method described below 
in Sec.\ref{sec-vortons} to integrate them, and  
we have succeeded in constructing the first two fundamental hopfions,
${\bf A}_{11}$ and ${\bf A}_{21}$.  For the $\Q=1$
solution the energy density is maximal at the origin and the energy
density isosurfaces are squashed spheres, while for the $\Q=2$ solution
they have toroidal structure (see Fig.\ref{FN-en}). 
For the solutions energies ${E}_{\Q}$ we obtained the 
values ${E}_1=1.22$,  ${E}_2=2.00$.
These features agree with the results of Gladikowski and Hellmund 
and with those of Refs.\cite{Battye:1998pe}, 
\cite{Battye:1998zn}, 
\cite{Sutcliffe:2007ui}, 
\cite{Hietarinta:1998kt}, 
\cite{Hietarinta:2000ci}.
In particular, the same values for the energy are quoted by Ward \cite{Ward:2000qj}.

\begin{figure}[ht]
\hbox to\linewidth{\hss%
	\resizebox{12cm}{2.7cm}{\includegraphics{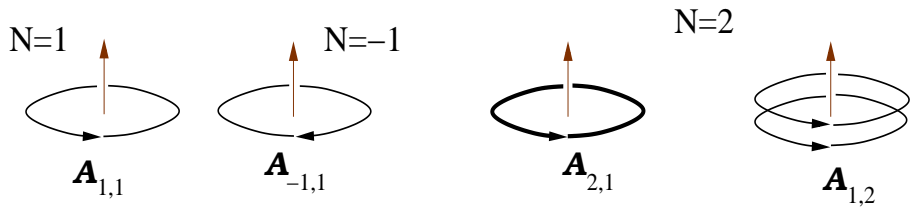}}
\hspace{5mm}%
\hss}
	\caption{Schematic representation of the hopfions as oriented circles}
\label{FigH}
\end{figure}

Ward also proposes a simple analytic approximation of the solutions based on the 
rational ansatz formula \eqref{WW-Hopf} \cite{Ward:2000qj}. For $n=m=1$ this formula gives
\be                                       \label{Hopf3}
W=\frac{x+iy}{f(r)+iz}\,e^{i\alpha}
\ee
where $f(r)=r\cot\chi(r)$. Here the constant phase factor has been introduced
to account for the residual O(2) symmetry \eqref{residual}. 
Minimizing the energy with respect $F(r)$ 
it turns out that choosing  
\be                                        \label{Hopf4}
f(r)=0.453\,(r-0.878)(r^2+0.705r+1.415)
\ee
gives for 
the energy  a value which is 
less than $1\%$ above the true minimum, ${E}_1=1.22$ \cite{Ward:2000qj}. 
Eqs.\eqref{Hopf3},\eqref{Hopf4} provide therefore a good analytic approximation
for the $\Q=1$ hopfion. They show, in particular, 
that the position curve of the hopfion
is a circle of radius $R=0.848$ and that for large $r$ the field shows a dipole type 
behaviour, 
\be
\n^1+i\n^2\approx 2W \approx  a\,\frac{x+iy}{r^3}. 
\ee 
Ward also suggests the moduli space approximation in which the $|\Q|=1$ 
hopfion is viewed as an oriented circle (see Fig.\ref{FigH}). 
There are six continuous moduli parameters: 
three coordinates of the circle center, two angles determining the position 
of the circle axis, and also the overall phase.
Choosing arbitrarily a direction along the axis (shown by the vertical arrow in 
Fig.\ref{FigH}), there are two possible
orientations corresponding to the sign of the 
Hopf charge, changing which is achieved by $W\to W^\ast$.  
 Although for one hopfion the phase is not important, 
the relative phases of several hopfions determine their interactions. 

Ward conjectures  that well-separated hopfions interact as dipoles with the 
maximal attraction/repulsion when they are parallel/antiparallel,
respectively, since the like charges attract in a scalar field theory. 
He verifies this conjecture numerically, and then numerically relaxes
a field configuration corresponding to two mutually attracting hopfions.  
He discovers two possible outcomes of this process. If the two hopfions are initially
located in one plane 
then they approach each other till the two circles merge to one
thereby forming the ${\bf A}_{2,1}$ hopfion with $\Q=2$.  
If they are initially
oriented along the same line then they approach each other 
but do not merge even in the energy minimum, 
where they remain separated by a finite distance. 
This corresponds to the   
${\bf A}_{1,2}$ hopfion, which is more energetic but locally stable. 

The ${\bf A}_{2,1}$ hopfion can be approximated by 
\be                                       \label{Hopf5}
W=\frac{(x+iy)^2 e^{i\alpha}}{f+i1.55\,z\,r},~~~
f=0.23(r-1.27)(r+0.44)(r+0.16)(r^2-2.15r+5.09)
\ee
whose energy is  $1.5\%$ above the true minimum, and there is also 
a similar approximation for the  ${\bf A}_{1,2}$ hopfion
\cite{Ward:2000qj}. 
 
\subsection{Unknots, links and knots}

Similarly to the $\Q=1,2$ solutions, 
hopfions can be constructed within the axial ansatz \eqref{Hopf:axial} 
also for higher values of $n,m$. However, for $\Q>2$ they will not 
generically correspond
to the {\it global} energy minima. This can be understood 
if we remember that the Hopf charge
measures the amount of the twist deformation. 
Twisting an 
elastic rod shows that if it is twisted too much then the loop made of it
will not be planar, since it will find it energetically favorable to bend 
toward  the third direction. It is therefore expected that 
the ground state hopfions for higher values of $\Q$ 
will not be axially symmetric planar
loops, but 3D loops, generically without  any symmetries. 

The interest towards this issue was largely stirred by the work of Faddeev and Niemi
\cite{Faddeev:1996zj}, \cite{Faddeev:1997pf},
who conjectured that higher $\Q$ hopfions should be not just
closed lines but {\it knotted} closed lines, with the degree 
of knotedness expressed in terms 
of the Hopf charge. In other words, they conjectured that there could be 
a field theory realization of stable knots -- the idea that had been put forward by 
Lord Kelvin in the 19-th century \cite{Kelvin}  
but never found an actual realization. 
This knot conjecture of Faddeev and Niemi had 
a large resonance and several groups had started large scale numerical simulations 
to look for knotted hopfions. Such solutions have indeed been found, 
although not with quite the same properties as had been originally predicted in 
Ref.\cite{Faddeev:1996zj}, 
\cite{Faddeev:1997pf}. 

The first, really astonishing set of results has been reported by Battye and Sutcliffe
\cite{Battye:1998pe}, 
\cite{Battye:1998zn},
who managed to construct hopfions up to $\Q=8$ and found the first non-trivial knot --
the trefoil knot -- for $\Q=7$. 
Similar analyses have been then  
independently carried out by 
Hietarinta and Salo 
\cite{Hietarinta:1998kt}, 
\cite{Hietarinta:2000ci} 
and also by Ward 
\cite{Ward:1998pj},
\cite{Ward:2000qj},
\cite{Ward}.
All groups performed the full 3D energy minimization starting from an initial 
field configuration with a given $\Q$. 
Various initial configurations were used,  as for example the ones 
given by  the rational ansatz \eqref{WW-Hopf}, supplemented by non-axially 
symmetric perturbations to break the exact toroidal symmetry. 
The value of $\Q$ being constant during 
the relaxation, the numerical iterations were found to converge  to non-trivial 
energy minima,
whose structure was sometimes completely different from that of 
the initial configuration (an online animation of the relaxation process  
is available in \cite{movies}). 
Several local energy minima typically exist  
for a given $\Q$,  
sometimes with almost the same energy,  so that it was not always easy to know 
whether the minima obtained were local or global. Different initial configurations
were therefore tried to see if the minimum energy configurations could be reproduced
in a different way.  
As a result, it appears that the global energy minima have now been identified
and cross-checked up to $\Q=7$ \cite{Hietarinta:2000ci}, \cite{Sutcliffe:2007ui}, 
after which the analysis has been extended up to $\Q=16$ \cite{Sutcliffe:2007ui}.
The properties of the solutions can be summarized as follows.  

For $\Q=1,2$ these are the toroidal hopfions of  Gladikowski and Hellmund  
\cite{Gladikowski:1996mb}, ${\bf A}_{11}$ and ${\bf A}_{21}$. 
Although initially obtained within the constrained, axially symmetric 
relaxation scheme, they also correspond to the global minima of the full 3D energy 
functional. Axially symmetric hopfions ${\bf A}_{m1}$  
exist also for higher $\Q=m$ \cite{Battye:1998pe}, 
\cite{Battye:1998zn}, but they no longer correspond to global energy minima. 
For $\Q=3$ the ground state hopfion is not planar and is called in \cite{Sutcliffe:2007ui}
$\tilde{{\bf A}}_{31}$,
which can be viewed as deformed ${\bf A}_{31}$,
with a pretzel-like position curve bent in 3D to brake the axial symmetry. 
However, for $\Q=4$ the axial symmetry is restored again in the ground state,
${\bf A}_{22}$, which seems to have a similar to the ${\bf A}_{12}$ two-ring structure 
\cite{Hietarinta:2000ci}, \cite{Ward}. 
The bent $\tilde{{\bf A}}_{41}$ also exists, 
but its energy it higher.

\begin{figure}[ht]
\hbox to\linewidth{\hss%
	\resizebox{10cm}{2cm}{\includegraphics{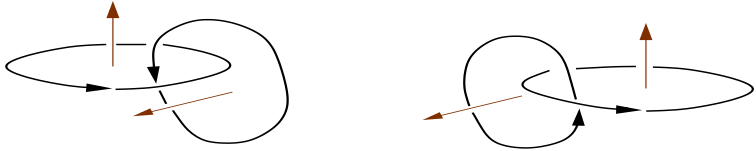}}
\hspace{5mm}%
\hss}
	\caption{Two possible ways to link $\Q=1$ hopfions. Left: the total linking
number is $2=1+1$ and the total charge if $\Q=1+1+2=4$.
Right: the total linking number is $-2=-1-1$ and the total charge is 
$\Q=1+1-2=0$.  
The numerical relaxation of these configurations gives therefore completely
different results \cite{movies}.}
\label{FigL}
\end{figure}

Up to now all hopfions have been the simplest knots
topologically equivalent to a circle, called {\it unknots}
in the knot classification. 
A new phenomenon arises for $\Q=5$, 
since the fundamental hopfion in this case consists of two 
linked unknots. This has nothing to do with the linking of 
preimages determining  the value of $\Q$.
This time the position curve itself consists of two disjoint loops,
corresponding to  a charge 2 unknot linked to a charge 1 unknot.  
The Hopf charge is not simply the sum of charges of each component, 
but contains in addition the sum of their linking numbers 
due to their linking with the other components. It is worth noting that 
the linking number of the oriented circles can be positive or negative, 
depending on how they are linked \cite{Hietarinta:1998kt} (see Fig.\ref{FigL}).  
Using the notation of Ref.\cite{Sutcliffe:2007ui},
the $\Q=5$ hopfion can be called 
${\bf L}_{1,2}^{1,1}$,  
where the subscripts label the Hopf charges of the unknot components of the link,
and the superscript above each subscript counts the extra linking number of that component.
The total Hopf charge is the sum of the 
subscripts plus superscripts. Similarly, for $\Q=6$ the ground state hopfion is 
a link of two charge 2 unknots, so that it can be called ${\bf L}_{2,2}^{1,1}$. 

For $\Q=7$ the true knot appears at last: the ground state configuration 
corresponds in this case to the simplest non-trivial torus knot: trefoil knot. 
Let us remind that a $(p,q)$ torus knot is formed by wrapping a circle around 
a torus $p$ times in one direction and $q$ times in the other, 
where $p$ and $q$ are coprime integers, $p>q$. One can explicitly parametrize it as
$\rho = R+\cos\left({p\varphi}/{q}\right)$,
$z = \sin\left({p\varphi}/{q}\right)$, where $R>1$. 
A $(p,q)$ torus knots can also be obtained as the intersection of the unit three sphere
$S^2\in\mathbb{C}^2$ defined by $|\phi|^2+|\sigma|^2=1$ with the complex algebraic curve
$\sigma^p+\phi^q=0$. The trefoil knot is the $(3,2)$ torus knot, and it determines the 
profile of the position curve of the $\Q=7$ fundamental hopfion denoted 
$7{\bf K}_{3,2}$ in Ref.\cite{Sutcliffe:2007ui}. 

Sutcliffe \cite{Sutcliffe:2007ui} extends the energy minimization up to $\Q= 16$
specially 
looking for other knots. For the input configurations in his numerical procedure  
he uses fields parametrized
by the rational map ansatz,
\be                                \label{pq}
W=\frac{\sigma^a \phi^b}{\sigma^p+\phi^q},
\ee
where $0<a\in\mathbb{Z}$, $0\leq b\in\mathbb{Z}$ and $W,\phi,\sigma$ are defined by 
Eqs.\eqref{W-Hopf},\eqref{W-Hopf1}. The position curve for such a field configuration 
coincides with the $(p,q)$ torus knot, since the condition $W=\infty$ reproduces the 
knot equation. The parameters $a,b$ determine the value of the 
Hopf charge, $\Q=aq+bp$ \cite{Sutcliffe:2007ui}. If $p,q$ are not coprime, then the 
denominator in \eqref{pq} factorizes and the whole expression describes
a link. 
As a result, fixing a value of $\Q$ one can construct many different knot or link
configurations compatible with this value. Numerically relaxing these 
configurations, Sutcliffe finds many new energy minima,
discovering  new knots and links \cite{Sutcliffe:2007ui}. 
He also obtains configurations that he calls $\chi$  
whose position curve seems to self-intersect and so it is not quite clear to what
type they belong, unless the self-intersections are only apparent and can be resolved
by increasing the resolution.

\begin{table} 
\caption{Known Hopf solitons 
according to Refs. 
\cite{Hietarinta:2000ci}, \cite{Sutcliffe:2007ui}, \cite{Battye:1998zn}. }
\label{tab1}
\begin{tabular}{c|c|c|c|c|c|c|c|c|c} \hline\hline
$\Q$ 
& {1} 
& {2} 
& 2 
& {3} 
& 3 
& {4} 
& 4 
& 4 
& {5} 
\\    
& $\underline{{\bf A}_{1,1}}$ 
& \underline{${\bf A}_{2,1}$}  
&  ${{\bf A}}_{1,2}$ 
& \underline{$\tilde{{\bf A}}_{3,1}$} 
& ${{\bf A}}_{3,1}$   
& \underline{${\bf A}_{2,2}$} 
& $\tilde{{\bf A}}_{4,1}$ 
& ${{\bf A}}_{4,1}$ 
& \underline{${\bf L}_{1,2}^{1,1}$ } 
\\ 
${\mcal E}$ 
& {1} 
& {0.97 }
& 0.98 
& {1.00 }
& 1.01 
& {1.01 }
& 1.03 
& 1.06 
& {1.02 }
\\   \hline       
$\Q$ 
& 5 
& 5 
& {6} 
& 6 
& 6 
& {7} 
& 7 
& {8} 
& 8 
\\    
& $\tilde{{\bf A}}_{5,1}$  
& ${{\bf A}}_{5,1}$  
& \underline{${\bf L}_{2,2}^{1,1}$}  
& ${\bf L}_{1,3}^{1,1}$ 
& ${\bf A}_{6,1}$ 
 & \underline{${\bf K}_{3,2}$ } 
& ${\bf A}_{7,1}$      
& \underline{${\bf L}_{3,3}^{1,1}$  }
& ${\bf K}_{3,2}$ 
\\ 
${\mcal E}$ 
& 1.06 
&  1.17 
& {1.01} 
& 1.09 
& 1.22 
& {1.01 } 
& 1.20
& { 1.02 }
& 1.02 
\\     \hline   
$\Q$ 
& 8 
& {9} 
& 9 
& {10} 
& 10 
& 10 
& {11} 
& 11 
& 11 
\\    
& ${\bf A}_{8,1}$ 
& \underline{ ${\bf L}_{1,1,1}^{2,2,2}$ }
& ${\bf K}_{3,2}$ 
& \underline{ ${\bf L}_{1,1,2}^{2,2,2}$ }
& ${\bf L}_{3,3}^{2,2}$   
& ${\bf K}_{3,2}$ 
& \underline{ ${\bf L}_{1,2,2}^{2,2,2}$ }
& ${\bf K}_{5,2}$  
& ${\bf L}_{3,4}^{2,2}$
\\ 
${\mcal E}$ 
& 1.40 
& { 1.02 }
& 1.02 
& { 1.02 }
& 1.02 
& 1.03 
& { 1.02 }
& 1.03 
& 1.04 
\\     \hline   
$\Q$ 
& 11 
& 12 
& 12 
& 12 
& 12 
& {13} 
& 13 
& 13 
& 13 
\\    
& ${\bf K}_{3,2}$                
& \underline{ ${\bf L}_{2,2,2}^{2,2,2}$ }
& ${\bf K}_{4,3}$ 
& ${\bf K}_{5,2}$ 
& ${\bf L}_{4,4}^{2,2}$
& \underline{  ${\bf K}_{4,3}$ }
&  ${\chi}_{13}$
&  ${\bf K}_{5,2}$ 
&  ${\bf L}_{3,4}^{3,3}$  
\\ 
${\mcal E}$ 
& 1.05           
& { 1.01 }
& 1.01 
& 1.04 
& 1.04 
& { 1.00 }
& 1.03 
& 1.04 
& 1.05 
\\     \hline   
$\Q$ 
& {14} 
& 14 
& 14 
& {15} 
& 15 
& 15   
& 16
\\    
& \underline{  ${\bf K}_{4,3}$ }
&  ${\bf K}_{5,3}$  
&  ${\bf K}_{5,2}$ 
&  \underline{ ${\mathbf \chi}_{15}$ }
&  ${\bf L}_{1,1,1}^{4,4,4}$ 
&  ${\bf K}_{5,3}$ 
&  \underline{$\chi_{16}$}    
\\ 
${\mcal E}$ 
& { 1.00 }
& 1.01 
& 1.05 
& { 1.01 }
& 1.02 
& 1.02 
& { 1.01 } 
\\   \hline      \hline   
 \end{tabular}
\end{table}

\begin{figure}[h]
\hbox to\linewidth{\hss%
  \resizebox{14cm}{6cm}{\includegraphics{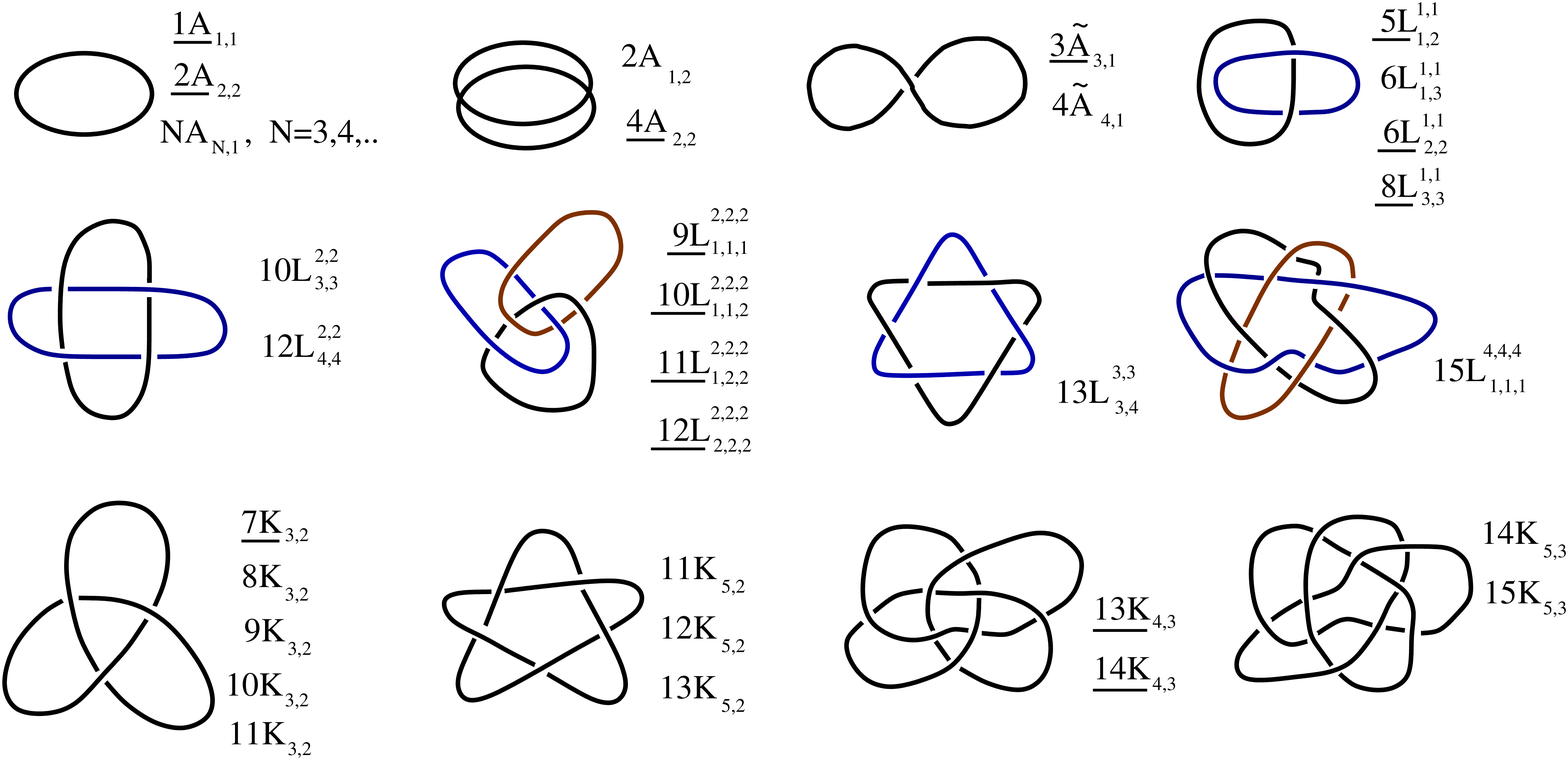}}%
\hss}
\caption{
Schematic profiles of the position curves for the known hopfions 
(excepting the 
$\chi$-solutions) 
according to the results of Refs.\cite{Hietarinta:2000ci},\cite{Sutcliffe:2007ui}. 
}
\label{Fig:knots}
\end{figure}

The properties of all known hopfions, according to the results of
Refs. \cite{Hietarinta:2000ci}, \cite{Sutcliffe:2007ui}, \cite{Battye:1998zn}, 
are summarized 
in Table \ref{tab1} and in Fig.\ref{Fig:knots}.  
Table \ref{tab1} shows the Hopf charge, 
the type of the solution, with the ground state configuration 
in each topological sector underlined,
and also the relative energy ${\mcal E}$ defined by the relation 
\be                              \label{reden}
E_\Q/E_1={\mcal E}\Q^{3/4},
\ee 
where $E_1$ is the energy of the $\Q=1$ hopfion. 
Of the two decimal places of values of ${\mcal E}$ shown in the table the second one 
is rounded.  
Different groups give slightly different values for the energy, 
but one can expect the relative energy to be less sensitive to this.  
The values of ${\mcal E}$ for $\Q\leq 7$ shown in the table correspond to the data 
of Hietarinta and Salo \cite{Hietarinta:2000ci} and of Sutcliffe 
\cite{Sutcliffe:2007ui}, and it appears  that for  solutions described by both of these 
groups these values are the same. The data for $8\leq \Q\leq 16$ are given 
by Sutcliffe \cite{Sutcliffe:2007ui}, apart from  those for the ${\bf A}_{\Q,1}$ hopfions
for $\Q=5,6,7,8$, which are found in the earlier  work of Battye and 
Sutcliffe \cite{Battye:1998zn}.
Although ${\bf A}_{\Q,1}$ solutions seem to exist
also for $\Q>8$, no data are currently available for this case. 
To obtain the energies from Eq.\eqref{reden} one can use the value $E_1=1.22$,
which is known to be accurate to the two decimal places \cite{Ward:2000qj}. 

As one can see, the hopfion energies follow closely the topological lower 
bound \eqref{Hopf}.
This suggests that the ground state hopfions actually attain this 
bound, so that they should be topologically stable. 
According to the 
data in Table \ref{tab1} one has $\inf\{{\mcal E}\}=0.97$ for $\Q\leq 16$, 
and if this is true  
for all values of $\Q$ then the optimal value for the constant 
in the bound \eqref{Hopf} is $c=E_1\inf\{{\mcal E}\}=1.18$. 

A rigorous existence proof for the hopfions was given by Lin and Yang \cite{Lin}, 
who demonstrate the existence of a smooth least energy configuration in every
topological sector whose Hopf charge value belongs to an infinite (but unspecified)
subset of $\mathbb{Z}$. 
This shows that ground state hopfions exist, although perhaps 
not for any $\Q\in\mathbb{Z}$.   
As shown in Ref.\cite{Lin}, their {energy} 
is bounded not only below but also {\it above} as
\be
E<C|\Q|^{3/4},
\ee
where $C$ is an absolute constant. This  
implies that knotted solitons are energetically preferred 
over widely separated unknotted multisoliton configurations when $\Q$
is sufficiently large. Indeed, for a 
decay into charge one elementary hopfions
the energy should grow at least as $\Q$ for large $\Q$, but it grows slower.

The position curves of the solutions, schematically shown in Fig.\ref{Fig:knots},
present an amazing variety of shapes. It should be stressed though that other 
characteristics of the solutions, as for example their energy density, do not 
necessarily show the same knotted pattern. 
Several types of links and 
knots appear, and  each particular type can appear several times, for different
values of the Hopf charge. Intuitively, one can view the position curves as wisps 
made of two intertwined lines corresponding to preimages of two infinitely close to 
$-\nn_\infty$ points on the target space \cite{Sutcliffe:2007ui}. Increasing the twist
of the wisp increases the Hopf charge, without necessarily 
changing the topology of the position curve. A more detailed inspection 
(see pictures in \cite{Sutcliffe:2007ui}) actually shows that configurations
appearing several times in Fig.\ref{Fig:knots}, as for example the trefoil knot 
${\bf K}_{3,2}$, become more and more distorted by the internal twist as $\Q$ increases. 
Finally, for some critical value of $\Q$, the excess of the intrinsic 
deformation makes it energetically favorable to change 
the knot/link type and pass to other, more complicated knot/link configurations. 
Estimating the length of the position curves shows that it grows as $\Q^{3/4}$,
so that the energy per unit length is approximately the same for all hopfions
\cite{Sutcliffe:2007ui}.

\subsection{Conformally invariant knots} 

Hopf solitons in the Faddeev-Skyrme model, also  
sometimes called in the literature  
knot solitons of Faddeev-Niemi,  
of Faddeev-Skyrme, or of Faddeev-Hopf
provide the best known
example of knot solitons in field theory.
However, there are also other field theory models admitting knotted solitons
with a non-zero Hopf index.
An interesting example proposed by Nicole \cite{Nicole} is obtained
by taking the first term in the  Faddeev-Skyrme model \eqref{action0} and
raising it to a fractional power,
\be                                      \label{Nicole}
{\mcal L}_{\rm Nicole}=\left(-\partial_\mu\nn\cdot\partial^\mu\nn\right)^{3/2}\,.
\ee 
A similar possibility, suggested by Aratyn, Ferreira and Zimerman (AFZ) 
\cite{Aratyn}, uses the second term in the  Faddeev-Skyrme Lagrangian,
\be                                      \label{AFZ}
{\mcal L}_{\rm AFZ}=\left({\mcal F}_{\mu\nu}{\mcal F}^{\mu\nu}\right)^{3/4}\,.
\ee 
Both of these models are conformally invariant in three spatial dimensions
and so the existence of static solitons is not excluded for them by the Derrick argument. 
In fact, static knot solitons in these models exist and can even be
obtained in close analytical form, which is achieved by simply using the 
axial ansatz \eqref{Hopf:axial} with the function $\Theta,\psi$ expressed 
in toroidal coordinates \eqref{toroidal} according to Eq.\eqref{toroidal1}
\cite{Aratyn:1999cf}, \cite{Adam:2006wg}. 
Curiously, this separates away the $v,\varphi$ 
variables in the field equations reducing the problem
to an {\it ordinary} differential equation for $\Theta(u)$ 
(in the Faddeev-Skyrme theory this does not work). In the AFZ model
solutions for $\Theta(u)$ can be expressed in terms of elementary functions 
for any $n,m$ \cite{Aratyn:1999cf}, while in the Nicole model they are obtained 
numerically \cite{Adam:2006wg}, 
apart from the $|n|=|m|=1$ case, where the solution turns out to be
 the same in both 
models  and is given by 
$\tan(\Theta/2)=\sinh u$ 
\cite{Nicole}, 
\cite{Aratyn}.  
These results remain, however,  interesting mainly from the purely
mathematical point of view at the time being, since it is difficult 
to justify physically 
the appearance of the fractional powers in Eqs.\eqref{Nicole},\eqref{AFZ}. 

Other examples of solitons with a Hopf charge 
will be discussed below in Sec.\ref{o} and Sec.\ref{magnet}.

\subsection{Can one gauge the knot solitons ?}

The Faddeev-Skyrme theory is a {\it global} field model, 
so that it cannot be a fundamental physical theory like gauge field theory models, 
but perhaps can be viewed as an effective theory. This suggests using the Faddeev-Skyrme 
knots for an effective description of some physical objects, and so it has
been conjectured by Faddeev and Niemi 
that they could be used for an effective  description 
of glueballs in  the strongly coupled Yang-Mills theory 
\cite{Faddeev:1998eq},
\cite{Faddeev:2006sw},
\cite{Shabanov:1999xy},
\cite{Shabanov:1999uv},
\cite{Wipf},
\cite{Faddeev:2003aw}. 
This conjecture is very interesting, quite in the spirit of the original
Lord Kelvin's idea to view atoms as knotted ether tubes 
\cite{Kelvin}, and perhaps it could apply in some form. In fact,  
when describing the $\eta(1440)$ meson, 
the Particle Data Group says  (see p.591 in Ref.\cite{Yao}) 
that ``the $\eta(1440)$ is an excellent candidate 
for the $0^{-+}$ glueball in the flux tube model \cite{Faddeev:2003aw}''. 

Some other physical applications of the global field theory knot solitons 
could perhaps be found. However, if they could be promoted to  {\it gauge} field theory
solutions, then they would be much more interesting physically, since in this case 
they would find many interesting applications, as for example in the theories of 
superconductivity and  of Bose-Einstein condensation
\cite{Babaev:2001jt},
\cite{Babaev:2001zy},
in the theory of plasma 
\cite{Faddeev:2000rp}, 
\cite{Faddeev:2000qw}, 
in Standard Model
\cite{Cho:2001gc},
\cite{Fayzullaev:2004xa},
\cite{Niemi:2000ny},
or perhaps even in cosmology, where they could presumably 
describe knotted cosmic strings \cite{Vilenkin}. 
For this reason it has been repeatedly conjectured in the literature 
that some analogs of the Faddeev-Skyrme knot solitons could also exist 
as static solutions   
of the gauge field theory equations of motion. This conjecture 
is essentially inspired
by the fact that the Faddeev-Skyrme theory already contains 
something like a gauge field:
${\mcal F}_{\mu\nu}$. Moreover, we shall now see that 
changing the variables one can rewrite the 
theory in such a form that it looks almost identical to a gauge field theory 
(or the other way round).

\subsection{Faddeev-Skyrme model versus semilocal Abelian Higgs model}

Let $\Phi$ be a doublet of complex scalar fields satisfying a constraint, 
\be                                                    \label{norm}
\Phi=
\left(\begin{array}{c}
\phi\\
\sigma\\
\end{array}
\right),~~~~~~~~~
\Phi^\dagger\Phi=|\phi|^2+|\sigma|^2=1,
\ee
such that $\Phi\in S^3$. 
Let us consider a  field theory defined by the Lagrangian density 
\be                                                        \label{action1}
{\mcal L}[\Phi]=-\frac{1}{4}\,{\ccal F}_{\mu\nu}{\ccal F}^{\mu\nu}
+({\ccal D}_\mu\Phi)^\dagger {\ccal D}^\mu\Phi
\ee
with 
${\ccal F}_{\mu\nu}=\partial_\mu {\ccal A}_\nu-\partial_\nu {\ccal A}_\mu$
and ${\ccal D}_\mu\Phi=(\partial_\mu-i{\ccal A}_\mu)\Phi$ and with
\be                                                     \label{A} 
{\ccal A}_\mu=-i\Phi^\dagger\partial_\mu\Phi\,.
\ee 
In fact, this  theory is again the  Faddeev-Skyrme model 
but rewritten in different variables, 
since upon the identification  
\be                                             \label{project}
n^a=\Phi^\dagger\tau^a\Phi
\ee 
($\tau^a$ being
the Pauli matrices) the fields ${\ccal A_\mu}$ and 
${\ccal F}_{\mu\nu}$ coincide with those in 
\eqref{action0} \cite{Ren} and the whole action 
\eqref{action1} reduces (up to an overall factor) 
to \eqref{action0}  
\cite{Faddeev:1997pf}. 
More precisely, \eqref{project} is the Hopf projection from $S^3$ 
parametrized by $(\phi,\sigma)$  
to $S^2$ parametrized by the complex projective coordinate $\phi/\sigma$.
For example, the axially symmetric fields \eqref{Hopf:axial}, \eqref{Hopf:axial1}
are obtained in this way by choosing the $CP^1$ variables 
\be                                        \label{axial-CP}
\phi=\cos\frac{\Theta}{2}\,e^{in\psi},~~~~~
\sigma=\sin\frac{\Theta}{2}\,e^{im\varphi},
\ee
such that the phases of $\phi$ and $\sigma$ wind, respectively, along the
two orthogonal direction  as shown in Fig.\ref{FigS} and the Hopf charge is 
$\Q=nm$.

The fields in the static limit, $\Phi=\Phi(\bx)$, can now
be viewed 
as maps $S^3\to S^3$, but their energy 
\be                                                                    \label{en1}
E[\Phi]=\int\left( |{\ccal D}_k\Phi|^2+
\frac{1}{4}\,({\ccal F}_{ik})^2
\right)
d^3 \bx 
\ee
is still bounded from below as in Eq.\eqref{Hopf}.
The  topological charge $\Q=\Q[\Phi]$, 
still expressed by Eq.\eqref{Q}, 
is now interpreted as the index 
of map $S^3\to S^3$,
 $\Q\in\pi_3(S^3)=\mathbb{Z}$. 
The theory  therefore  
admits the same knot solitons as the original Faddeev-Skyrme model. 
However, in the 
new parametrization the theory looks almost like a gauge field theory,
in particular it exhibits a local 
U(1) gauge invariance under $\Phi\to e^{i\alpha}\Phi$, 
$A_\mu\to A_\mu+\partial_\mu\alpha$.

Let us now compare the model \eqref{action1} to a genuine gauge field theory 
with the Lagrangian  density 
\be                                                           \label{action2}
{\mcal L}[\Phi,A_\mu]=-\frac{1}{4}\,F_{\mu\nu}F^{\mu\nu}
+(D_\mu\Phi)^\dagger D^\mu\Phi-\frac{\lambda}{4}\,(\Phi^\dagger\Phi-1)^2\,.
\ee
Here $\Phi$ is again a doublet of complex scalar fields, but this time 
without the normalization condition \eqref{norm}, the condition  \eqref{A}
being also relaxed, so that $A_\mu$ is now an independent field. One has 
${F}_{\mu\nu}=\partial_\mu {A}_\nu-\partial_\nu {A}_\mu$
and ${D}_\mu\Phi=(\partial_\mu-i{A}_\mu)\Phi$. 
This semilocal \cite{Achucarro:1999it}
Abelian Higgs model with the SU(2)$_{\rm global}\times$U(1)$_{\rm local}$
internal symmetry arises in different contexts, in particular it can be viewed 
as the Weinberg-Salam model in the limit where the weak mixing angle is $\pi/2$
and the SU(2) gauge field decouples. 
The non-relativistic limit of this theory
is the two-component Ginzburg-Landau model \cite{Ginzburg}. 

Let us now consider the limit $\lambda\to\infty$. In this sigma model limit the
constraint \eqref{norm} is enforced and the potential term in \eqref{action2} vanishes.  
The theories \eqref{action1} and \eqref{action2} then look identically the same,
the only difference being that in the first case the vector field ${\ccal A}_\mu$ 
is defined by Eq.\eqref{A} and so is composite, 
while in the second case $A_\mu$ in an independent field. 

\subsection{Energy bound in the Abelian Higgs model}

The question now arises: does the gauge field theory \eqref{action2},
at least in the limit where $\Phi^\dagger\Phi=1$, admit 
knot solitons analogues to those of the global model \eqref{action1} ? 
If exist, such solutions would correspond to minima 
of the energy in the theory \eqref{action2} 
in static, purely magnetic sector,  
\be                                                                    \label{en2}
E[\Phi,A_k]=\int\left( |{D}_k\Phi|^2+
\frac{1}{4}\,({F}_{ik})^2
\right)
d^3 \bx \,.
\ee
At first thought, one may think that the answer to this question should be affirmative. 
Indeed, the energy functionals \eqref{en1} and \eqref{en2} look identical. 
They have the same internal symmetries and the same scaling behaviour under
$\bx\to \Lambda \bx$. The two theories also have the same topology 
associated to the field $\Phi$,
since in both cases $\Phi(\bx)$ defines a 
mapping $S^3\to S^3$ with the topological 
index \eqref{Q}. 

The gauged model \eqref{en2} contains in fact even more 
charges than the global theory \eqref{en1}, 
since it actually has two vector fields: the independent gauge field 
$A_k$  and the composite field 
${\ccal A}_k=\frac{i}{2}(\partial_k\Phi^\dagger\Phi-\Phi^\dagger\partial_k\Phi)$. 
It is convenient to introduce their difference
$C_k=A_k-{\ccal A}_k$. 
Defining the linking number between two vector fields, 
\be
I[A,B]=\frac{1}{4\pi^2}\int \epsilon_{ijk}{A}_i \partial_j{B}_k\, d^3\bx\,,  
\ee
one can construct three different charges,
\be                             \label{charges}
\Q[\Phi]\equiv I[{\ccal A},{\ccal A}],~~~~~
{\rm L}=I[C,{A}],~~~~~
N_{\rm CS}[A]\equiv I[{A},{A}],
\ee
which are, respectively, the topological charge \eqref{Q}, 
the linking number between $A_k$ and $C_k$, 
and the Chern-Simons number of the gauge field.
The following 
inequality, established by Protogenov and Verbus \cite{Protogenov:2002bt},
holds:
\be                                                           \label{Proto} 
E[\Phi,A_k]\geq c_1|\Q|^{3/4}\left(1-\frac{|{\rm L}|}{|\Q|}\right)^2\,,
\ee
where $c_1$ is a positive constant. This can be considered 
as the generalization of the 
topological bound \eqref{Hopf} of Vakulenko-Kapitanski. 

Given all these, one may believe that the local  
 theory \eqref{en2} admits topologically 
stable knot solitons similar to those of the global theory \eqref{en1}.

\subsection{The issue of charge fixing}

The difficulty with implementing the Protogenov-Verbus  bound 
\eqref{Proto} in practice 
is that it contains two different charges, $\Q$ and L,
but without invoking additional 
physical assumptions  there is no reason why L should be fixed while minimizing 
the energy.  

Let us consider first the charge $\Q=\Q[\Phi]$. Its variation vanishes identically,
so that it 
does not change under smooth deformations of $\Phi$. It is therefore a genuine  
topological charge whose value is completely determined by the boundary 
conditions, and so it should be fixed when minimizing the energy. 

Let us now consider the linking number L=$I[C,A]$. The analogs of this quantity have been
studied in the theory of fluids, where they are known to be integrals 
of motion \cite{Zakharov}. In the context of gauge field theory,
this quantity is gauge invariant. However, it is not 
a topological invariant,
since its variation does not vanish identically and so it does change 
under arbitrary smooth deformations of $C_k=A_k-{\ccal A}_k$.
On cannot fix L using only continuity arguments,
because there are no topological conditions imposed on  
$A_k$, whose  arbitrary deformations are allowed.  
The only topological quantity associated to $A_k$ 
could be a magnetic charge related to a non-trivial U(1) bundle structure.
However, since we are interested in globally regular solutions, 
the bundle base space
is $\mathbb{R}^3$ (or $S^3$) without removed points, 
in which case the bundle is trivial.  

As a result,  on continuity grounds only,
one can fix $\Q$  but not L. But then, 
as is obvious from Eq.\eqref{Proto},  there is no 
non-trivial lower bound for the energy, since one can always choose L=$\Q$
in which case the expression on the right in \eqref{Proto} vanishes. 
More precisely, since there are no constraints  for $A_k$, 
nothing prevents one from smoothly deforming it to zero, 
after which one can scale away the 
rest of the configuration. Explicitly, given fields $\Phi,A_k$ 
one can reduce $E[\Phi,A_k]$ to {\it zero} via a continues 
sequence of smooth field deformations
preserving the value of the  topological charge  $\Q[\Phi]$,  
\be
E[\Phi(\bx),A_k(\bx)]\to E[\Phi(\Lambda \bx),\gamma A_k(\bx)]
\ee
by taking first the limit $\gamma\to 0$ and then $\Lambda\to \infty$ \cite{Forgacs}. 

The conclusion is that without constraining $A_k$, with only $\Q[\Phi]$ fixed, 
the absolute minimum of $E[\Phi,A_k]$ is zero, so that 
there can be no absolutely stable knots. To have such solutions, 
one would need to constraint
somehow the vector field $A_k$, for example 
 it would be enough
to insure that $C_k=A_k-{\ccal A}_k$ be 
zero or small.
Such a condition 
is often assumed in the literature 
\cite{Babaev:2001zy},
\cite{Babaev:2001jt},
\cite{Cho:2001gc},
\cite{Fayzullaev:2004xa},
but usually simply {\it ad hoc}. 
Unfortunately, it cannot be justified
on continuity grounds only, without an additional physical input.

\subsection{Searching for gauged knots} 

The above arguments do not rule out
 all solutions in the theory. Even though the 
global minimum of the energy is zero, there could still 
be non-trivial local minima or saddle points. 
The corresponding static solutions would  be metastable or unstable. 
One can  therefore wonder whether  such solutions exist. This question has 
actually a long 
history, being first addressed by de Vega \cite{deVega:1977rk} and by 
Huang and Tipton \cite{Huang:1980bz} over 30 years ago,
and being then repeatedly reconsidered by different authors 
\cite{Koma:1999sm},
\cite{Niemi:2000ny},
\cite{Forgacs},
\cite{Ward:2002vq},
\cite{Jaykka:2006gf},
\cite{Doudoulakis:2006iw},
\cite{Doudoulakis:2007ti},
\cite{Doudoulakis:2007xz}.
However, the  answer is still unknown. 
No solutions have been found up to now, neither has it been shown 
that they do not exist. 

Trying to find the answer, all the authors were minimizing  the energy 
given by the sum of $E[\Phi,A_k]$ 
and of a potential term that can be either of the 
form contained in \eqref{action2} 
or  a more general one. A theory 
with two vector fields with a local U(1)$\times$U(1) 
invariance -- Witten's model of superconducting strings \cite{Witten:1984eb} --
has also been considered in this context 
\cite{Doudoulakis:2006iw},
\cite{Doudoulakis:2007ti},
\cite{Doudoulakis:2007xz}. 
The energy was minimized within classes of fields with a 
given topological charge $\Q$ and having 
profiles of a  loop of radius $R$. 
The resulting minimal value of energy was always found to 
be a monotonously growing  
function of $R$, thus always showing the tendency of the loop to shrink  
thereby reducing
its energy.

These results render the existence of solutions somewhat implausible. 
However, they do not yet prove
their absence. Indeed, local energy minima  may be difficult to 
detect via energy minimization, 
as this would require starting the numerical iterations 
in their close vicinity, because otherwise the minimization procedure
converges to the trivial global minimum. 
In other words, one has to choose a good initial configuration. 
However, since `there is a lot of room in function space', 
chances to make the right choice are not high. 

It should be mentioned that a positive result was once reported 
in Ref.\cite{Niemi:2000ny},
were the energy was minimized in the $\Q=2$ sector 
and an indication 
of a convergence to a non-trivial minimum was observed. However, 
this result was not confirmed in an independent analysis 
in Ref.\cite{Ward:2002vq}, so that  
it is unclear whether it should be attributed to a 
lucky choice of the initial configuration or to 
some numerical artifacts.  

\section{Knot solitons in gauge field theory}

As we have seen, fixing only the topological charge does not guarantee 
the existence of knot solitons in gauge field theory. 
In order to obtain such solutions
one needs to constraint  the gauge field in order that it could not 
be deformed to zero. 
Below we describe two known examples of such solutions.  

\subsection{Anomalous solitons}

One possibility to constraint the gauge field is to fix 
its Chern-Simons number.
An example of how this can be done was suggested long ago by 
Rubakov and Tavkhelidze 
\cite{Rubakov:1986am}, 
\cite{Rubakov:1985nk}, 
who showed that the Chern-Simons number 
can be fixed by including fermions into the system.  
They considered the 
Abelian Higgs model \eqref{action2}, but with a {\it singlet} 
and not doublet Higgs field, augmented by including 
chiral fermions.
In the weak coupling limit at zero temperature this theory contains 
states with $N_F$ non-interacting fermions and with the bosonic fields
being in vacuum, $A_\mu=0$, $\Phi=1$. The
energy of such states is  $E\sim N_F$. Rubakov and Tavkhelidze argued that 
this energy could be decreased  via exciting the 
bosonic fields in the following way. 

Owing to the axial anomaly, when the gauge field $A_\mu$ varies, 
the fermion energy levels can cross zero and dive into 
the Dirac see. 
The fermion number can therefore change, but 
the difference $N_F-N_{\rm CS}$ is conserved. As a result, starting from
a purely fermionic state and increasing the gauge field, 
one can smoothly deform this state  
to a purely
bosonic state, whose 
Chern-Simon number will be fixed by the initial conditions,
\be                                                     \label{CS}
N_{\rm CS}=N_F.
\ee
Now, the energy of this state, 
\be                                                                    \label{en3}
E[\Phi,A_k]=\int\left( |{D}_k\Phi|^2+
\frac{1}{4}\,({F}_{ik})^2+\frac{\lambda}{4}\,(|\Phi|^2-1)^2
\right)
d^3 \bx \,,
\ee
can be shown to be bounded from {\it above} by $E_0(N_{\rm CS})^{3/4}$
where $E_0$ is a constant, and so for large $N_F=N_{\rm CS}$ it grows slower 
than the energy of the original fermionic state,  
$E\sim N_{\rm CS}$ \cite{Rubakov:1986am}, \cite{Rubakov:1985nk}.
Therefore, for large enough $N_F$, it is energetically favorable for 
the original purely fermionic state 
to turn into a purely bosonic state. The latter is called {\it anomalous} 
\cite{Rubakov:1986am}, \cite{Rubakov:1985nk}.
The energy of this anomalous state can be obtained by minimizing the 
functional \eqref{en3} with the Chern-Simons number fixed by the condition \eqref{CS}.  

Such an energy minimization was carried out in the recent work of Schmid and 
Shaposhnikov \cite{Schmid:2007dm}. 
First of all, they established the following inequality,
\be                                  \label{CS:bound}
E[\Phi,A_k]\geq c(N_{\rm CS})^{3/4},
\ee  
which reminds very much of the Vakulenko-Kapitanski bound \eqref{Hopf} 
for the Faddeev-Skyrme model,
but with the topological charge replaced by $N_{\rm CS}$. 
This gives a very good example of how constraining the gauge field
can stabilize the system: even though in the theory 
with a singlet Higgs field there is no
topological charge similar to the Hopf charge, 
 its role can be taken over by the Chern-Simons charge.

\begin{figure}[h]
\hbox to\linewidth{\hss%
  \resizebox{7cm}{5cm}{\includegraphics{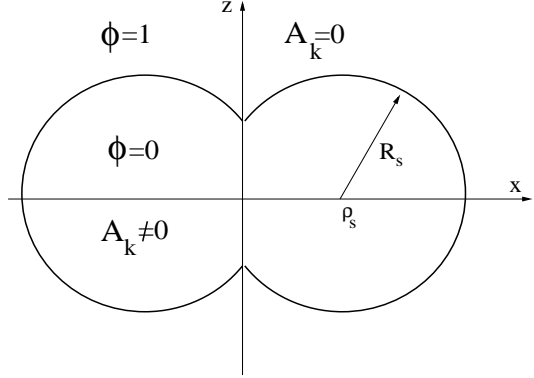}}%
\hss}
\caption{\small 
Spindle torus shape of the anomalous solitons for large $N_{\rm CS}$. 
}
\label{Fig1}
\end{figure}

In order to numerically minimize the energy, 
Schmid and Shaposhnikov considered the Euler-Lagrange equations
for the functional
\be
 E[\Phi,A_k]+\mu\int\epsilon_{ijk}A_i\partial_j A_k d^3\bx
\ee
where $E[\Phi,A_k]$ is given by \eqref{en3} and 
$\mu$ is the Lagrange multiplier. In the gauge where $\Phi=\Phi^\ast\equiv\phi$
these equations read (with $\vec{A}=A_k$)
\begin{subequations}               
\begin{align}
&\Delta\phi-\vec{A}^{\,2}\phi-\frac{\lambda}{2}\,(\phi^2-1)\phi=0, \label{e1} \\
&\vec{\nabla}\times(\vec{\nabla}\times\vec{A})
+2\mu\vec{\nabla}\times\vec{A}+2\phi^2\vec{A}=0,\label{e2}
\end{align}
\end{subequations} 
where $\vec{\nabla}$ and $\Delta=(\vec{\nabla})^2$ 
are the standard gradient and Laplace operators,
respectively. 
Multiplying Eq.\eqref{e2} by $\vec{A}$ and integrating gives the expression for the 
Lagrange multiplier,
\be
\mu=\frac{1}{32\pi^2 N_{\rm CS}}\left.\left.\int\right(
(\nabla\times \vec{A})^2+|\vec{A}|^2\phi^2\right)d^3x.
\ee  
Solutions of the elliptic equations \eqref{e1},\eqref{e2} 
were studied in \cite{Schmid:2007dm}
in the 
axially symmetric sector, where $\phi=\phi(\rho,z)$ and 
$\vec{A}=\vec{A}(\rho,z)$, with the boundary conditions 
\be
\phi=1,~~~~~\vec{A}=0
\ee
at infinity and 
\be
\partial_\rho\phi=\partial_\rho A_z=0,~~~~
A_\rho=A_\varphi=0
\ee
at the symmetry axis $\rho=0$. The solutions obtained are quite interesting.
They 
are very strongly localized in a compact region, $\Omega$, of the $(\rho,z)$
plane centered around a point $(\rho_s,0)$. Inside $\Omega$ the 
field ${A}_k$ is non-zero, while $\phi$ is almost constant 
and very close to zero. As one approaches the boundary of the region,
$\partial\Omega$, the field ${A}_k$ tends to zero, while $\phi$ is still almost 
zero.
Finally, in a  small neighborhood of 
$\partial\Omega$ whose thickness is of the order of the Higgs boson wavelength,
$\phi$ starts varying and quickly increases up to its asymptotic value. 
Outside $\Omega$ one has everywhere $\phi\approx 1$ and $A_k\approx 0$,
such that the energy density is almost zero.  The energy for these solutions scales as 
$(N_{\rm CS})^{3/4}$. 
For large $N_{\rm CS}$ the region $\Omega$ 
can be described by the simple analytic formula,
\be
 (\rho-\rho_s)^2+z^2<R_s^2,
\ee
where $R_s>\rho_s$. 
In other words, the 3D domain where the soliton energy is concentrated 
can be obtained by rotating around the $z$-axis a disc centered at a point 
whose distance from the $z$-axis is less than its radius (see Fig.\ref{Fig1}). 
Such a geometric figure is called spindle torus \cite{Schmid:2007dm}. 
The solutions for large $N_{\rm CS}$ are 
very well approximated by setting $\phi=1$ and $A_k=0$
outside the spindle torus, while inside it one has $\phi=0$ and $A_k$ is obtained
by solving the linear equation \eqref{e2}. The spindle torus approximation
becomes better and better for large $N_{\rm CS}$, in which  limit 
Schmid and Shaposhnikov obtain the following asymptotic formulas for the energy 
and parameters of the torus, which agree very well with their
numerics,
\be
E=118\,\lambda^{1/4}M_{W}(N_{\rm CS})^{3/4},~~~~~
\rho_s=\frac{1.7}{M_{W}}\left(\frac{N_{\rm CS}}{\lambda}\right)^{1/4},~~~~~
R_s=1.49\rho_s\,,
\ee
where $M_W$ is the vector field mass.

It is likely that these solution attain the lower energy bound \eqref{CS},
which means that they should be topologically {\it stable}. 
It should, however, be emphasized that the anomalous solitons require a rather exotic physical 
environment, since for them to be energetically favoured  
as compared to the free fermion condensate, the density of the latter should attain
enormous values possible perhaps only in the core  of neutron stars.

Summarizing, fixing the Chern-Simons number forbids deforming 
the gauge field to zero
and gives rise to  stable knot solitons in gauge 
field theory, even if
 the Higgs field 
is topologically trivial. It is unclear whether
this result can be generalized within the context of the model \eqref{action2} 
with two component Higgs field -- since the Protogenov-Verbus
formula \eqref{Proto} contains 
not the Chern-Simons number but the linking number L. 

\subsection{Non-Abelian rings \label{rings}} 

Another interesting class of objects arises in the non-Abelian gauge
field theory, where one can have smooth, finite energy loops stabilized by 
the magnetic energy.

\subsubsection{Yang-Mills-Higgs theory}
Let us parametrize 
the non-Abelian Yang-Mills-Higgs theory for a compact and simple  gauge group 
${\mathcal G}$ as
\be                                           \label{YMH}
{\mcal L}[A_\mu,\Phi]=-\frac14 \langle F_{\mu\nu}F^{\mu\nu}\rangle
+({D}_\mu\Phi)^\dagger {D^\mu\Phi}-
U(\Phi)\,.
\ee
Here the gauge field strength is 
$F_{\mu\nu}=\partial_\mu A_\nu-\partial_\nu A_\mu
-ig[A_\mu, A_\nu]\equiv F^a_{\mu\nu}{\bf T}_a$
where $A_\mu=A^a_\mu{\bf T}_a$ is the gauge field, 
$a=1,2,\ldots ,{\rm dim} ({\mathcal G})$,
and $g$ is the gauge coupling. The Hermitian gauge group
generators ${\bf T}_a$ satisfy the relations
\be
[{\bf T}_a,{\bf T}_b]=i C_{abc}{\bf T}_c,~~~~~ 
{\rm tr}({\bf T}_a{\bf T}_b)={\rm K}\delta_{ab}\,.
\ee
The Lie algebra inner product
is defined as $\langle AB\rangle =\frac{1}{{\rm K}}\,{\rm tr}(AB)=A^a B^a$. 
The Higgs field $\Phi$ is a vector in the representation space of 
${\mathcal G}$ where the generators ${\bf T}_a$ act; this space
can be complex or real. The covariant derivative of the Higgs field
is ${D}_\mu\Phi=(\partial_\mu-igA_\mu)\Phi$ and the Higgs
field potential can be  chosen as 
$U(\Phi)=\frac{\lambda}{2}\,(\Phi^\dagger\Phi-1)^2$. 
The Lagrangian is invariant under the local gauge transformations,
\be                              \label{UUU}
\Phi\to {\rm U}\Phi,~~~~A_\mu\to {\rm U}
(A_\mu+\frac{i}{g}\partial_\mu){\rm U}^{-1},
\ee
where ${\rm U}=\exp(i\alpha^a(x^\mu) T_a)\in{\mathcal G}$. 
The field equations read
\begin{align}                                         \label{YMHeqs}
{\hat {\mcal D}}_\mu F^{\mu\nu}&=ig\left\{({D}^\nu\Phi)^\dagger {\bf T}_a\Phi
-\Phi^\dagger{\bf T}_a{D}^\nu\Phi\right\}{\bf T}_a \,,\notag \\
{D}_\mu{D}^\mu\Phi&=
-\frac{\partial U}{\partial(\Phi^\dagger\Phi)}\,\Phi\,,
\end{align}
where $\hat{{\mcal D}}_\mu=\partial-ig[A_\mu,~~]$ is the covariant derivative in the 
adjoint representation. The energy momentum tensor is 
\be                              \label{Tmunu}
T^\mu_\nu=-\langle F_{\nu\sigma}F^{\mu\sigma}\rangle 
+({D}_\nu\Phi)^\dagger {D^\mu\Phi}
+({D}^\mu\Phi)^\dagger {D_\nu\Phi}-
\delta^\mu_\nu{\mcal L}\,.
\ee

\subsubsection{Monopole rings}

The ring solitons in the theory \eqref{YMH} have been first constructed by 
Kleihaus, Kunz and Shnir \cite{Kleihaus:2003xz},\cite{Kleihaus:2004is}
in the case where ${\mathcal G}$=SU(2) and the Higgs field 
is in its adjoint, $\Phi\equiv\Phi^a$, such that the gauge group
generators are $3\times 3$ matrices with components 
$({\bf T}_a)_{bc}=-i\epsilon_{abc}$. The fundamental solutions  in this theory 
are the magnetic monopoles of 't Hooft and Polyakov 
\cite{'tHooft:1974qc},\cite{Polyakov:1974ek},
while the ring solitons are more general 
solutions. Specifically, 
in the static, axially symmetric and purely magnetic case 
it is consistent to choose the following ansatz for 
the fields in spherical coordinates:
\begin{align}                                   \label{anz0}
A_\mu dx^\mu&=(K_1 dr+(1-K_2)d\vartheta){\bf T}_\varphi
+m(K_3{\bf T}_r+(1-K_4){\bf T}_\vartheta)\sin\vartheta\, d\varphi, \nonumber \\
{\bf T}_a\Phi^a&=\phi_1{\bf T}_r+\phi_2{\bf T}_\vartheta\,,
\end{align}
where functions $K_1,K_2,K_3,K_4,\phi_1,\phi_2$ depend on $r,\vartheta$
and are subject of suitable boundary conditions 
at the symmetry axis and at infinity 
\cite{Kleihaus:2003xz},\cite{Kleihaus:2004is}. Here 
\begin{align}                                   \label{mn}
{\bf T}_r&=\sin({k}\vartheta)\cos({m}\varphi) {\bf T}_1+
\sin({k}\vartheta)\sin({m}\varphi){\bf T}_2
+\cos({k}\vartheta){\bf T}_3, \nonumber \\
{\bf T}_\vartheta&=\frac{1}{k}\,\frac{\partial}{\partial \vartheta}{\bf T}_r\,,
~~~~~~~
{\bf T}_\varphi=\frac{1}{m\sin\vartheta}\,
\frac{\partial}{\partial \varphi}{\bf T}_r\,,~~~
\end{align}
with ${k},{m}\in Z$. Using 
the gauge invariant tensor
\be                                            \label{hooft}
{\mcal F}_{\mu\nu}=\Phi^a F^a_{\mu\nu}-\epsilon_{abc}\Phi^a D_\mu\Phi^b D_\nu\Phi^c
\ee
and its dual, $\tilde{{\mcal F}}_{\mu\nu}=
\frac12\epsilon_{\mu\nu\alpha\beta}{\mcal F}_{\alpha\beta}$,
one can define the  
electric and magnetic currents, respectively, as 
\be                                              \label{charge-current}
j_\mu=\partial^\alpha{\mcal F}_{\alpha\mu},~~~~~~~~
\tilde{j}_\mu=\partial^\alpha\tilde{{\mcal F}}_{\alpha\mu}\,.
\ee   
It turns out that the solutions depend crucially on values of $k,m$
in \eqref{anz0}. 
\begin{figure}[h]
\hbox to\linewidth{\hss%
\resizebox{7cm}{5cm}{\includegraphics{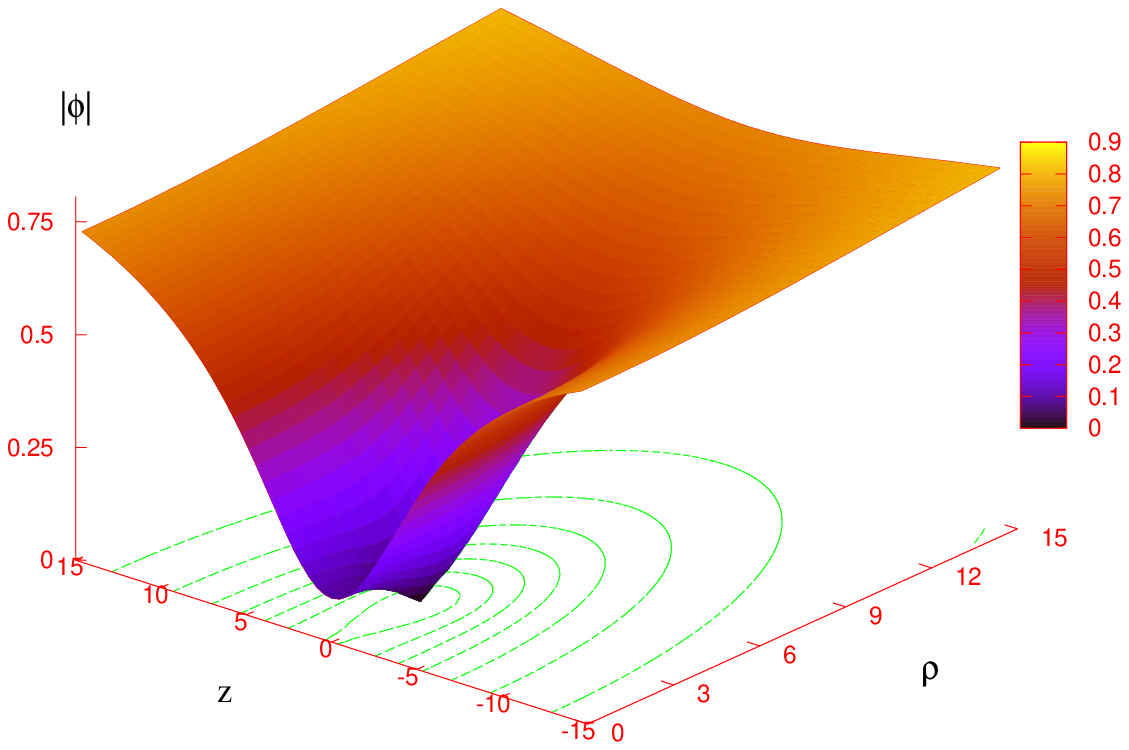}}%
\hspace{5mm}%
\resizebox{6cm}{5cm}{\includegraphics{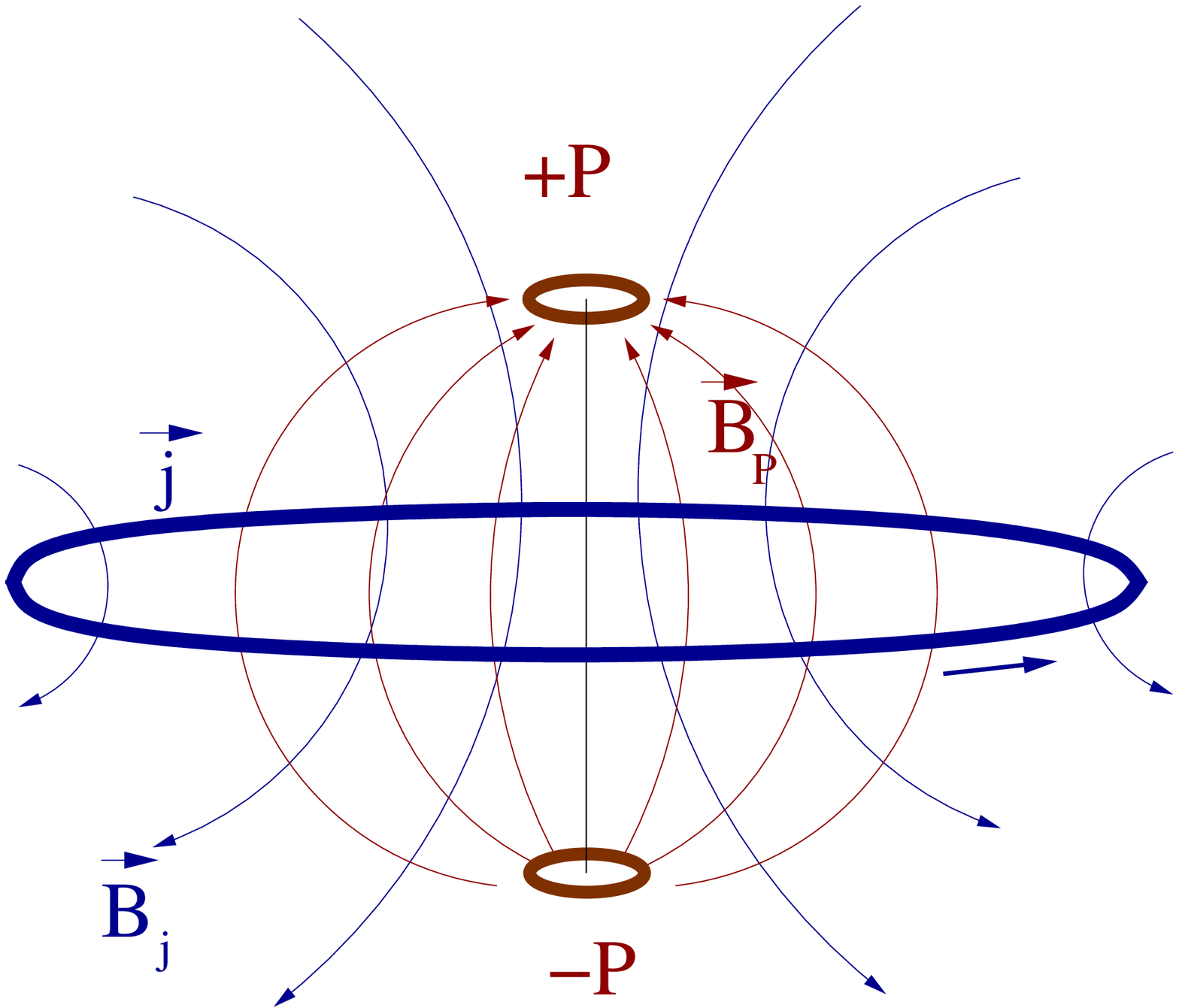}}%
\hss}
\caption{\small Left: Higgs field amplitude for the $k=2$, $m=3$ monopole ring solution
in the  limit where the Higgs field potential is zero. $|\Phi|$ vanishes
at a point in the $(\rho,z)$ plane away from the $z$-axis, which corresponds to a ring.   
Right: schematic shape of charge/current distribution for this solution. 
}
\label{Fig2}
\end{figure}
In particular, their 
magnetic charge is given by 
\be                                 \label{Qmagn}
{\mcal Q}=\frac{m}{2}\,[1-(-1)^k].
\ee
The following solutions are known 
in the limit of vanishing Higgs potential
(for a generic potential their structure is more complicated)
\cite{Kleihaus:2003xz},\cite{Kleihaus:2004is}: \\
${k}=1,{m}=1$ -- the spherically symmetric 't Hooft-Polyakov monopole.  \\
${k}=1,{m}>1$ -- multimonopoles.\\
${k}>1,{m}=1,2$ -- monopole-antimonopole sequences.\\
${k}=2l\geq2,{m}\geq 3$ -- monopole {vortex rings}. \\
In the first three cases the Higgs field has discrete zeros  
located at the $z$-axis. 
Solutions of the last type are  especially interesting in the 
context of our discussion, since 
zeros of the Higgs fields in this case are not discrete but continuously 
distributed along a circle (for $k=2$)
around the $z$-axis (see Fig.\ref{Fig2}). Solutions in this 
case can be visualized as stationary rings
stabilized by the magnetic energy.  The mechanism of their 
stabilization is quite interesting
and can be elucidated as follows \cite{Shnir:2005te,Shnir}.
If one studies the profiles of the currents \eqref{charge-current} for these solutions,
it turns out that  both  
the magnetic charge density $\tilde{j}_0$ and the 
electric current density $j_k$ have ring shape distributions,
as qualitatively shown in Fig.\ref{Fig2}. 

  Although the total magnetic charge is zero, 
locally the charge density is non-vanishing and 
the system can be visualized  as a pair of 
magnetically charged rings with opposite charge located at 
$z=\pm z_0$, accompanied by a circular
electric current in the $z=0$ plane.  
The two magnetic rings create a magnetic field orthogonal to the $z=0$ plane. 
This magnetic field forces the electric charges in the 
plane to Larmore orbit, which 
creates  a circular current. The Biot-Savart magnetic field produced by this current
acts, in its turn, on the magnetic rings keeping them away from each other, so that 
the whole system is in a self-consistent equilibrium  \cite{Shnir:2005te}
(assuming the magnetic rings to be rigid). 

It is, however, unlikely that this sophisticated balance  mechanism 
stabilizing the rings against contraction 
could also guarantee their stability with respect to all possible 
deformations. In fact, the monopole-antimonopole solution is  
known to be unstable \cite{Taubes:1982ie}, while the monopole rings 
can be viewed as generalizations of this solution. They are therefore likely
to be saddle points of the energy functional and so they should be unstable as well.

\subsubsection{Sphaleron rings}

Very recently, a similar ring construction was carried out by 
Kleihaus, Kunz and Leissner \cite{Kleihaus:2008gn} 
within the context of the Yang-Mills-Higgs theory \eqref{YMH}
with ${\mathcal G}$=SU(2) and with the Higgs field 
in its  fundamental complex doublet representation, where 
${\bf T}_a =\frac12 \tau_a$. 
This theory can be viewed as the SU(2)$\times$U(1) 
Weinberg-Salam theory in the limit where 
the weak mixing angle vanishes and the U(1) gauge field decouples. 
Kleihaus, Kunz and Leissner used exactly the same field ansatz \eqref{anz0},
only modifying the Higgs field as  
 \begin{align}                                   \label{anz0a}
\Phi&=(\phi_1{\bf T}_r+\phi_2{\bf T}_\vartheta)\Phi_0\,,
\end{align}
where $\Phi_0$ is a constant 2-vector.  As in the monopole case,
in this case too the solutions depend strongly on the choice of 
the integers $k$ and $m$ in Eq.\eqref{mn}. The fundamental solutions in this case are 
the sphalerons -- 
unstable saddle point configurations that can be smoothly deformed to vacuum. 
They can be characterized by the Chern-Simons number, given   
by the same formula as the magnetic charge in the monopole case, 
up to the factor $1/2$, so that half-integer 
values are now allowed: ${\mcal Q}=m[1-(-1)^k]/4$.

Setting $k=m=1$ gives 
the Klinkhamer-Manton sphaleron \cite{Klinkhamer:1984di}, in which case the 
Higgs field vanishes at one point. Choosing $k=1$, $m>1$  or $k>1$, $m=1$ gives
multisphalerons or sphaleron-antisphaleron
solutions for which  the Higgs field has several isolated zeros
located at the symmetry axis \cite{Kleihaus:2008gn}.
A new type of solution arises for $k\geq 2$, $m\geq 3$, in which case the Higgs field
vanishes on one or more rings centered around the symmetry axis. 
In this respect these sphaleron rings are quite 
analogues to the monopole rings. It is unclear at the moment whether 
their existence can be qualitatively explained by a mechanism similar to that  
for the monopole rings, shown in Fig.\ref{Fig2}. 

Since sphalerons are unstable objects, it is very likely that sphaleron rings
are also unstable. However, similar to the sphalerons, they could perhaps
be interesting physically as mediators of baryon number violating processes 
\cite{Klinkhamer:1984di}.


\section{Angular momentum and radiation in field systems}

\begin{figure}[ht]
\hbox to\linewidth{\hss%
	\resizebox{7cm}{4cm}{\includegraphics{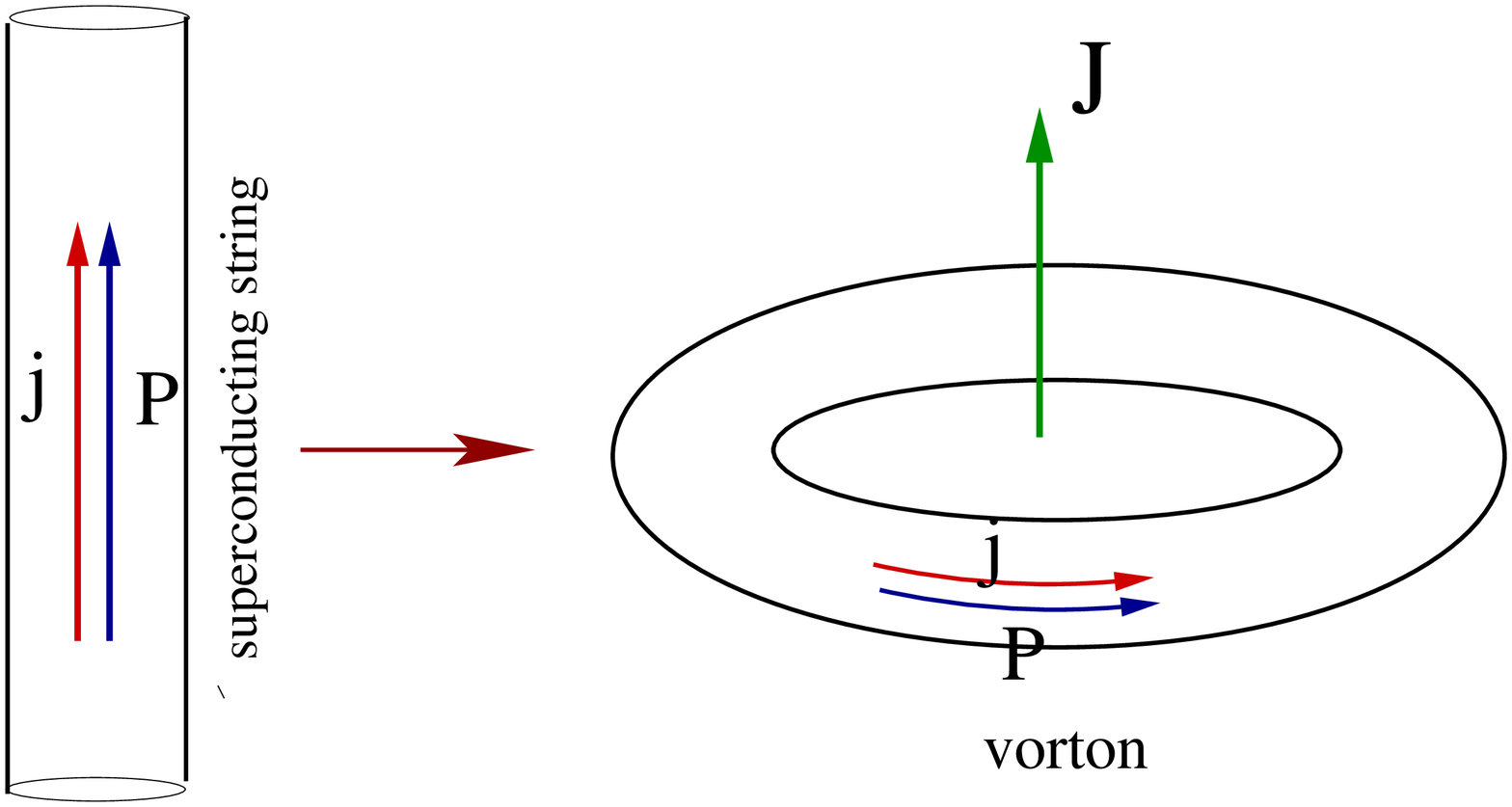}}
\hspace{5mm}%
        \resizebox{7cm}{4cm}{\includegraphics{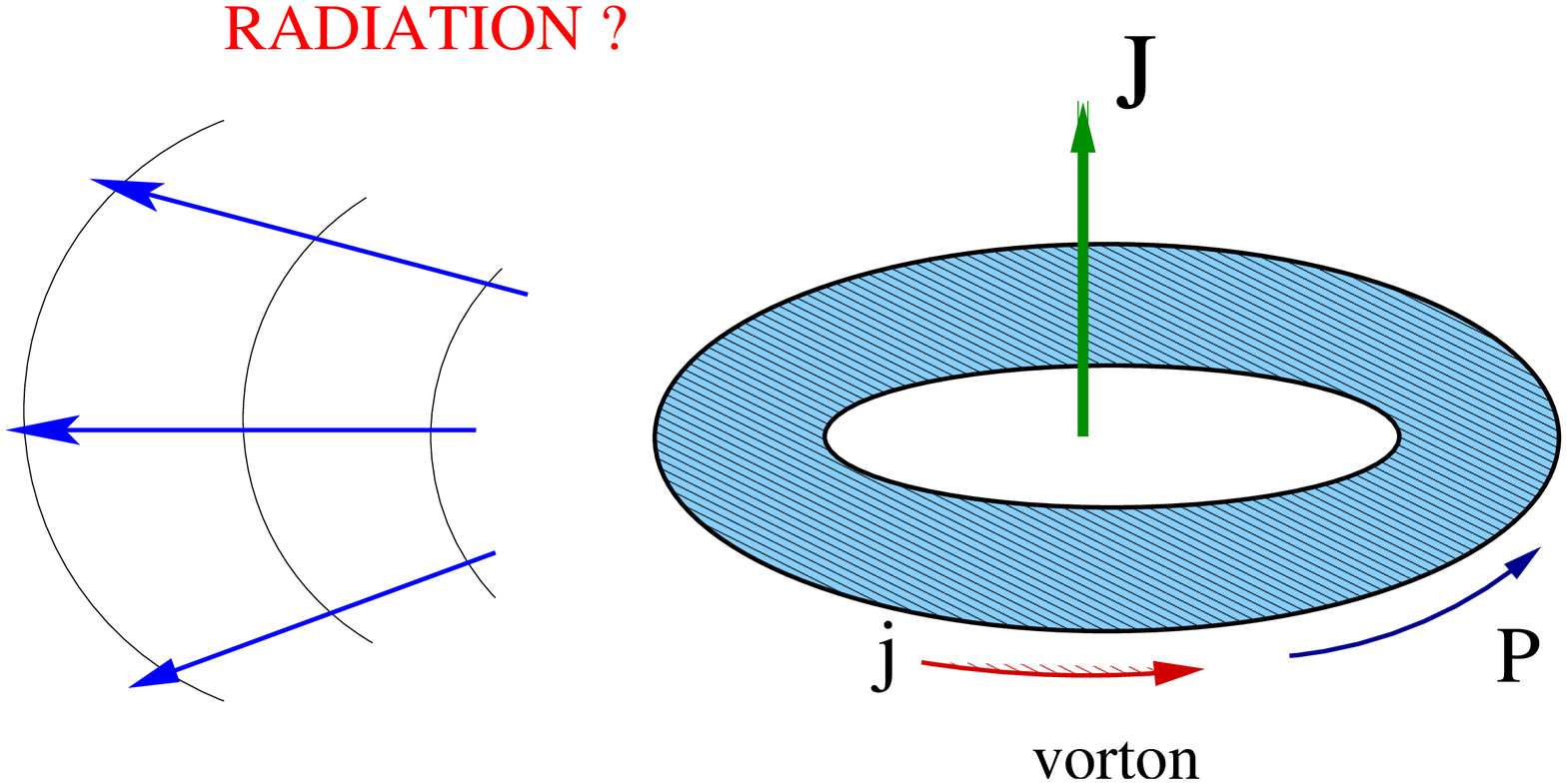}}	
\hss}
\caption{One can make a loop from a vortex carrying a current $j$ and momentum $P$. 
It will have an angular momentum, but it may be radiating.}
\label{FigV}
\end{figure}

We are now passing to the spinning systems with the ultimate intention to discuss 
spinning vortex loops stabilized by the 
centrifugal force -- vortons. As was already said in the Introduction, 
usually vortons are considered within a qualitative macroscopic description 
as loops made of vortices and stabilized by rotation. 
This description is suggestive, but it does not take into account
the radiation damping. At the same time, 
the presence of the vorton angular momentum requires 
some internal motions in the system (see Fig.\ref{FigV}),  
as for example circular currents, and these  are likely to  
generate radiation carrying away both the energy and angular momentum. 
It is therefore plausible that macroscopically constructed vortex loops will not
be stationary field theory objects, but at best only quasistationary, with a 
finite lifetime determined by the radiation rate. 

The best way to decide whether vortons are truly stationary or only 
quasistationary is to explicitly resolve the corresponding field theory equations. 
The current situation  in this direction is not, however, very suggestive. 
Within the original local U(1)$\times$U(1) Witten's
model of superconducting cosmic strings 
\cite{Witten:1984eb} vorton solutions have never been constructed.    
To the best of our knowledge, the only explicit vorton solutions 
have been presented by Lemperier and Shellard \cite{Lemperiere:2003yt} 
within the global 
version of Witten's model, and also by Battye, Cooper and Sutcliffe  
\cite{Battye:2001ec} in a special limit of the same global model. 
In addition, vortons in a $2+1$ dimensional field theory toy model 
have been recently analyzed  \cite{Battye-kink}. 

Lemperier and Shellard \cite{Lemperiere:2003yt} considered the full 
hyperbolic evolution problem for the fields, with the initial data corresponding to 
a vortex loop. Evolving dynamically  this loop in time, 
they saw it oscillate  around a visibly stationary equilibrium position, 
and they could follow these oscillations for several dozens 
characteristic periods of the system. This suggests that vortons exist and are stable
against perturbations. 
However, one cannot decide on these grounds  whether the equilibrium configurations 
are truly stationary or only quasistationary, since the radiation damping 
could be non-zero but too small to be visible in their numerics. 
In addition, Lemperier and Shellard did not actually consider precisely 
the global version of Witten's model (see Eq.\eqref{lag} below),
but, in order to improve the numerics, 
added to it a $Q$-ball type interaction 
term of the form $|\phi|^6 |\sigma|^2$, where $\phi,\sigma$ are the two scalars 
in the model.

Battye, Cooper and Sutcliffe \cite{Battye:2001ec} did not study 
Witten's model but minimized the energy of a non-relativistic Bose-Einstein condensate, 
which  seems to be mathematically equivalent to solving equations 
of Witten's model in a special limit.  
They found non-trivial energy minima
saturated by configurations of vorton type, which again suggests that
vortons exist, at least in this limit. Moreover, this suggests   
that they are indeed non-radiative --
since being already in the energy minimum they cannot loose energy anymore.  
It would therefore be interesting to construct these solutions in a different way,
extending the analysis to the full Witten's model.

The method we shall employ below to study truly non-radiating vortons
 will be to construct them as stationary solutions of the elliptic boundary 
value problem obtained by separating the time variable. However, 
first of all we need to  
understand how in principle a non-radiating field system can have a non-zero 
angular momentum. Both angular momentum and radiation are associated to some 
internal 
motions in the system, and it is not completely  clear how to reconcile 
the presence of the former with the  absence of the latter.

\subsection{Angular momentum for stationary solitons}

In what follows we shall be considering field theory systems obeying
the following four conditions: 

\noindent 
(1) stationarity \\
(2) finiteness of energy \\
(3) axial symmetry   \\
(4) non-vanishing angular momentum

The angular momentum is defined as the Noether charge
associated to the global spacetime symmetry generated by the axial Killing
vector $K=\partial/\partial\varphi$,
\be                                    \label{JJ}
J=\int T^0_\varphi d^3x\,.
\ee
Let us discuss the first three conditions. 

(1) A system is stationary if its energy momentum tensor $T^\mu_\nu$ does not depend on time.  
According to the standard definition 
of symmetric fields \cite{Forgacs:1979zs}, for stationary fields 
the action of time translations can be compensated by 
internal symmetry transformations.  

If all internal symmetries of the theory are local, then there is a gauge where
the compensating symmetry transformation is trivial, so that the stationary  fields
are time-independent.  
We shall call them 
{\it manifestly} stationary. 
If the theory contain also global 
internal symmetries and if the compensating symmetry transformation is global,  
then its action cannot be trivialized and so the action of time translations
will  be non-trivial. The fields will explicitly depend on
time in this case, typically via time-dependent phases, and we shall call them 
{\it non-manifestly} stationary. 

For example, in a system with two complex scalars coupled to a U(1) gauge field 
one cannot gauge away simultaneously 
phases of both scalars. A non-manifestly stationary field configuration will then be $\phi_1(\bx)$,
$\phi_2(\bx)e^{i\omega t}$, $A_\mu(\bx)$.  However, if there is  only one
scalar, then it is always possible to gauge away its time-dependent phase. 

(2) Even if $T^\mu_\nu$ is time-independent, 
one can still have a constant radiation flow compensated by the 
energy inflow from infinity. However, if the energy is finite, then 
the fields fall-off fast enough at infinity to eliminate this 
possibility.  

In principle, one can also
have situations where  $T^\mu_\nu$ is time-dependent, but radiation 
is nevertheless absent, as for the breathers in 1+1 dimensions \cite{Rajaraman}.    
However, such cases are probably less typical and we shall not discuss them.

(3)  It is intuitively clear that asymmetric spinning systems will more likely 
radiate than symmetric ones. 
It is therefore most natural to assume spinning 
non-radiating solitons to be axially symmetric. 
The axially symmetry can  be manifest or non-manifest. 
In fact, it appears that the axial symmetry condition can sometimes be relaxed, 
but such a possibility seems to be more exotic and will be discussed below
only very briefly in Sec.\ref{mon-pairs}.

Let us now analyze possibilities for condition (4) 
to coexist  with (1)--(3).

\subsubsection{The case of manifest symmetries -- no go results}
In theories where {\it all} 
internal symmetries are local the  
stationary field are time-independent. 
Can one have  an angular momentum in this case ? 

In fact, 
even without an explicit time dependence 
one can have 
a non-vanishing field momentum expressed by the Poynting vector, 
$\vec{\mcal E}\times\vec{\mcal B}$, 
and this could give a contribution to the angular momentum,  
\be                                  \label{Poynt}
\int \vec{r}\times(\vec{\mcal E}\times\vec{\mcal B}) \, d^3 x \;.
\ee 
If the theory contains gauged scalars, they will give an additional 
contribution. If there exists a stationary, globally regular, finite energy
on-shell configuration for which the
integral \eqref{JJ} is non-zero, this would correspond to a rotating and non-radiating 
soliton.  The existence of such solutions is  not {\it a priori} excluded.  
However, it turns out that for a number of physically interesting cases
they can be ruled out.  

Specifically, since the time translations and spatial rotations commute, 
the system is not only manifestly stationary but also manifestly axially symmetric
\cite{Forgacs:1979zs}. 
It turns out that the latter condition allows one to transform the volume integral in  
Eqs.\eqref{JJ}  to a {surface integral},
which is often enough to conclude that $J=0$. 

To me more precise,   let  us consider the Yang-Mills-Higgs theory \eqref{YMH}. 
If the fields $(A_\mu,\Phi)$ 
are manifestly invariant under the action of a Killing symmetry generator 
$K=\partial/\partial s$, then there is a gauge where they do not depend
on the corresponding spacetime coordinate $s$ and 
their Lie 
derivatives along $K$ vanish, ${\mcal L}_K A_\mu={\mcal L}_K\Phi=0$.
In some other gauge the fields could depend on $s$, but only in such a way that 
\cite{Forgacs:1979zs}
\be                                         \label{sym}
{\mcal L}_K A_\mu=\hat{{\mcal D}}_\mu W(K)\,,~~~~~~
{\mcal L}_K\Phi= ig W(K)\Phi\,.
\ee
Here 
$W(K)$ takes its values in the Lie algebra of the gauge group and
transforms as connection under gauge transformations.
If $K=\partial/\partial\varphi$ is the axial Killing vector and 
$W_\varphi\equiv W(K)$, then in spherical coordinates these conditions reduce to 
\be                                         \label{sym1}
\partial_\varphi A_\mu=\hat{{\mcal D}}_\mu W_\varphi\,,~~~~~~
\partial_\varphi \Phi= ig W_\varphi\Phi\,.
\ee
If these conditions are fulfilled, then it is 
straightforward to check with the field equations  \eqref{YMHeqs}
that the $T^0_\varphi$ component of the energy-momentum tensor \eqref{Tmunu}
has a total derivative structure 
\cite{VanderBij:2001nm}, \cite{Volkov:2003ew}, 
 \be                              \label{Tmu}
T^0_\varphi=\partial_k\langle (W_\varphi-A_\varphi)F^{0k}\rangle\,.
\ee
Since both $A_\varphi$ and $W_\varphi$ transform as connections 
under gauge transformations, their difference is gauge covariant, so that 
this formula is gauge invariant. 
For globally regular solutions the 
electric field ${\mathcal E}^k=F^{0k}$ 
is everywhere bounded, which allows one to transform  
the volume integral
of $T^0_\varphi$ to a surface integral
over the boundary of $\mathbb{R}^3$ at infinity. This gives the surface 
integral representation for the angular momentum
\be                                             \label{JJJ}
J=\oint \langle
(W_\varphi-A_\varphi)\vec{{\mcal E}}\rangle\, \vec{dS}\,.
\ee
This formula imposes rather strong restrictions on the existence of spinning
solitons. 
 For example, it shows 
that if the electric field 
${\mcal E}$ decays at infinity faster than $1/r^2$, such that the electric charge 
is zero, then $J=0$. A non-zero electric charge is therefore necessary 
to have a non-zero angular momentum \cite{VanderBij:2001nm}.   
Similarly, 
using the asymptotic conditions in the far field zone, 
one can show that for a number of important cases 
the asymptotic behavior of the fields does not allow for 
the surface integral to be non-zero. 

\begin{figure}[h]
\hbox to\linewidth{\hss%
  \resizebox{10cm}{5cm}{\includegraphics{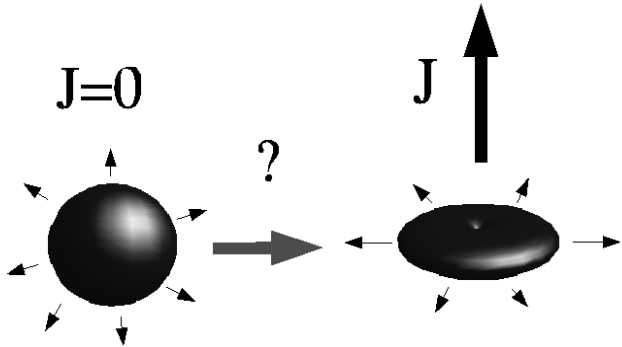}}%
\hss}
\caption{\small 
Do static solitons admit stationary, spinning generalizations ?
}
\label{FigDo}
\end{figure}

Specifically, one can consider any known static, spherically symmetric soliton,
as for example the 't Hooft-Polyakov monopole, and ask if it admits stationary,
spinning generalizations (see Fig.\ref{FigDo}) ? For example, 
the static Schwarzschild black hole solution of Einstein's equation in General Relativity
can be generalized to the manifestly stationary, spinning Kerr black hole. One can wonder
if similar generalizations are possible for solitons 
in non-linear field theories in Minkowski space.
If they exist, then in the far field zone their 
fields should approach those of the original static soliton, so that they should be
 spherically symmetric, up to small corrections which could 
contribute to the surface integral \eqref{JJJ}.  Since the 
corrections are small, their most general form can be determined
from the linearized field equations \cite{Volkov:2003ew}. 

Surprisingly, 
one discovers in this way 
that none of the well known solitons in the gauge field theory \eqref{YMH}
with gauge group SU(2), as
for example 
the magnetic monopoles of 't~Hooft-Polyakov 
\cite{'tHooft:1974qc}, \cite{Polyakov:1974ek}, 
dyons of Julia-Zee \cite{Julia:1975ff}, and sphalerons
of Klinkhamer-Manton \cite{Klinkhamer:1984di},
admit spinning generalizations
 within the manifestly stationary and manifestly axisymmetric sector 
\cite{VanderBij:2001nm},
\cite{vanderBij:2002sq},
\cite{Volkov:2003ew}.
The picture in Fig.\ref{FigDo} thus does not apply for these solitons. 
If the gauge symmetry is completely broken in the Higgs vacuum,
as for the sphalerons, then the fields approach their vacuum values
exponentially fast and the surface integral vanishes. For 
the monopoles and dyons there is a long-range field associated with the unbroken
U(1), so that some additional analysis is required to show  
that in the asymptotic region there are actually no field modes  giving 
a non-zero contribution to the surface integral \cite{Volkov:2003ew}. 

As a result, the best known SU(2) solitons cannot spin in the 
manifestly stationary and manifestly axisymmetric sector. 
This restricts rather strongly 
the existence of spinning solitons with manifest symmetries,
although does not rule them out completely. 
Such solutions might exist in theories with other gauge groups.  
Their explicit 
examples in U(1) gauge field theories will be presented below. 
Non-manifestly symmetric rotational excitations  could perhaps
exist for the sphalerons, since only an SU(2) part of their  U(2) 
internal  symmetry is gauged. A more exotic possibility to have stationary
rotation without axial symmetry will be briefly discussed in Sec.\ref{mon-pairs}.

\subsubsection{The case of non-manifest symmetries}
More general possibilities to have spinning solitons arise in theories where 
{\it not all} internal symmetries are 
local, since in this case the conditions \eqref{sym} can be generalized as 
\be                                         \label{sym1a}
{\mcal L}_K A_\mu=\hat{{\mcal D}}_\mu W(K)\,,~~~~~~
{\mcal L}_K\Phi= igW(K)\Phi+iT(K)\Phi\,,
\ee
where $T(K)$ is a function of the {\it global} 
symmetry generators. One can gauge away $W(K)$ 
but not $T(K)$, so that the action of the Killing symmetry 
will always be non-trivial, since $\Phi$ will 
depend explicitly on the corresponding spacetime 
coordinate $s$ as
\be
\Phi(s)=e^{is T(K)}\Phi_0\,.
\ee
This means that the $s$-dependence is equivalent
to a sequence  of internal symmetry transformations, in which case
the invariant objects like $T^\mu_\nu$ will not 
depend on $s$ at all. Such a symmetry can be called 
{\it non-manifest}. Non-manifestly stationary and non-manifestly 
axisymmetric fields would typically depend on 
$t,\varphi$ via the complex phase factor 
\be                                                               \label{om}
\exp\{i(\omega t + m\varphi)\},
\ee
in which case one can say that it is the phase that spins. 
The angular momentum 
in this case cannot be totally 
expressed by a surface integral 
and will contain a volume integral contribution, in which case
typically  $J\sim \omega m$.

As a result, one can have $J\neq 0$. However, 
the absence of radiation is not yet guaranteed. 
Separating the time variable makes 
the equations elliptic, but with the mass term(s) modified as 
\be
M^2\to M^2-\omega^2,
\ee
where $M^2$ collectively denotes masses of the 
field excitations in the asymptotic zone.  
Now, if the equations admit globally regular solutions with 
\be                                       \label{rad}
\omega^2<M^2
\ee
 then these
solutions will behave asymptotically as $\exp\{-\sqrt{M^2-\omega^2}~ r\}$ 
and there will be no radiation.
On the other hand, if $\omega^2>M^2$ then solutions in the 
asymptotic region will oscillate as 
$\exp\{\pm i\sqrt{\omega^2-M^2}~ r\}$ thus showing the presence
of the ingoing and outgoing radiation, even though $T^\mu_\nu$ is time-independent,
so that 
the total energy will be infinite.  
As a result, the no-radiation condition \eqref{rad} becomes crucial in this case.

\subsubsection{Spinning solitons as solutions of elliptic equations}

Summarizing the above discussion, one can conclude that the 
existence of non-radiating spinning solitons, although not forbidden,  
is not guaranteed either.  
Even for known static solitons their spinning 
generalizations may or may not exist,
and if they do exist, this should be considered 
as something exceptional rather than the general rule.
At first view, such a conclusion may seem to contradict our experience, since normally 
one knows that compact objects can spin. 
However, we are not saying
that generic field theory solitons cannot spin. They always can, but
it seems that  they should generically radiate at the same time. 
 One can always `give a kick' 
to any static soliton, as for example to the magnetic monopole
or to any of the knotted solitons of Faddeev-Skyrme shown in Fig.\ref{Fig:knots},
so that they will start spinning. However, at the same time 
they will start radiating away all the received energy and angular momentum,
till they relax back to the original static, non-spinning configuration.
It seems that 
`spinning and radiating'  represents the generic way the  
field systems behave, while `spinning without radiating' should rather be
considered as something exceptional, possible only  
in some special field theory models.   

Perhaps  the best way to really establish the
existence of spinning and non-radiating solitons 
would be to construct them as solutions of the
elliptic boundary value problem obtained by separating the time variable. 
In gauge field theory, as for example in the local 
U(1)$\times$U(1) Witten's model 
\cite{Witten:1984eb}, one could look for manifestly stationary 
and axisymmetric vortons, 
in which case there is no radiation.
However, since the angular momentum in this case can be 
expressed by the surface integral \eqref{JJJ} and so is
determined only by the asymptotic behavior of the fields, 
there are higher chances that it could  vanish.  
For global field theories 
one could consider systems with non-manifest symmetries, 
in which cases
there are better chances to have a non-zero angular momentum, 
but also higher chances to have
radiation, unless the no-radiation condition \eqref{rad} 
is fulfilled. 

\section{Explicit examples of stationary spinning solitons}

Before coming to vortons, one can wonder, in view of the above 
discussion, if there exist at all any 
known examples of non-radiating spinning 
solitons. As we said, for the SU(2) magnetic monopoles,
dyons and sphalerons there are no spinning generalizations,
at least in the manifestly stationary and 
manifestly axisymmetric case. Nevertheless, explicit examples of 
spinning solitons in Minkowski space in $3+1$ dimensions exist
and below we shall review all known solutions of this type. These are
the spinning $Q$-balls, spinning Skyrmions and also rotating 
monopole-antimonopole pairs, apart from the vortons. 
Interestingly, it seems that spinning 
solitons are in some sense more easily constructed in curved space,
because the spinning degrees of freedom can be naturally associated
to the non-radiative  dipole moment of 
gravitational field. For this reason 
a number of articles cited below actually describe the  Minkowski space 
spinning solitons 
only as a special limit of the more general, self-gravitating 
configurations.

\subsection{$Q$-balls \label{Qballs}}

This is an important for our discussion example  which 
shares many features with the vortons. At the same time, 
spinning $Q$-balls are simpler than vortons,
and so 
they can be used to introduce a number of notions to be applied later. 
We shall therefore discuss them  in some detail. 

$Q$-balls have been introduced by Coleman \cite{Coleman:1985ki}.
These 
are non-topological solitons \cite{Lee:1991ax} 
found in a theory  with 
 a single complex scalar field with the Lagrangian 
density 
\be                   \label{lQ}
{\mathcal L}_{Q}[\Phi]=  
\partial_\mu \Phi^\ast \partial^\mu \Phi  -U(|\Phi|),
\ee 
the corresponding field equation being 
\be                                   \label{Qe00}
\partial_\mu\partial^\mu\Phi+
\frac{\partial U}{\partial |\Phi|^2}\,\Phi=0,
\ee
while the energy-momentum tensor 
\be
T_{\mu\nu}=\partial_\mu\Phi^\ast\partial_\nu\Phi+
\partial_\nu\Phi^\ast\partial_\mu\Phi-g_{\mu\nu}{\mcal L}_{Q}.
\ee
The  potential $U$ should have the absolute minimum, $U(0)=0$, and 
should also satisfy the condition 
\be                                \label{omega+}
\omega_{\rm -}^2= \min_{f}\frac{U(f)}{f^2}<
\omega_{\rm +}^2= \frac12\left.\frac{d^2U}{df^2}\right|_{f=0}, 
\ee
whose meaning will be explained below. 
The value of $\omega_{+}$ determines the mass of the 
field quanta, $M=\omega_{+}$.  
For the condition \eqref{omega+} to be fulfilled
$U$ should (if it is even in $|\Phi|$)  
contain powers of $|\Phi|$ higher than four, 
which means that the theory cannot be renormalizable \cite{Coleman:1985ki}. 
A convenient choice is
\be                           \label{pQ}
 U=\lambda |\Phi|^2(|\Phi|^4-a |\Phi|^2+b),
\ee 
where $\lambda$, $a$, $b$ are positive 
constants, so that $\omega_{+}^2=\lambda b$
and $\omega_{-}^2=\omega_{+}^2(1-a^2/4b)$. 
In our numerics below we shall always 
choose $\lambda=1$, $a=2$, $b=1.1$ \cite{Volkov:2002aj}, which is not a restriction 
(for $\omega_{-}\neq 0$) 
since $\lambda,a$ can be changed by rescaling the coordinates and field, while the mass 
$M=\sqrt{\lambda b}$ enters 
the field equation \eqref{Qe} only in the combination $M^2-\omega^2$ where
$\omega$ is another free parameter.

The global invariance of the theory under 
$\Phi\to \Phi e^{i \alpha }$ implies the 
conservation of the Noether charge 
\be                                                      \label{Noether}
Q=i\int (\partial_t\Phi^\ast\Phi-\Phi^\ast\partial_t\Phi) d^3x\,.
\ee
The scaling argument of Derrick  applies for the theory \eqref{lQ} and 
rules out all static solutions with finite energy. Therefore, in order to  
circumvent this argument, solutions should depend  on time,
\be                    \label{QQQ}
\Phi=\phi(\bx)e^{i\omega t},
\ee 
in which case 
the Noether charge is 
\be                               \label{QQ}
Q=2\omega \int \phi^2 d^3 {\bf x} \equiv 2\omega \ssigma.
\ee
In what follows we shall assume $\omega$ to be positive.  
More explicitly, with 
\be
E_2=\int(\nabla\phi)^2d^3 {\bf x},~~~~~ 
E_0=\int U d^3 {\bf x},
\ee 
the Lagrangian reads
\be
L= \int {\mcal L}_{Q}\, d^3 {\bf x} =\omega^2\ssigma-E_0-E_2.
\ee
Under scale transformations $\bx\to \Lambda\bx$ one has  
$
L\to \Lambda^3(\omega^2\ssigma-E_0)-\Lambda E_2
$
so that $L$ will be stationary for $\Lambda=1$ if the virial relation 
is fulfilled,
\be                                        \label{Qvirial}
3\omega^2\ssigma=3E_0+E_2,
\ee
which is only possible (for $E_0\neq 0$, $E_2\neq 0$) if $\omega\neq 0$. 

$Q$-balls are finite energy solutions of the Lagrangian 
field equations for the ansatz \eqref{QQQ},
\be                                   \label{Qe}
\left(\Delta+\omega^2\right)\phi=
\frac{\partial U}{\partial |\phi|^2}\,\phi.
\ee
Even though $\Phi$ depends on time, 
$T_{\mu\nu}$ is time independent --
the system is non-manifestly stationary. 
Equivalently, $Q$-balls can be obtained 
by minimizing the total energy
\be                       \label{EneQ}
E=\int T^0_0 d^3\bx=\int(\omega^2|\phi|^2+|\nabla\phi|^2+U) 
d^3 {\bf x}\equiv \omega^2\ssigma+E_0+E_2
\ee
by keeping fixed the charge $Q$. 
Indeed, rewriting the energy as 
\be                       \label{EneQ1}
E=\frac{Q^2}{4\ssigma}+E_0+E_2
\ee
shows that minimizing   $E$ with $Q$ fixed is equivalent to 
extremizing $L$,
since 
\be                         \label{deltaE}
\delta E=-\frac{Q^2}{4\ssigma^2}\,\delta\ssigma+\delta (E_0+E_2)=
-\omega^2\delta\ssigma+\delta (E_0+E_2)=
-\delta L\,,
\ee
so that the on-shell condition $\delta L=0$ follows from $\delta E=0$. 
It is also instructive to see how the same thing comes about within the 
Lagrange multiplier method. Introducing 
\begin{align}                      \label{EQ}
E_Q&=\int(\omega^2|\phi|^2+|\nabla\phi|^2+U) 
d^3 {\bf x}
+\mu\left(2\omega\int|\phi|^2 d^3\bx-Q\right), 
\end{align}
the condition $\partial E_Q/\partial\mu=0$ fixes the charge,   
the condition $\partial E_Q/\partial\omega=0$ 
gives $\mu=-\omega$ insuring  that $E_Q=L+const.$, so that the condition 
$\delta E_Q/\delta\phi=0$ reproduces the field equation. 

Equivalently, one can minimize the potential energy of the field, 
$E_0+E_2$, by keeping fixed 
\be                           \label{NN}
\ssigma=\int |\phi|^2 d^3\bx
\ee
via extremizing the functional 
\begin{align}                    \label{NE}
E_\ssigma&=\int(|\nabla\phi|^2+U) 
d^3 {\bf x}
+\mu_1\left(\int|\phi|^2 d^3\bx-\ssigma\right).
\end{align}
The condition $\partial E_\ssigma/\partial\mu_1=0$ imposes the constraint \eqref{NN},    
while the condition 
$\delta E_\ssigma/\delta\phi=0$ reproduces the field equation \eqref{Qe} with 
$\omega^2=-\mu_1$. The time dependence of the solutions is very implicit 
in this approach, since $\omega$ appears only as a Lagrange multiplier. 
It follows from Eq.\eqref{deltaE} that 
\be                             \label{omega2}
\omega^2=\frac{\partial(E_0+E_2)}{\partial \ssigma}, 
\ee
so that 
fixing $\ssigma$ fixes also the charge 
$Q=2\omega(\ssigma)\ssigma$.  
In the non-relativistic theory  $\ssigma$ 
can be viewed  as the particle number (see Sec.\ref{VS} below).

Let us also note that 
using the charge definition 
\eqref{QQ} and the virial relation \eqref{Qvirial}
the energy \eqref{EneQ} can be rewritten   as
\be                             \label{Eomega}
E=\omega Q+\frac23\,E_2\,.
\ee

$Q$-balls could exist in the supersymmetric extensions of Standard Model 
\cite{Kusenko:1997zq} and could perhaps contribute to    
the dark matter \cite{Kusenko:1997si}.

\subsubsection{Non-spinning $Q$-balls}

Let us briefly  consider the simplest  spherically symmetric 
$Q$-balls \cite{Coleman:1985ki}, since this will help to understand 
more complex, spinning solutions.  
Setting 
$\Phi=e^{i\omega }f(r)$, the 
real amplitude $f(r)$ satisfies the equation 
\be                                   \label{Qeq0}
f^{\prime\prime}+\frac2r\,f^\prime+\omega^2f=
\frac12\frac{\partial U}{\partial f}\,.
\ee
For the energy to be finite
$f$ should vanish at infinity, so that  
for large $r$ one has 
\be                                    \label{larger}
f\sim\frac{1}{r}\,\exp\{-\sqrt{M^2-\omega^2}\,r\}.
\ee

\begin{figure}[h]
\hbox to\linewidth{\hss%
  \psfrag{x}{$r$}
  \psfrag{n=0}{$n=0$}
  \psfrag{n=1}{$n=1$}
  \psfrag{n=2}{$n=2$}
  \resizebox{7cm}{6cm}{\includegraphics{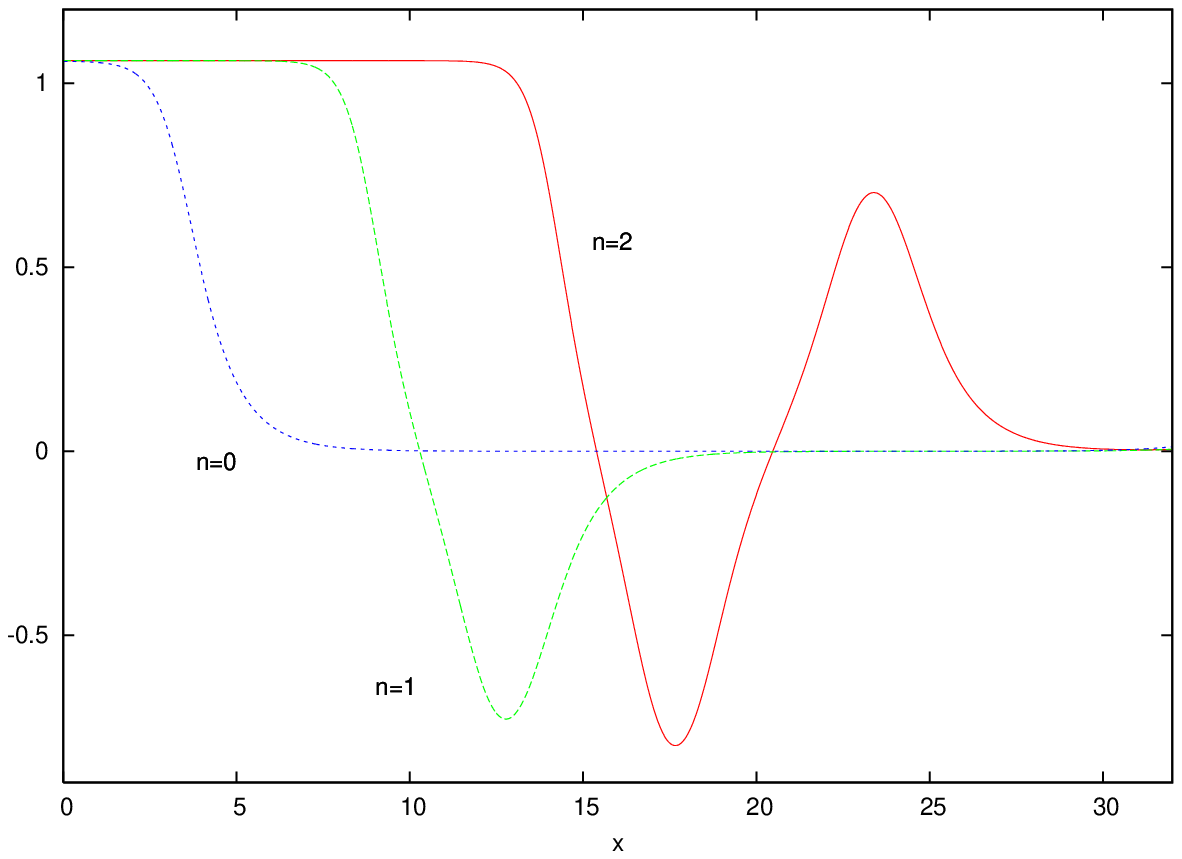}}%
\hspace{1mm}
  \psfrag{x}{$f$}
  \psfrag{f0}{$f_0$}
  \psfrag{fi}{$f_\infty=0$}
  \psfrag{y}{$U_{\rm eff}$}
  \psfrag{o0}{$A$}
 \psfrag{B}{$B$}
 \psfrag{C}{$C$}
 \psfrag{D}{$D$}
  \psfrag{o1}{$\omega_{+}$}
  \psfrag{o2}{$\omega_{-}$}
  \resizebox{7cm}{6cm}{\includegraphics{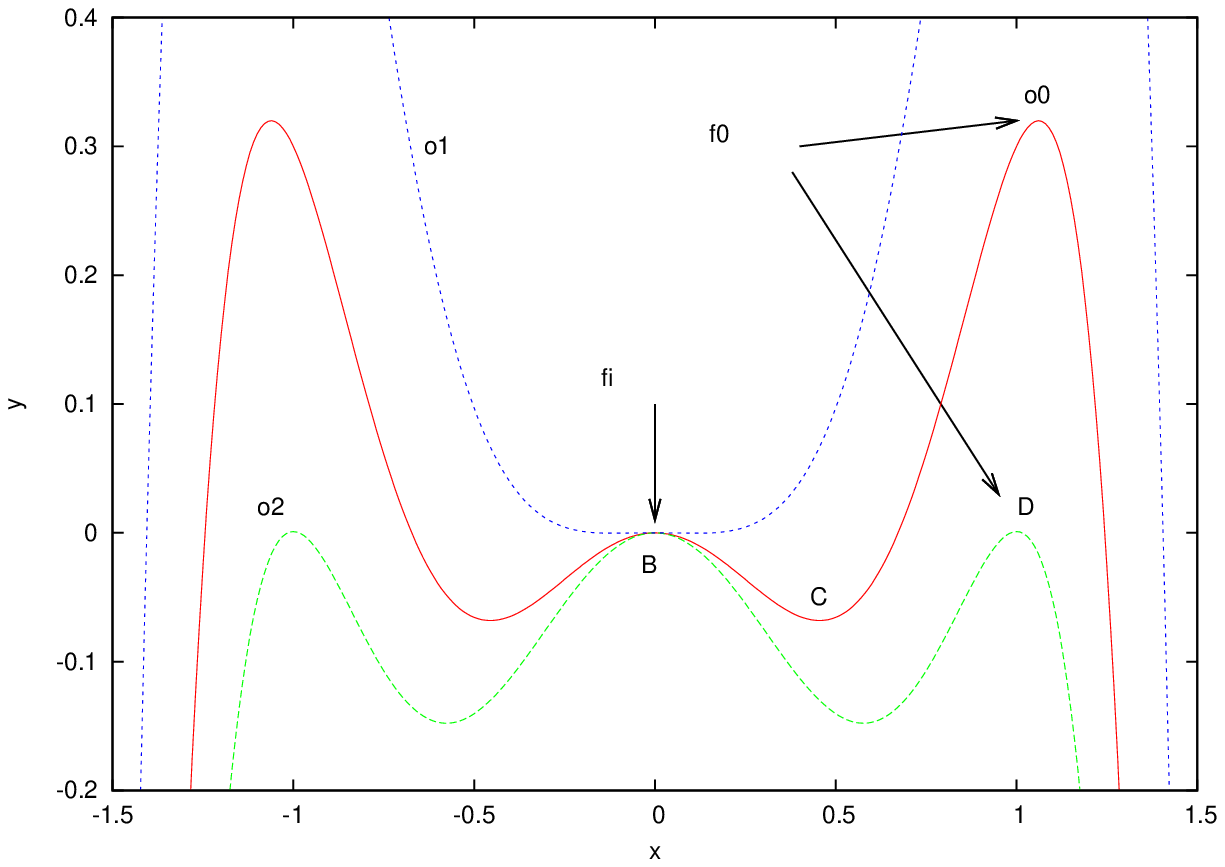}}%
\hss}
\caption{Left: the amplitude $f(r)$ for the $n=0,1,2$ spherically symmetric $Q$-balls 
with $\omega=0.2$. Right: the effective potential $U_{\rm eff}$ in Eq.\eqref{fric}
for three values of $\omega$.}
\label{FIG:Q}
\end{figure}

Solutions of Eq.\eqref{Qeq0} comprise an infinite family 
labeled by an integer 
$n=0,1,\ldots $ counting the 
nodes of $f(r)$ \cite{Volkov:2002aj} (see Fig.\ref{FIG:Q}). 
 The energy increases with $n$. 
These solutions admit a simple qualitative interpretation 
\cite{Coleman:1985ki}, 
since Eq.\eqref{Qeq0} is then equivalent to 
\be                                   \label{fric}
f^{\prime 2}+U_{\rm eff}(f)=\tilde{E}-4\int_0^r f^{\prime 2}\,\frac{dr}{r}\,.
\ee
This describes a particle moving with friction in one-dimensional potential
$U_{\rm eff}(f)=\omega^2 f^2-U(f)$ shown in Fig.\ref{FIG:Q}, 
the integration constant $\tilde{E}$ playing a role of the total energy.
At the `moment' $r=0$ the `particle' rests 
at a point with coordinate $f_0$ 
close to the potential maximum $A$ (see Fig.\ref{FIG:Q}), so that its energy is 
$\tilde{E}=U_{\rm eff}(f_0)$.  Then it starts moving to the left,  
dissipating  its energy as it goes. One can  adjust the value of $f_0$ so that 
for $r\to\infty$ it dissipates all its energy and arrives at the local maximum $B$ 
of the potential with zero total energy to rest there -- either directly or after 
$n$ oscillations between the 
two potential hills. 

It follows that $\omega$ should belong to the interval 
$\omega_{-}<\omega < \omega_{+}$, the condition 
Eq.\eqref{omega+} making  sure that this interval is non-empty. 
 $Q$-balls become large as $\omega\to\omega_{\pm}$, their charge and energy
growing without bounds. 

As $\omega\to\omega_{-}$ the maximum $A$ of the potential descends
towards the position $D$. The `particle' then stays for a `long time' at $D$,
till the friction term becomes suppressed by the $1/r$ factor. 
Then it starts moving, crosses the potential well in a finite `time' $\Delta r$ 
and asymptotically approaches the potential maximum $B$. In the simplest $n=0$ case 
solutions in this limit  
can be described by a smoothed step-function, 
\be                       \label{thin}
f(r)\approx f_0\Theta(R-r),
\ee 
where $R\to\infty$ as $\omega\to\omega_{-}$ and  
$f^\prime\approx f_0/\Delta r$ 
in a region of a fixed size $\Delta r\ll R$ around $r=R$.  
This is sometimes called 
$Q$-balls in the thin wall approximation \cite{Coleman:1985ki}. 
Their charge $Q\sim R^3$
while $E_2\sim R^2$ so that for large $R$ 
one can neglect the second term in Eq.\eqref{Eomega},
which gives for their energy 
$E=\omega_{-} Q$.
The same value is obtained inserting \eqref{thin} to Eq.\eqref{EneQ1},
neglecting the $E_2$ term and minimizing with respect to $R,f_0$.

When $\omega\to\omega_{+}$ the local minimum $C$
of the potential approaches the local maximum $B$. The potential 
in their vicinity becomes approximately 
$U_{\rm eff}\approx f^2(f^2-M_\omega^2)$ with $M_\omega^2=\omega_{+}^2-\omega^2$. 
Inserting this to Eq.\eqref{fric} gives $Q$-balls in the thick wall
approximation \cite{Kusenko}, 
\be                          \label{thick}
f(r)\approx \sqrt{2}M_\omega\, y(M_\omega r),
\ee
where $y(x)$ fulfills 
$y''+(2/x)y'+(y^2-1)y=0$. Solutions of this 
equation satisfy 
$y(0)=y_n$ and 
$y\sim e^{-x}$ for large $x$ (one has $y_0=4.33$).
Although $f\sim M_\omega$ is small in this case, 
the typical configuration radius $R\sim 1/M_\omega$ is large. The charge 
is also large, $Q\sim   1/M_\omega$, while $E_2\sim  M_\omega$,
so that  
one can again neglect the second term in Eq.\eqref{Eomega},
which gives for the energy 
$E=\omega_{+} Q$.

\subsubsection{Spinning $Q$-balls}

The spinning, axially symmetric generalizations for the 
spherically symmetric $Q$-balls 
have been constructed 
 by Volkov and Wohnert \cite{Volkov:2002aj} 
and have been analysed also 
in Refs.\cite{Axenides:2001pi},
\cite{Kleihaus:2005me},
\cite{Kleihaus:2007vk},
\cite{Brihaye:2007tn}. 
These {non-manifestly} stationary and non-manifestly axisymmetric
solutions provide the first explicitly constructed
example of stationary, spinning solitons in Minkowski 
space in a relativistic field theory in $3+1$ spacetime dimensions. 
They have the typical ring structure, very similar to that  for vortons,
and they can also have the same topology 
as vortons. They admit simple generalizations to the gauged case. 
Spinning $Q$-balls thus provide a simple prototype example of vortons. 
Their analogs also exist  in the context of non-linear optics
\cite{mihalache}, where they describe spinning light pulses
(see Sec.\ref{light} below).

Spinning $Q$-balls have a spinning phase:  
the field being non-manifestly stationary and non-manifestly
axisymmetric, 
\be
\Phi=e^{i(\omega t +m\varphi)}f(r,\vartheta).
\ee
The real field amplitude $f(r,\vartheta)$ satisfies the equation 
\be                                   \label{Qeq}
\left(\Delta-\frac{m^2}{r^2\sin^2\vartheta}+\omega^2\right)f=
\frac12\frac{\partial U}{\partial f}\,.
\ee
For $m\neq 0$ the energy will be finite if only  $f$ vanishes at the symmetry axis,
while at infinity the asymptotic behavior \eqref{larger} still applies. 
The energy momentum $T^\mu_\nu$ depends only on $r,\vartheta$
and the angular momentum is 
\be
J=\int T^0_\varphi\, d^3 {\bf x} =2m\omega\int f^2 d^3 {\bf x}=2m\omega\ssigma,
\ee
so that it is classically quantized as 
\be                               \label{JQ}
J=mQ\,.
\ee
In view of this relation, 
spinning $Q$-balls correspond to minima of 
energy with fixed angular momentum. They could therefore be obtained
by extremizing 
\be                                                   \label{EJ}
E=\frac{J^2}{4m^2\int f^2 \,d^3\bx}+\int\left((\nabla f)^2
+\frac{m^2 f^2}{r^2\sin^2\vartheta}+U\right)d^3\bx
\ee
with constant $J,m$, the corresponding extremum condition being given by
 Eq.\eqref{Qeq}. A direct minimization of this functional 
carried out in Ref.\cite{Axenides:2001pi} 
(although for a different choice of the potential $U$)
suggests  that its extrema exist but seem to have negative directions. 
One should therefore integrate the field equation to construct these solutions.

\begin{figure}[ht]
\hbox to\linewidth{\hss%
	\resizebox{6cm}{6cm}{\includegraphics{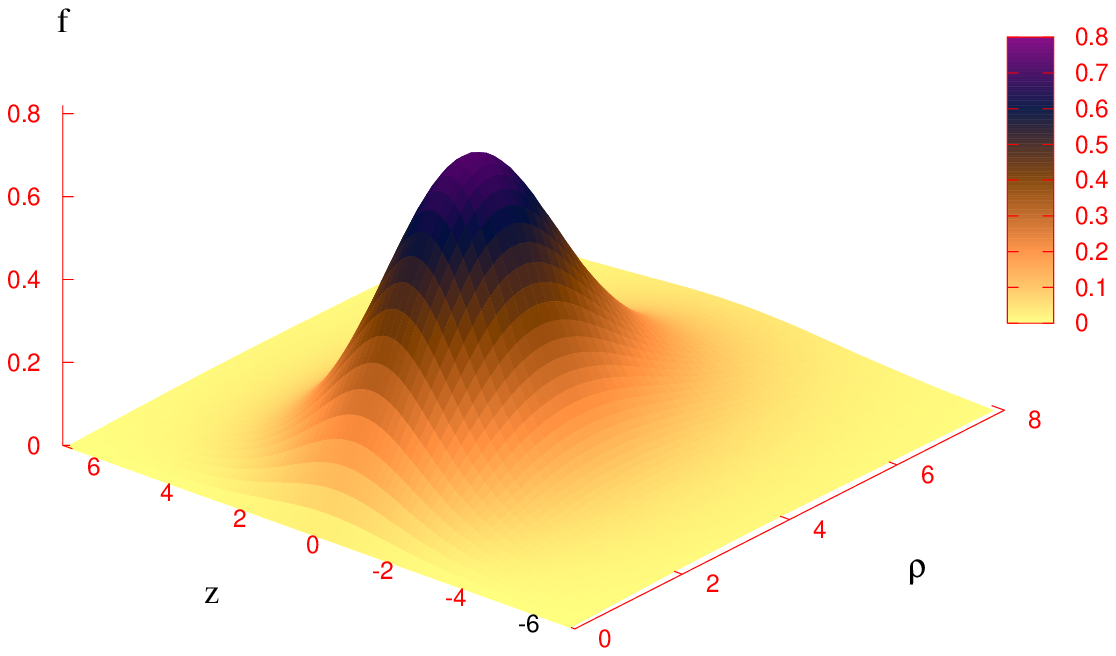}}
\hspace{1mm}%
        \resizebox{6cm}{6cm}{\includegraphics{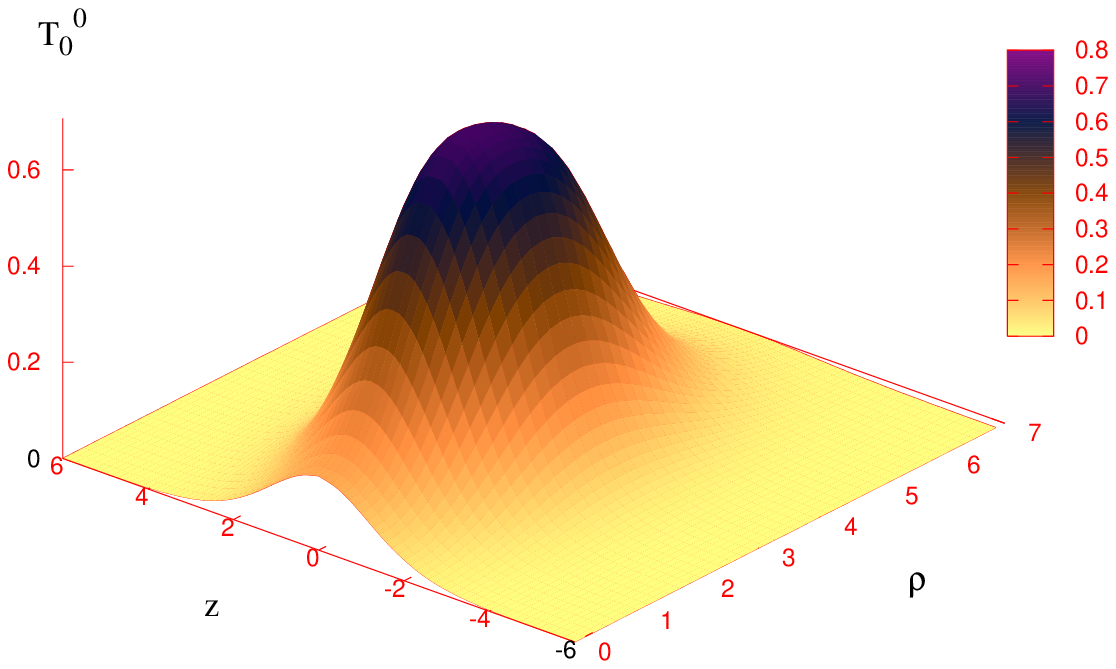}}
\hspace{1mm}%
        \resizebox{4cm}{4cm}{\includegraphics{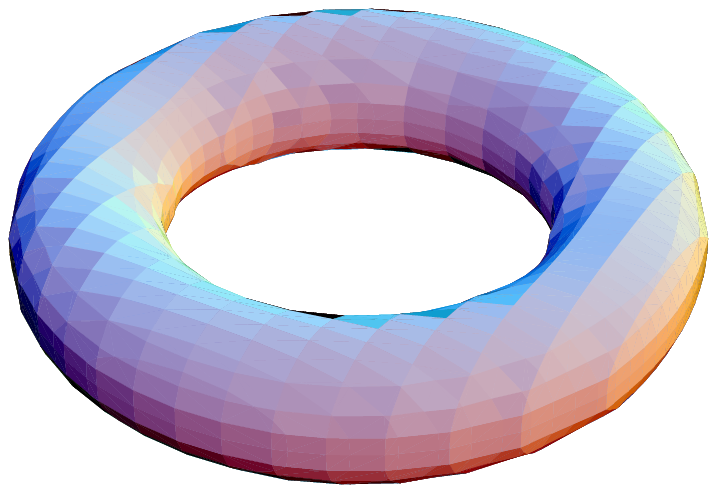}}	
\hss}
	\caption{The amplitude $f(r,\vartheta)$, energy density $T^0_0$
and the energy isosurface with $T^0_0=0.65$  
for the $1^{+}$ spinning $Q$-ball with $\omega=0.9$.}
\label{FigQ1}
\end{figure}

Taking into account the boundary conditions at the $z$-axis and at infinity, 
Eq.\eqref{Qeq} with $m\neq 0$ admits 
two different types of solutions determined by the 
behaviour of $f$ under $z\to-z$ \cite{Volkov:2002aj}. 

For even-parity solutions, called $m^{+}$, 
the amplitude $f(r,\vartheta)$, energy-momentum and charge densities are 
maximal in the equatorial plane and the energy is  concentrated
in a toroidal region encircling the $z$-axis (see Fig.\ref{FigQ1}). 

\begin{figure}[ht]
\hbox to\linewidth{\hss%
	\resizebox{6cm}{6cm}{\includegraphics{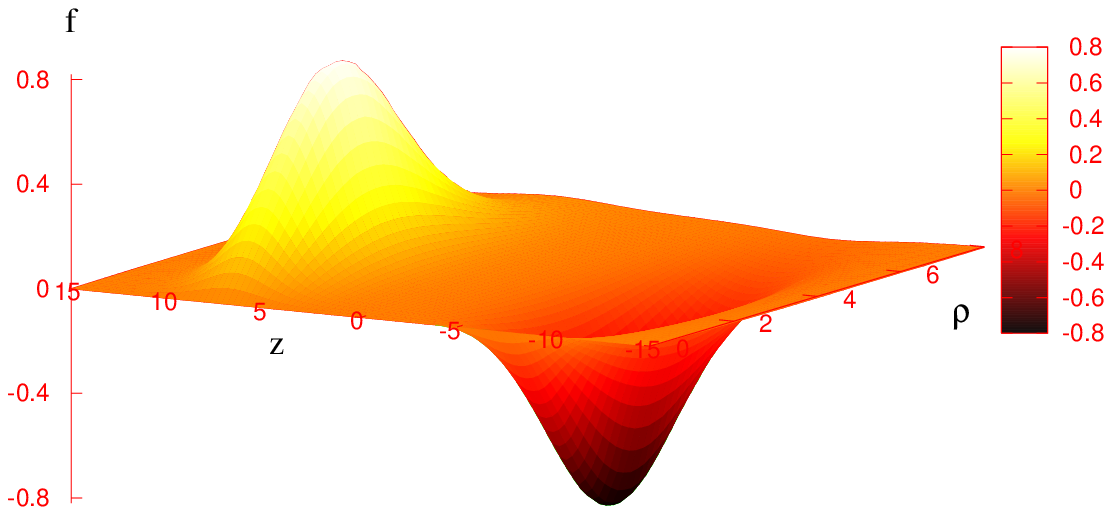}}
\hspace{1mm}%
        \resizebox{6cm}{6cm}{\includegraphics{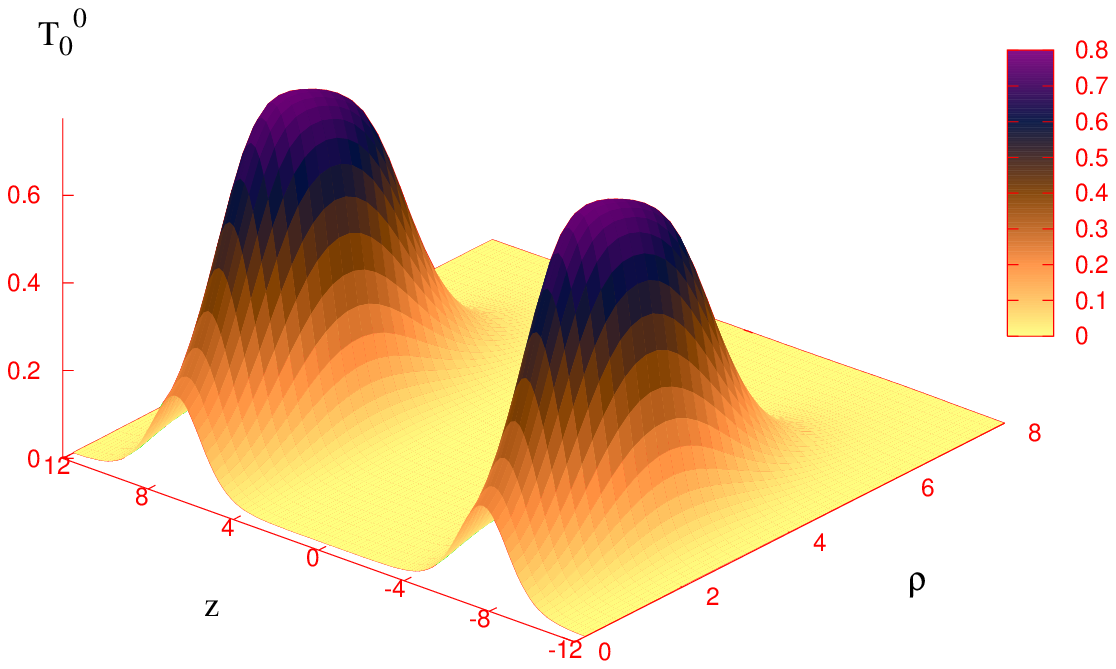}}
\hspace{1mm}%
        \resizebox{4cm}{5cm}{\includegraphics{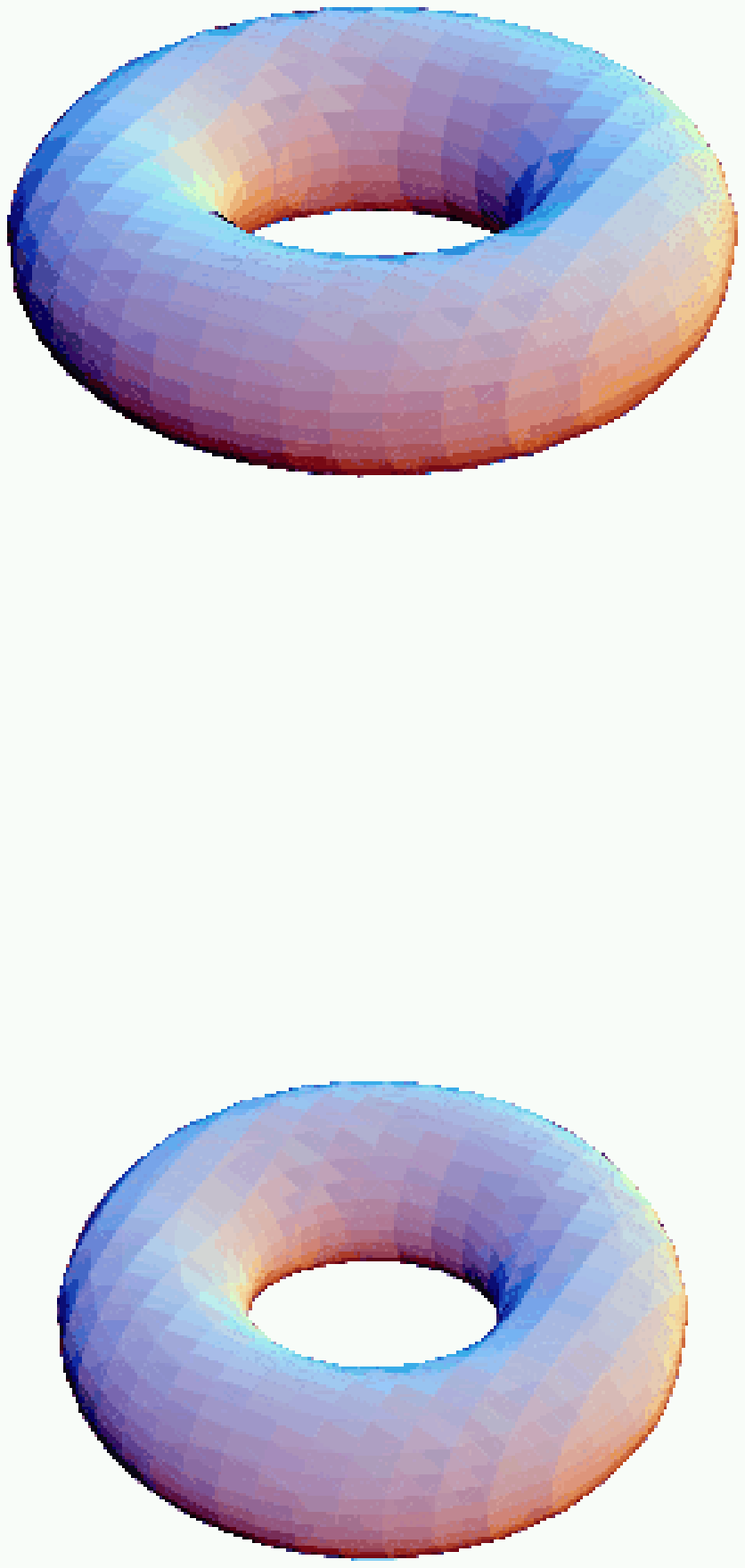}}	
\hss}
	\caption{The amplitude $f(r,\vartheta)$, energy density $T^0_0$
and the energy isosurface with $T^0_0=0.65$  
for the $1^{-}$ spinning $Q$-ball with $\omega=0.9$.}
\label{FigQ2}
\end{figure}

For odd-parity solutions, called $m^{-}$, 
the amplitude $f(r,\vartheta)$
vanishes in the equatorial plane, while the 
energy-momentum and charge densities  
show two maxima located symmetrically with 
respect to the plane, so that the solutions 
exhibit in this case a double torus or dumbbell-like 
structure (see Fig.\ref{FigQ2}).

\begin{figure}[ht]
\hbox to\linewidth{\hss%
	\resizebox{7cm}{6cm}{\includegraphics{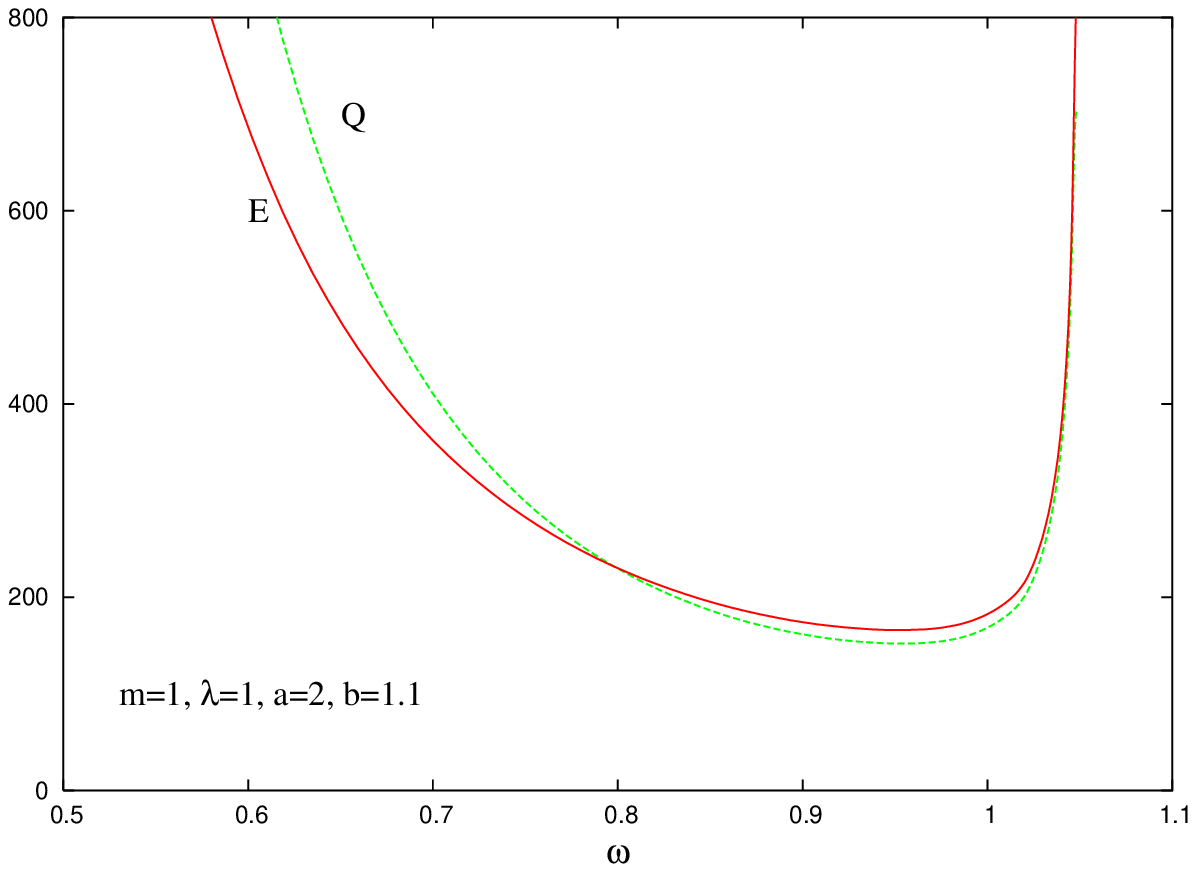}}
\hspace{5mm}%
        \resizebox{7cm}{6cm}{\includegraphics{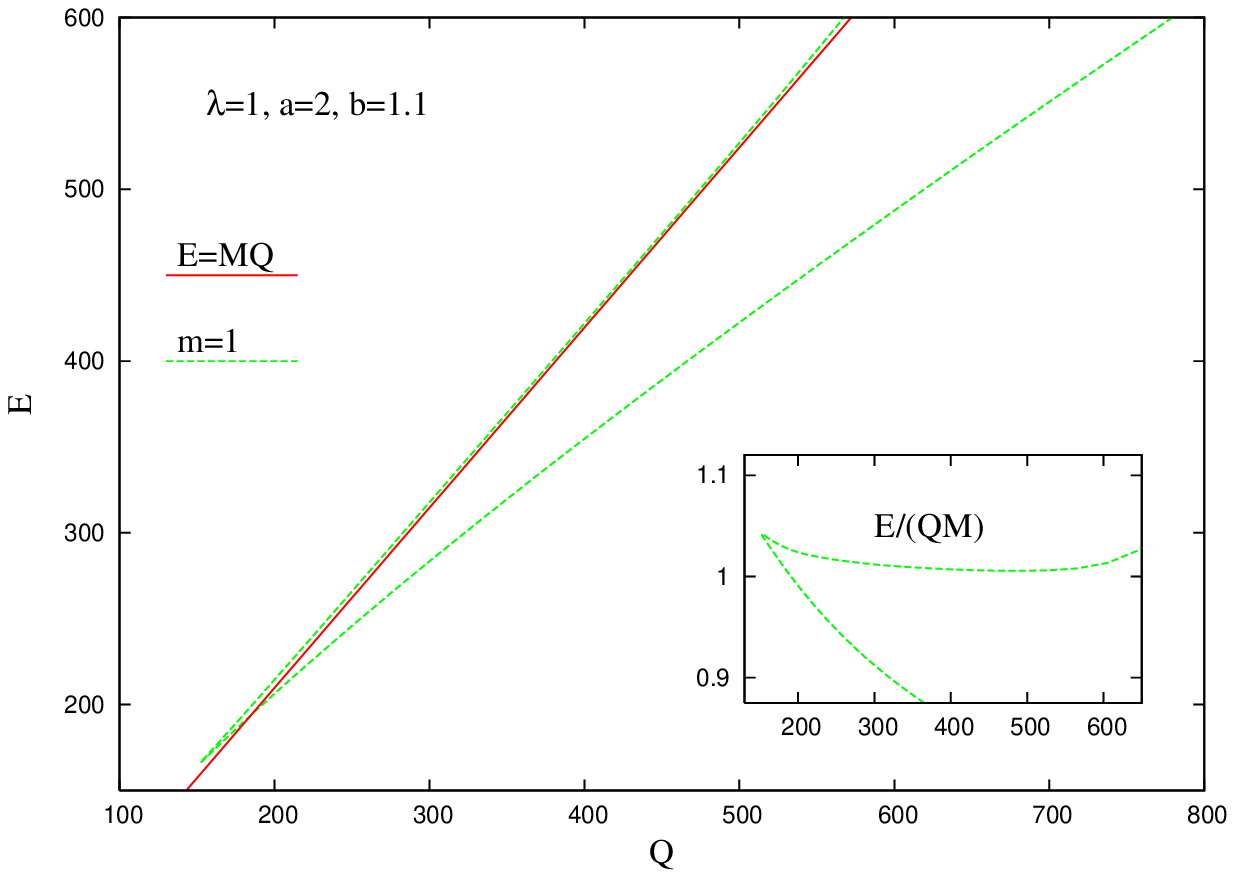}}	
\hss}
	\caption{Energy and charge 
$E(\omega),Q(\omega)$ (left) and $E(Q)$ (right)
for the $1^{+}$ spinning $Q$-balls. For large $Q$ one has 
$E(Q)=\omega_{-} Q$ and $E(Q)=\omega_{+} Q$ for the lower
an upper branch, respectively. The upper branch solutions are unstable,
since $E/(MQ)>1$, while the lower branch solutions are stable for large enough $Q$.}
\label{FigQ3}
\end{figure}

Apart from choosing the parity and the value of $m$, 
solving the differential equation \eqref{Qeq} also 
requires choosing $\omega^2$ as an input parameter,
the energy and charge being computed from the numerical output.
As in the spherically symmetric case,  
spinning solutions are also found to 
exist only in a finite frequency range, 
$\omega_{-} < \omega< \omega_{+}$. 
Here $\omega_{+}=M$ as in Eq.\eqref{omega+}, while $\omega_{-}$ seems to be
$m$-dependent. 
As $\omega$ approaches the limiting values, 
the energy and charge seem to grow
without bounds (see Fig.\ref{FigQ3}), while somewhere in between there is 
a critical value of frequency, $\omega_{\rm crit}$, 
for which  both $E(\omega)$ and $Q(\omega)$ attain 
their minimal values. This behaviour is found for all $m$ \cite{Kleihaus:2005me} 
and it is  qualitatively the same as in the $m=0$ case. 

The existence of a minimal value of the $Q$-ball charge implies that 
the angular momentum \eqref{JQ} cannot be arbitrarily small. $Q$-balls
cannot therefore rotate slowly. They  show a {\it discrete} spectrum of spinning 
excitations.

Although it is difficult  to  qualitatively analyse  the behaviour  
of the $m\neq 0$ solutions, there are some analogies with the $m=0$ case. 
Specifically, $Q$-balls become large as $\omega\to\omega_{\pm}$.
For $\omega\to\omega_{-}$ they can be viewed as squashed spheroids,
homogeneously filled inside, which reminds of the thin wall approximation.  
Unfortunately, an analog of the step function solution \eqref{thin} does not 
directly apply for $m\neq 0$, 
since the function $f$ cannot be constant 
inside the ball because it must vanish at the symmetry axis.  Instead,
$f$ increases as one moves away from the  axis and reaches maximal values at the 
surface of the spheroids, after which it rapidly goes to zero. 
The energy density is approximately constant inside the spheroid, with a slight
increase at its surface, and rapidly vanishes outside it.   
 For $\omega\to\omega_{+}$ solutions also become large spheroids,
but this time they are hollow, with the maximal energy density concentrated
at the surface and being close to zero everywhere else. It seems that the thick wall
approximation \eqref{thick} can be directly generalized to the axially 
symmetric case to give   
$
f(r,\theta)\approx \sqrt{2}M_\omega\, y(M_\omega r,\theta)
$
where $M_\omega\to 0$ and $y(r,\theta)$ fulfills 
$(\Delta-\frac{m^2}{r^2\sin^2\theta} +y^2-1)y=0$.

If one considers $Q$ and not $\omega$ as the solution parameter 
then, using $E(\omega)$ and $Q(\omega)$  to express $E$ in terms of $Q$,
one discovers that the function $E(Q)$ is double-valued with a cusp,
as shown in  Fig.\ref{FigQ3} 
\cite{Kleihaus:2005me}.
For a given $Q$ there are thus two different 
spinning $Q$-ball solutions with different energies.   
Solutions from the less energetic branch correspond to the 
$\omega<\omega_{\rm crit}$ parts of the $E(\omega)$, $Q(\omega)$ curves
in Fig.\ref{FigQ3}. As in the $m=0$ case, the plots shown in Fig.\ref{FigQ3} 
demonstrate for large $Q$ the 
linear dependence, $E=\omega_{\pm}Q$. This can again be explained
by the general relation \eqref{Eomega}, implying  that for large $Q$
one can neglect
the gradient energy $E_2$ as compared to $Q$. 

The existence of two different solutions with the same $Q$ suggests that the 
more energetic of them is unstable. In fact, it follows from Eq.\eqref{Eomega}
that the energy-to-charge 
ratio for $Q$-balls is $E/Q=\omega+2E_2/3Q$.
For the upper branch solutions  one has $\omega\to\omega_{+}=M$ for large $Q$, 
implying that $E/Q=M+O(Q^{-2})$ where the subleading term can in principle 
be positive or negative, depending on the details of the $Q$-ball potential. 
In the present case it is positive, as can be seen in Fig.\ref{FigQ3}, so that 
the upper branch $Q$-balls are unstable with respect 
to decay into free particles.     
The same argument for the lower branch solutions gives 
$E/Q=\omega_{-}<M$ for large $Q$, so that they cannot decay into free particles. 
However, there can be other decay modes, and so the stability analysis is needed.
In fact, the $m=0$ lower branch $Q$-balls are known to be stable 
\cite{Correia}.       
The stability analysis of the even parity solutions with $m\neq 0$ will be described below 
in Sec.\ref{light}. 
It seems that for  $|m|=1$ the lower branch solutions are stable only for large
enough $Q$, while already for $|m|=2$ all of them are unstable -- 
 they seem to decay by splitting
into several non-spinning $Q$-balls.

It is plausible that spinning generalizations could also be constructed 
for the excited spherically symmetric $Q$-balls with $n>0$. 
The complete family of spinning $Q$-balls should therefore contain 
not only $m^\pm$ solutions, but also the excited $(n,m^{\pm})$ solutions with  
$n=1,2,\ldots $ for which the amplitude $f(r,\vartheta)$ exhibits nodes. 
More precisely, parametrizing the $(\rho,z)$ plane by a complex variable
$w=\rho+i z$, one can expect the 
amplitude $f(w)$ of the $(n,m^{+})$ and $(n,m^{-})$ solutions to have the same zeros as,
respectively,  $\Re(F^{+}_n(w))$ and $\Im(F^{-}_n(w))$,
where 
\be                                     \label{complex} 
F^\pm_n(w)=\prod_{j=1}^n \alpha^\pm_j(w-\rho^\pm_j),
\ee
with some suitably chosen 
$\alpha^\pm_j\in\mathbb{C}$ and $\rho^\pm_j>0$. 
However, very little is known about such excited solutions at present.

\subsubsection{Twisted $Q$-balls}

Here and in the next subsection we briefly describe our 
new results on further generalizations of the spinning $Q$-balls,
not yet discussed in the literature. Let us consider again 
the theory \eqref{lQ}, but 
generalize the field ansatz to include an independent phase,
\be
\Phi=e^{i(\omega t +m\varphi -n\psi(r,\theta))}f(r,\theta)\equiv 
(X(r,\vartheta)+iY(r,\vartheta))e^{i(\omega t +m\varphi)}.
\ee
We require the phase function $\psi(r,\vartheta)$ to increase by $2\pi $
after one revolution around the contour $C$  shown
in Fig.\ref{FigS}. The overall phase 
$\omega t +m\varphi-n\psi(r,\theta)$ then winds 
around the circle $S$ and along the contour $C$,
exactly as for the Faddeev-Hopf field Eq.\eqref{Hopf:axial}. 
For regular fields $f(r,\vartheta)$ 
vanishes  at $C$ and has  $n$ zeros inside $C$. 
The solutions are thus characterized by two integers $(n,m)$ giving rise 
to the `topological charge' $\Q=nm$, 
although now this does not represent a genuine  topological invariant.  
We shall call $Q$-balls with $\Q\neq 0$ twisted, by analogy with twisted loops
in the Faddeev-Hopf theory,
while those with $\Q=n=0$ described above will be called simply spinning
or `non-twisted'. 

\begin{figure}[ht]
\hbox to\linewidth{\hss%
	\resizebox{7cm}{6cm}{\includegraphics{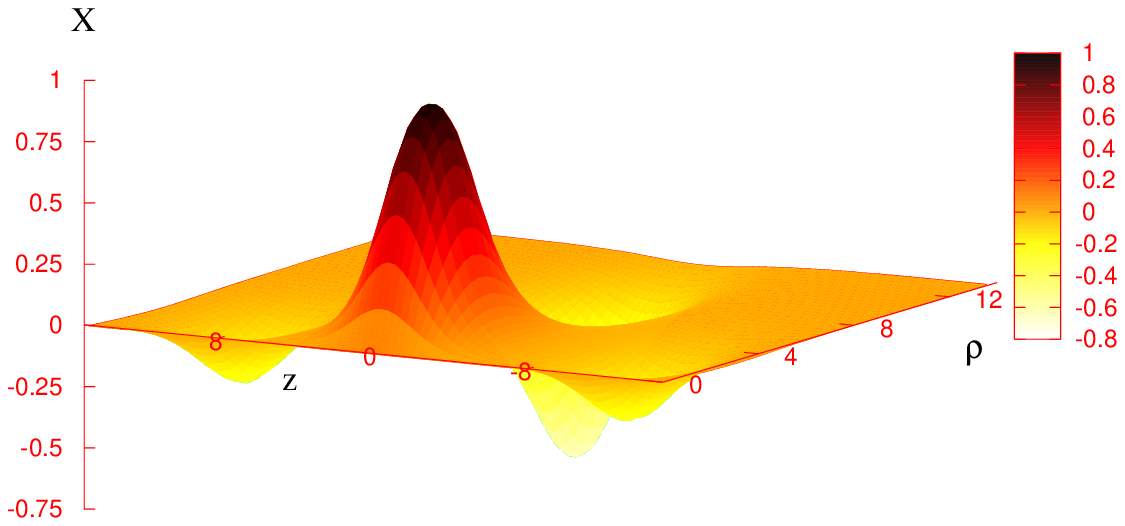}}
\hspace{5mm}%
        \resizebox{7cm}{6cm}{\includegraphics{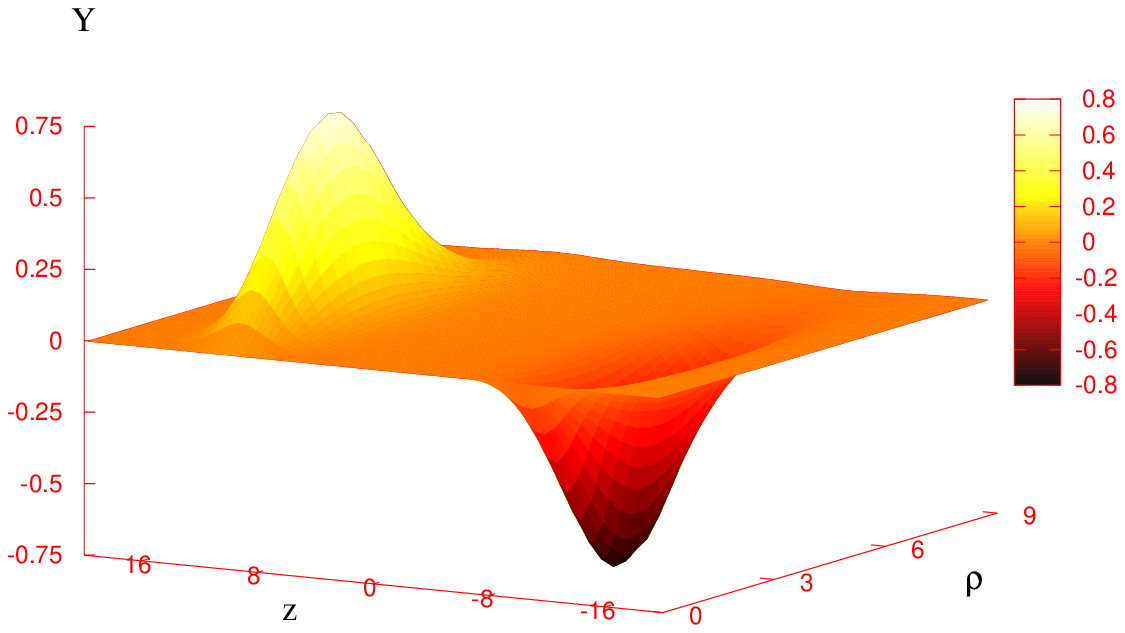}}	
\hss}
	\caption{Profiles of the $(1,1)$ twisted $Q$-ball with $\omega=0.9$.
}
\label{FigQ0}
\end{figure}

The field equations read  
\begin{align}                                   \label{Qeq1}
\left(\Delta-\frac{m^2}{r^2\sin^2\vartheta}+\omega^2\right)X&=
\frac12\frac{\partial U(\sqrt{X^2+Y^2})}{\partial X}\,, \notag \\
\left(\Delta-\frac{m^2}{r^2\sin^2\vartheta}+\omega^2\right)Y&=
\frac12\frac{\partial U(\sqrt{X^2+Y^2})}{\partial Y}\,,
\end{align}
where one can require $X,Y$ to be symmetric and antisymmetric, respectively, 
with respect to reflections in the equatorial plane. 
The simplest twisted solutions are obtained 
for $n=m=1$ and are shown in Fig.\ref{FigQ0}.

\begin{figure}[ht]
\hbox to\linewidth{\hss%
	\resizebox{7cm}{6cm}{\includegraphics{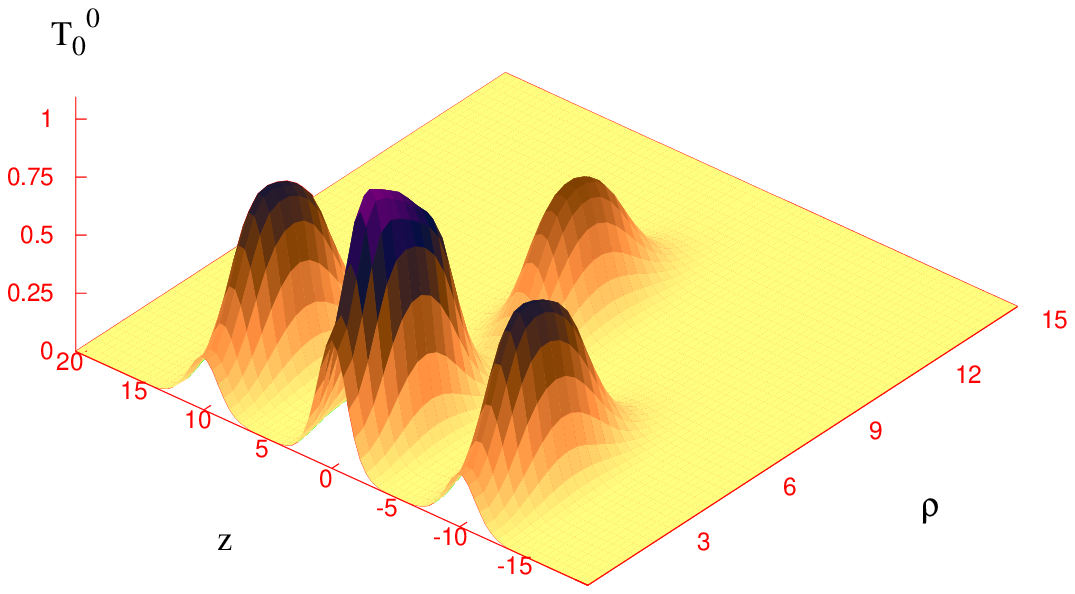}}
\hspace{5mm}%
        \resizebox{7cm}{6cm}{\includegraphics{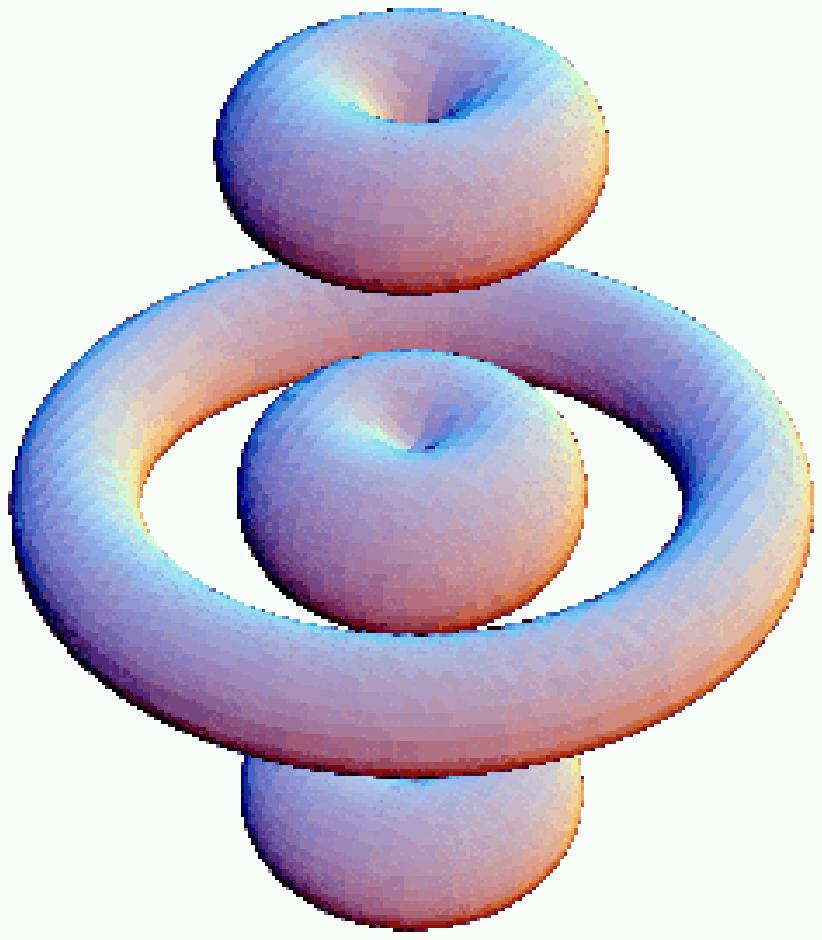}}	
\hss}
	\caption{Energy density for the $(1,1)$ twisted $Q$-ball 
with $\omega=0.9$ (left) and the isosurfaces of constant
energy density with 
$T^0_0=0.3$ (right).}
\label{FigQQ}
\end{figure}
The profiles of the energy density look spectacular; see Fig.\ref{FigQQ}.
It is unclear  at present 
whether these solutions are dynamically
stable, but they certainly exist
as solutions of the elliptic system  \eqref{Qeq1}.
The twisted $Q$-balls are much more heavy 
than the non-twisted ones. For example, for $\omega=0.9$   
we found for the twisted $(1,1)$ solution the energy
$E=929.34$,
while the non-twisted  $1^{+}$ and $1^{-}$ spinning
$Q$-balls have, respectively, 
$E=169.27$ and $E=338.66$.

\subsubsection{Spinning gauged $Q$-balls}

Another possibility to generalize the spinning $Q$-balls 
is to couple them to a gauge field 
within the theory 
\be                                                     \label{lQg}
{\mcal L}_{\rm gQ}[A_\mu,\Phi ]=
-\frac14\,F_{\mu\nu}F^{\mu\nu}+
{\mcal L}_{\rm Q}[\Phi ].
\ee
Here $F_{\mu\nu}=\partial_\mu A_\nu-\partial_\nu A_\mu$ while 
${\mcal L}_{\rm Q}[\Phi ]$ is the same $Q$-ball Lagrangian as in \eqref{lQ} but with 
the derivatives of $\Phi$  
replaced by the covariant derivatives, 
$\partial_\mu\Phi\to D_\mu\Phi=(\partial_\mu-i g A_\mu)\Phi$ 
where $g$ is the gauge coupling constant. 
This theory is in fact the U(1) version  of the general  gauge field model \eqref{YMH}.
The local gauge transformations \eqref{UUU} now read
$\Phi \to \Phi e^{i g \alpha }$,
$A_\mu\to A_\mu +\partial_\mu \alpha$ 
while the field equations \eqref{YMHeqs} assume the form
\begin{align}
\label{gauged-Q-ball-eqs}  
 \partial^\mu   F_{\mu\nu}&= 
ig \left \{({D}_\mu\Phi)^\ast\Phi- \Phi^\ast ({D}_\mu\Phi) \right\}
\equiv g j_\nu,  \notag \\
{D}_\mu {D}^\mu \Phi &=-\frac{\partial U}{\partial |\Phi|^2}\,\Phi.
 \end{align} 
The conserved Noether charge  analogues to the $Q$-ball charge 
Eq.\eqref{Noether} is 
\be
Q=\int j_0\, d^3\bx=\frac{1}{g}\oint 
\vec{\mcal E}d\vec{S}\equiv \frac{4\pi Q_{\rm el}}{g},
\ee 
where $Q_{\rm el}$ is the electric charge. 
Spherically symmetric solutions of this model were discussed in 
 Ref.\cite{Lee:1988ag}.

\begin{figure}[ht]
\hbox to\linewidth{\hss%
	\resizebox{7cm}{6cm}{\includegraphics{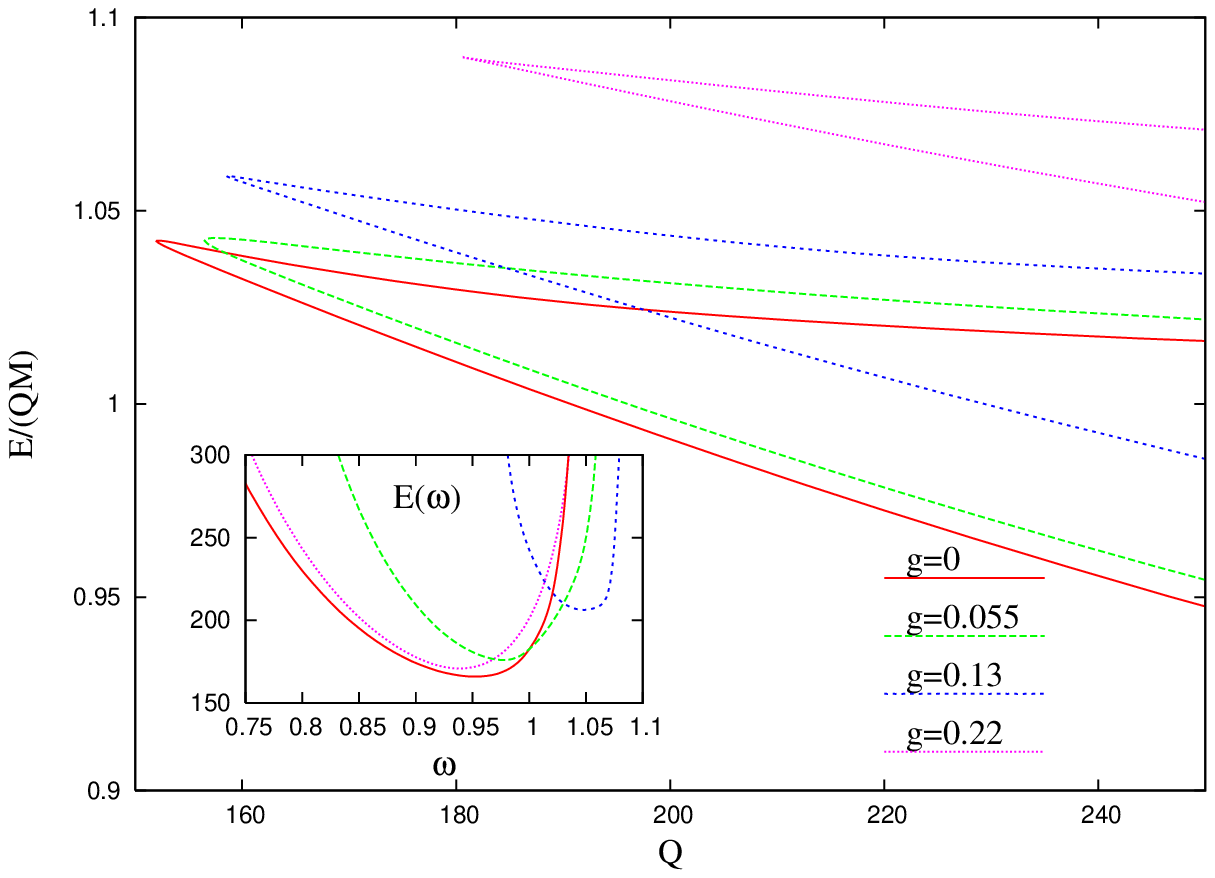}}
\hspace{5mm}%
        \resizebox{7cm}{6cm}{\includegraphics{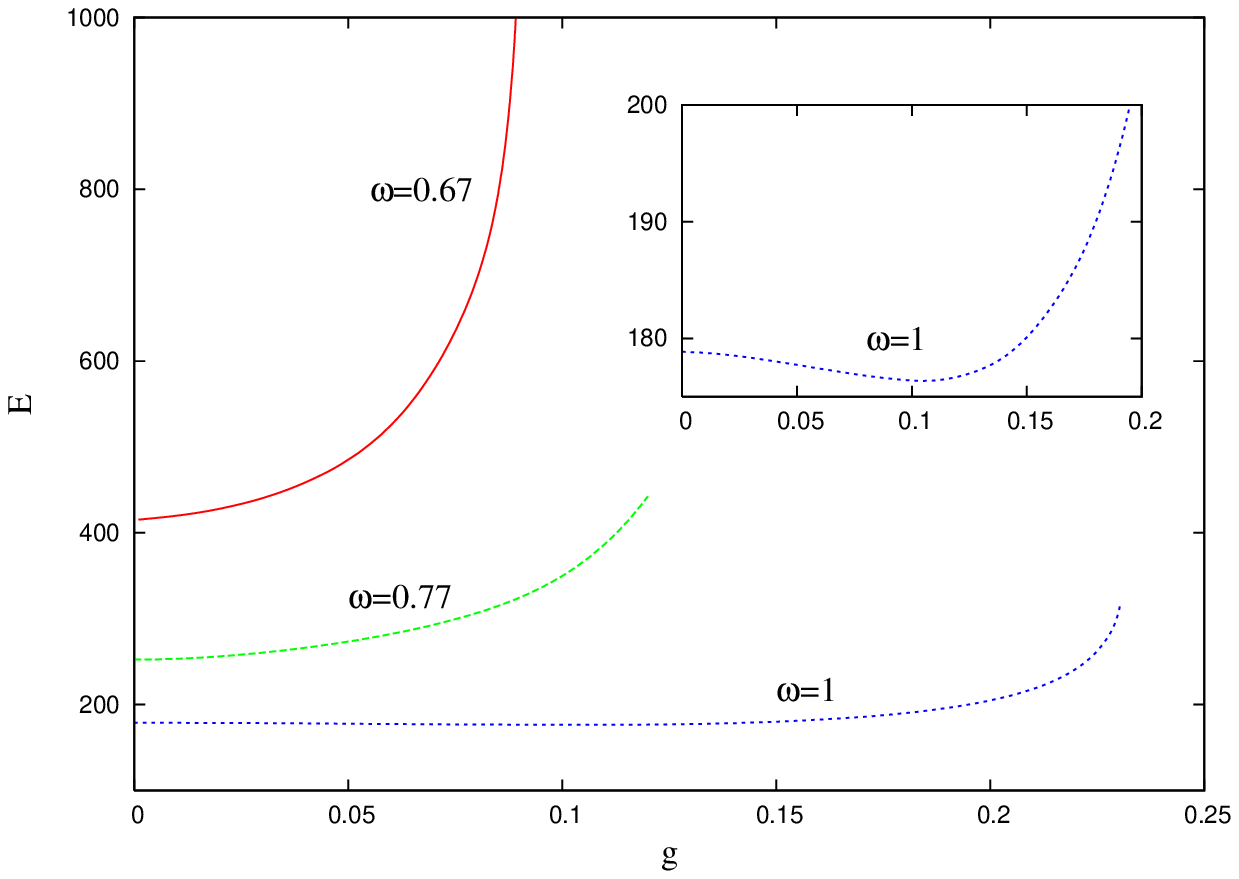}}	
\hss}
	\caption{Left: the energy  $E(\omega)$ and $E(Q)/(QM)$ 
for several values of $g$ for the $1^{+}$ gauged $Q$-balls.
Right: the energy $E(g)$ for several values of $\omega$
for the $1^{+}$ solutions. The plots of $Q(\omega)$ and $Q(g)$
look qualitatively similar.
}
\label{FigQg}
\end{figure}

We make the ansatz
\be                          \label{gQfields}        
A_\mu dx^\mu=A_0(r,\vartheta)dt+A_\varphi(r,\vartheta)\sin \theta d\varphi, ~~~~  
\Phi=f(r,\vartheta)e^{i (m\varphi+\omega t) },
\ee
with real $f(r,\vartheta)$, and require  
the gauge field to vanish at infinity, while at the $z$-axis 
$A_\varphi=\partial_\theta A_0=0$. 
Although $\Phi$ depends on $t,\varphi$, this dependence can be gauged away,
so that the system is 
{manifestly} stationary and {manifestly} axially symmetric. 
Numerically solving the field equations gives spinning $Q$-balls with 
a long range gauge field which behaves for large $r$ as   
(${\bf m}$ being  the magnetic dipole moment)
\be                           \label{Ainf}
A_0=\frac{Q_{\rm el}}{r}+\dots,~~~~~~~
A_\varphi=\frac{{\bf m}\sin \theta }{r^2}+\dots \,.
\ee 
In the limit $g\to 0$ they reduce to the non-gauged spinning $Q$-balls. 
It seems that solutions exist if only $g$ does not exceed 
a certain maximal value $g_{\rm max}(\omega)$ (see Fig.\ref{FigQg}). 
This feature can be understood qualitatively \cite{Lee:1988ag}: since 
$Q$-balls can be viewed as condensate states of mutually attracting scalar particles,
gauging them creates  an electric repulsion that destroys  
 the condensate for large enough $g$.

Since the symmetries of the solutions are manifest, one can use the 
surface integral formula \eqref{JJJ} for the angular momentum. 
Applying the symmetry equations 
\eqref{sym1} to the ansatz \eqref{gQfields} gives $W_\varphi=m/g$, 
inserting which to \eqref{JJJ} and using \eqref{Ainf} yields 
\be                                             \label{JJJ1}
J=\oint 
\left(\frac{m}{g}-A_\varphi\right)\vec{{\mcal E}}\, \vec{dS}
=\frac{4\pi Q_{\rm el} m}{g}=Qm.
\ee
This shows that spinning is possible in manifestly stationary and {manifestly} axially symmetric systems where all the spinning phases can be gauged away.

As seen in Fig.\ref{FigQg},  the dependence of solutions on $\omega$ 
is similar to that in the ungauged case: they exist  
for a limited range of $\omega$. 
Expressing $E$ in terms of $Q$ gives again a two-branch 
function $E(Q)$, the solutions from the lower branch being stable
for large enough $Q$ with respect to decay into free particles. 
$E,Q$ cannot be arbitrarily large for $g\neq 0$, since 
the electric and scalar field contributions to the energy grow as $Q_{\rm el}^2\sim Q^2$
and $Q$, respectively, and so for large $Q$ the electric term dominates
and destroys the soliton.

For small $g$  solutions can be represented as 
$\Phi=\Phi^{(0)}+g^2\Phi^{(2)}+\ldots$ and 
$A_\mu=gA^{(1)}_\mu+\ldots$ where 
$\Phi^{(0)}$ is the non-gauged $Q$-ball. 
The energy   is  
$
E(g)=E^{(0)}+g^2E^{(2)}+\ldots 
$
and calculating $E^{(2)}$ reveals  that it is not sign definite -- 
due to the electric field contribution. 
The plots in Fig.\ref{FigQg} show that it can be both 
positive and negative, depending on the solution.

\subsubsection{Spinning interacting $Q$-balls}
\begin{figure}[ht]
\hbox to\linewidth{\hss%
	\resizebox{6cm}{4cm}{\includegraphics{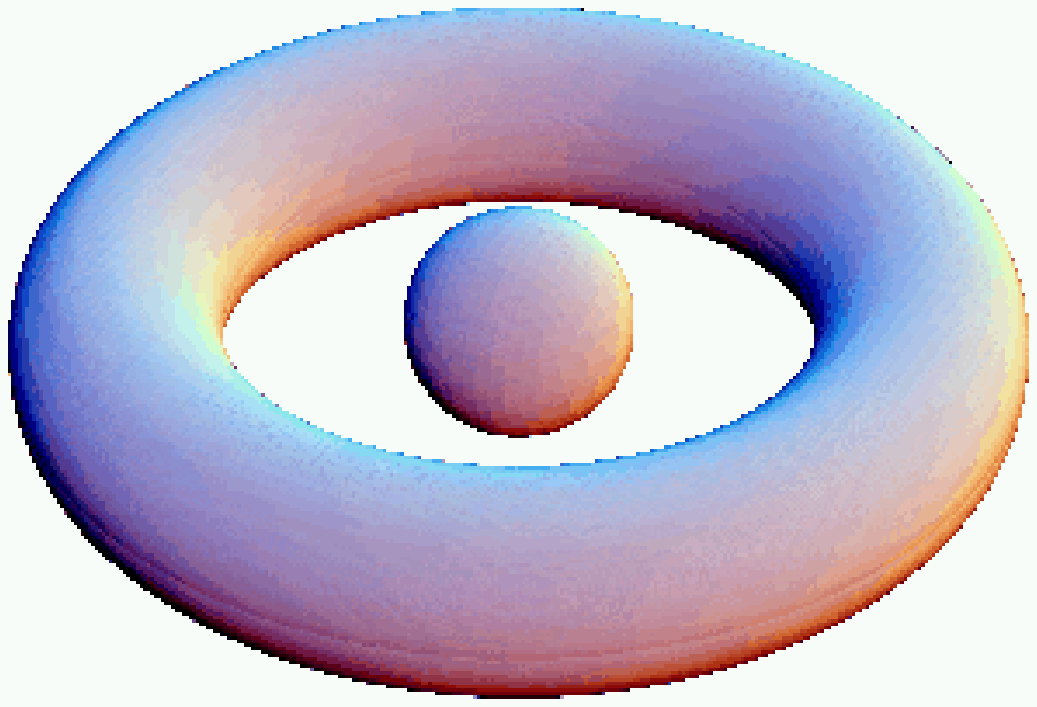}}
\hspace{5mm}%
        \resizebox{6cm}{4cm}{\includegraphics{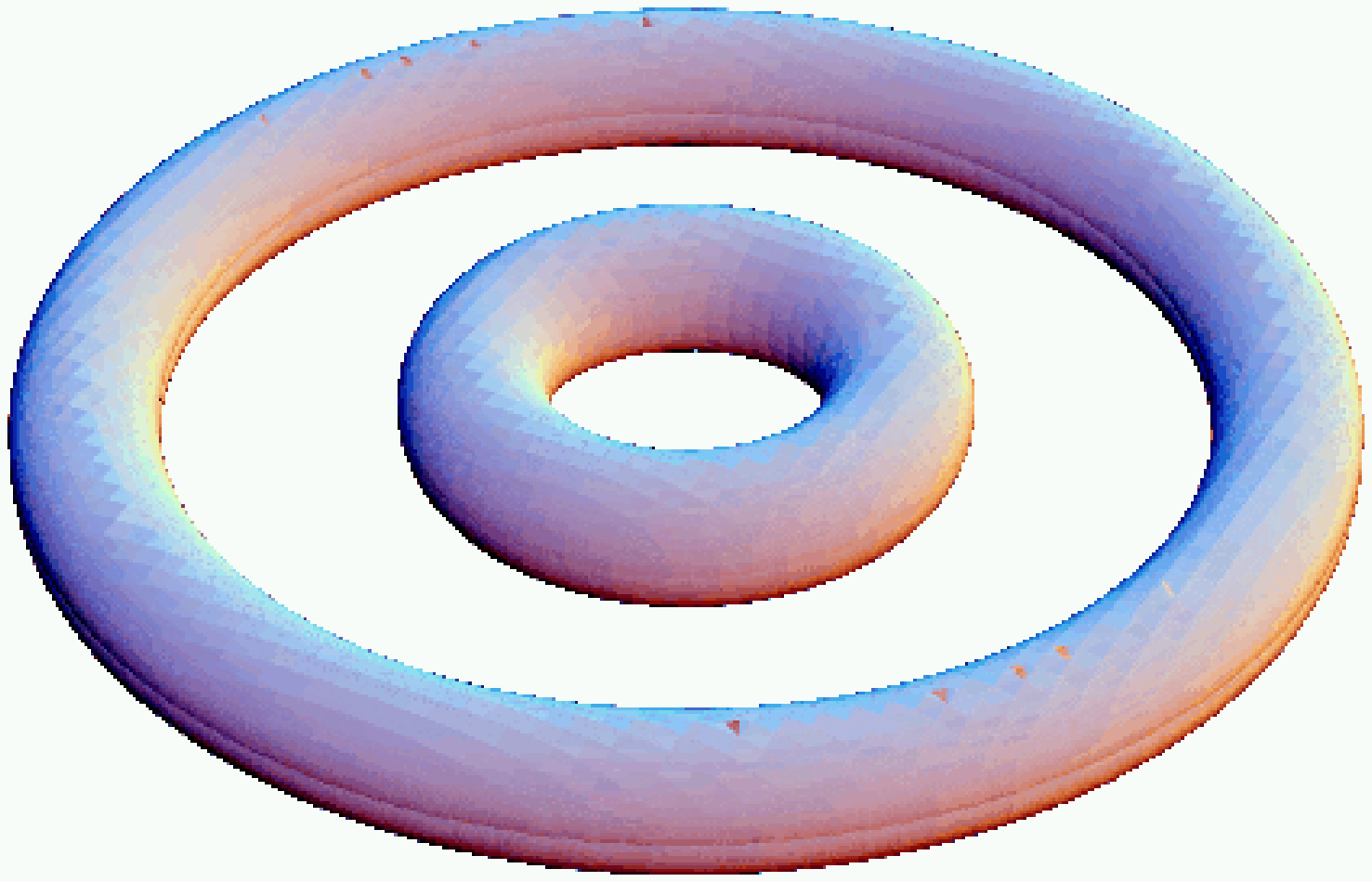}}	
\hss}
	\caption{Energy isosurfaces with $T_0^0=0.1$ for 
the $m_1=0,~m_2=3$ {`Saturn'}  (left) and    
with  $T_0^0=0.3$ for
the $m_1=1,~m_2=3$ {bi-ring}  (right) solutions in the theory \eqref{2Q}. }
\label{Fig:saturn}
\end{figure}

Yet another way to generalize $Q$-balls is to couple to each other
several copies of the theory. This allows one to consider `non-linear superpositions'
of individual $Q$-balls. 
Although this idea has already been considered
in the literature \cite{Brihaye:2007tn}, we have reconsidered it 
and found some curious solutions which could be interesting in the context of
our discussion. In the simplest case one can choose 
\be
\label{2Q}
 L[\Phi_1,\Phi_2]=L_Q(\Phi_1)+L_Q(\Phi_2) -\gamma |\Phi_1|^2|\Phi_2|^2,
\ee
where $L_Q(\Phi_1)$  and $L_Q(\Phi_2)$ are two copies of the $Q$-ball
Lagrangian (\ref{lQ}). Setting 
\be
\Phi_1=e^{i(\omega_1 t +m_1\varphi)}f_1(r,\theta),~~~~~~
\Phi_2=e^{i(\omega_2 t +m_2\varphi)}f_2(r,\theta),
\ee
it is interesting to consider solutions 
with $m_1\neq m_2$. 
\begin{figure}[ht]
\hbox to\linewidth{\hss%
	\resizebox{7cm}{4.5cm}{\includegraphics{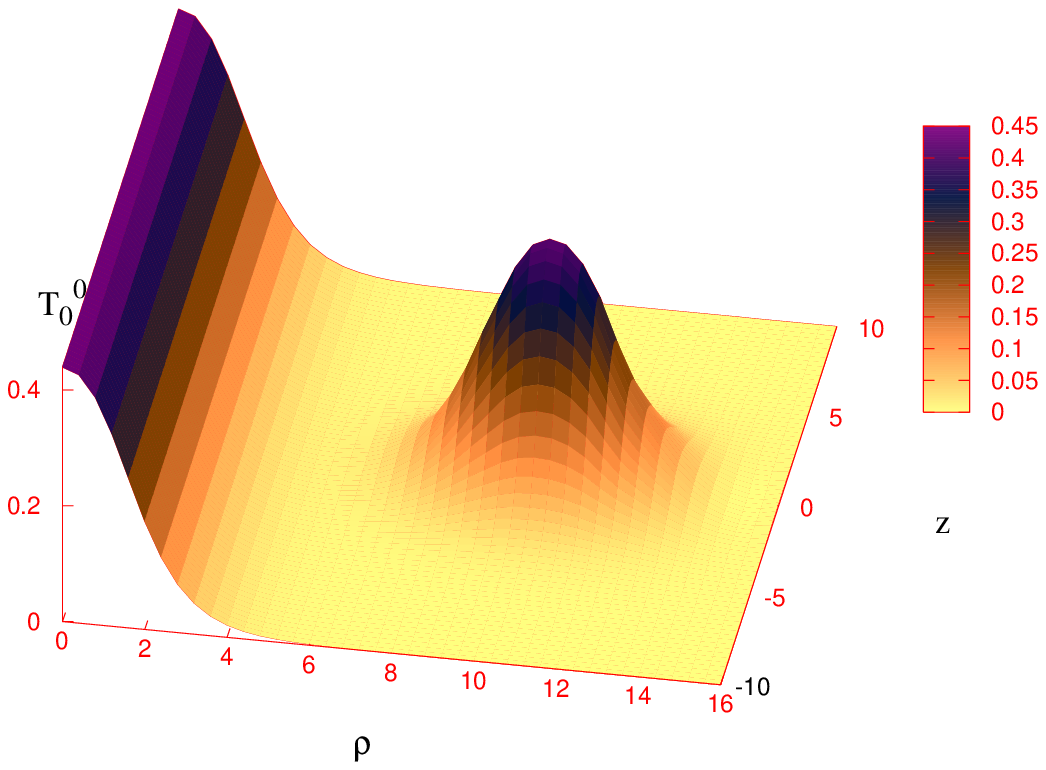}}
\hspace{5mm}%
        \resizebox{7cm}{4.5cm}{\includegraphics{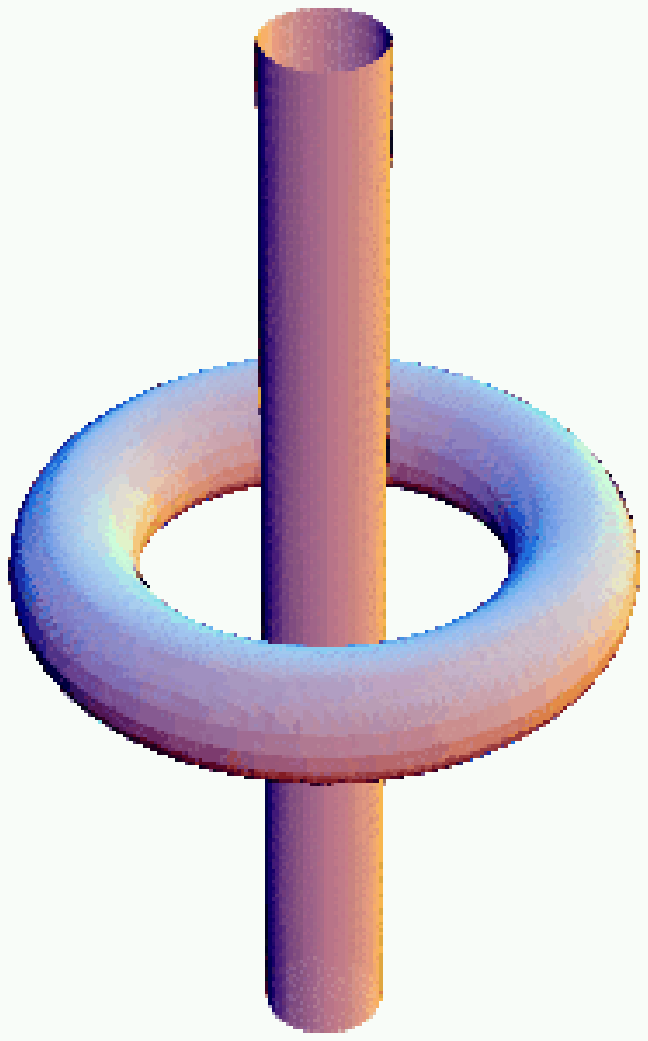}}	
\hss}
	\caption{The energy density $T_0^0$ and the $T_0^0=0.1$ energy  
isosurface for the $m_1=0$, $m_2=3$ `hoop' solution   
in the theory \eqref{2Q}. }
\label{hoop}
\end{figure} 
Choosing $\omega_1=\omega_2=1$, $\gamma=1$ and 
restricting to the even parity sector, we find solutions
of the `Saturn type', with a central concentration of 
the energy density produced by the first scalar with $m_1=0$ and 
surrounded  by a ring created 
by the second scalar with $m_2>0$ (Fig.\ref{Fig:saturn}), 
solutions for $m_2>m_1>0$ with bi-ring profiles (Fig.\ref{Fig:saturn}),
as well as solutions describing  a superposition of 
a straight   `$Q$-ball vortex' with $m_1=0$
`hooped'  by a $Q$-ball ring with $m_2>0$  (Fig.\ref{hoop}).

\subsection{Skyrmions}

The first known example of topological solitons in $3+1$ 
dimensions was suggested  almost 50 years ago
within the non-linear relativistic field theory model 
proposed by T.H.R.~Skyrme 
\cite{Skyrme:1961vq}. These
solitons are now called  skyrmions. 
Skyrme himself considered them as field theoretic realizations of baryons. 
Nowadays the Skyrme model is regarded as an effective, low energy approximation 
of  QCD 
\cite{Adkins:1983ya},\cite{Adkins:1983hy}.
In this approximation the static skyrmions are 
promoted to spinning objects by making use of the effective rigid body 
approximation that will be 
considered below.  The masses of the spinning skyrmions obtained in this way
are then compared to the hadron masses.  

The question of whether spinning skyrmions 
really exist as stationary field theory
objects was addressed only very recently in Ref.\cite{Battye:2005nx},
whose authors constructed spinning skyrmions by applying 
the same mechanism as for the $Q$-balls and arrived 
at conclusions which differ considerably from those obtained 
within the rigid 
body approximation. 

Since there exist excellent descriptions of the Skyrme model and its solutions 
in the literature \cite{Makhankov},\cite{Manton:2004tk}, 
we shall 
very briefly summarize only the features essential for our discussion. 
The fundamental field variables in the theory  
comprise an SU(2)-valued matrix $U(x^\mu)$
which satisfies field equations for  the Lagrangian
\be                                                      \label{Skyrme}
{\mcal L}_{\rm S}[U]={\rm tr}
\left(\frac{1}{2}\,\partial_\mu U^\dagger \partial^\mu U+
\frac{1}{8}\,[\partial_\mu U^\dagger,\partial_\nu U]
[\partial^\mu U^\dagger,\partial^\nu U]\right).
\ee
It is convenient to introduce a pure gauge SU(2) connection, 
\be                                       \label{puregauge}
{\bf A}_\mu=iU^\dagger\partial_\mu U\equiv \tau^a{\bf A}^a_\mu,~~~~~ 
\partial_\mu{\bf A}_\nu-\partial_\nu{\bf A}_\mu=
i[{\bf A}_\mu,{\bf A}_\nu],
\ee 
in terms of which 
the field equations read
\be                                       \label{eqs-skyrme}
\partial^\mu\left({\bf A}_\mu-\frac{1}{4}\,[[{\bf A}_\mu,{\bf A}_\nu] ,{\bf A}^\nu]\right)=0.
\ee
The energy for static fields is
\be                                                      \label{Skyrme1}
E[U]=\int {\rm tr}\left(\frac12({\bf A}_k)^2
-\frac{1}{8}([{\bf A}_i,{\bf A}_k])^2\right) d^3\bx\,.
\ee
For the energy to be finite
the field  $U(\bx)$ should approach a constant value at infinity, which can be
chosen to be the unit matrix, so that $\lim_{|\bx|\to\infty}U(\bx)=1$.   
This allows one to replace $\mathbb{R}^3$ by its one-point compactification 
$S^3$. Since SU(2) is topologically also $S^3$ it follows that any 
finite energy field configuration can be viewed as a map,
$
U(\bx): S^3\to S^3,
$
and can therefore be characterized by the integer degree of map,
\be                                  \label{baryon}
\Q[U]=\frac{i}{24\pi^2}\int 
{\rm tr}\left(\epsilon_{ijk}{\bf A}_i {\bf A}_j {\bf A}_k\right)d^3\bx \,.
\ee
This topological charge is called in Skyrme theory `baryon number'. 
Fixing it the energy obeys the Bogomol'nyi type inequality
\be
E[U]\geq 12\sqrt{2}\pi^2|\Q[U]|,
\ee
which can be easily obtained by rearranging terms in the integrand in \eqref{Skyrme1}.  
The existence of this lower bound suggests looking for  energy
minimizers in each topological sector. 

It will be convenient for what follows to parametrize the matrix $U$ in terms 
of two complex scalar fields $\phi,\sigma$ 
subject to the normalization condition
$
|\phi|^2+|\sigma|^2=1$
as 
\be 						\label{U}
U=
\left(
\begin{array}{cc}
\phi & i\sigma^\ast \\
i\sigma & \phi^\ast 
\end{array}
\right). 
\ee
Simplest skyrmions are spherically symmetric, 
$U=\exp(i\chi(r)\,{\bx\cdot\mbox{\boldmath$\tau$}}/{r})$, hence 
\be                               \label{SU}
\phi=\cos\chi(r)+i\sin\chi(r)\cos\vartheta ,~~~~~
\sigma=\sin\chi(r)\sin\vartheta e^{i\varphi}.
\ee
Inserting this to Eqs.\eqref{eqs-skyrme} the variables separate
and the equations reduce to an  
 ODE for $\chi(r)$. This equation admits globally regular solutions
with the boundary conditions $\chi(0)=\Q\pi$ and $\chi(\infty)=0$
for any value of the topological charge $\Q$. 
However, it seems that only for $\Q=\pm 1$ these solutions correspond to the
absolute  energy minimum, while those for $|\Q|>1$ are local minima or saddle
points. Global energy minima for $|\Q|>1$ are not spherically symmetric and can be 
obtained by directly minimizing the energy \eqref{Skyrme1}. 
For $|\Q|=2$ they are
axially symmetric, with the constant baryon number isosurfaces having 
toroidal shape, so they are somewhat similar to knots. 
However, higher charge skyrmions do not resemble knots at all and 
look like polyhedral shells. Their detailed description 
can be found in the monograph  \cite{Manton:2004tk}. 

\subsubsection{Skyrme versus Faddeev-Skyrme \label{versus}}

It is worth emphasizing the similarity between the Skyrme and Faddeev-Skyrme 
models. Both can be represented in the form (up to normalization) 
\be						\label{vers}
{\mcal L}[\mbox{\boldmath$\phi$} ]=
\partial_\mu \mbox{\boldmath$\phi$}\cdot\partial^\mu\mbox{\boldmath$\phi$}
-\frac14\,(\partial_\mu \mbox{\boldmath$\phi$}\cdot\partial^\mu\mbox{\boldmath$\phi$})^2
+\frac14\,(\partial_\mu \mbox{\boldmath$\phi$}\cdot\partial_\nu\mbox{\boldmath$\phi$})
(\partial^\mu \mbox{\boldmath$\phi$}\cdot\partial^\nu\mbox{\boldmath$\phi$})
\ee
where \mbox{\boldmath$\phi$} is a vector normalized by the condition
$
\mbox{\boldmath$\phi$}\cdot \mbox{\boldmath$\phi$}=1.
$ 
In the Faddeev-Skyrme case one has $\mbox{\boldmath$\phi$}=\phi^a$ with $a=1,2,3$ so that 
the target space is $S^2$.
In the Skyrme model one has  $\mbox{\boldmath$\phi$}=(\phi^0,\phi^a)$ with $a=1,2,3$
so that the target space is $S^3$, the Skyrme field being 
$U=\phi^0+i\phi^a\tau^a$. 
One can also  
construct a generalized theory whose target space `interpolates' between $S^2$ and $S^3$
and so it includes both models as special cases \cite{Ward:2004gr}.

The Faddeev-Skyrme model can be regarded as a consistent
truncation of the Skyrme theory.  
Specifically, if $\nn=\n^a$ is a solution of the 
Faddeev-Skyrme model, then $\mbox{\boldmath$\phi$}=(0,\n^a)$ solves the equations 
of the Skyrme model. 
 As a result, 
knot solitons can be embedded into the Skyrme model \cite{Meissner}, \cite{Cho},
although their stability properties will then be different. 

Another embedding of knot solitons into Skyrme theory is obtained by 
expressing the fields  ${\mcal A}_\mu$ and $\nn$ 
in terms of the $CP^1$ coordinates 
$\phi,\sigma$ using Eqs.\eqref{norm},\eqref{A},\eqref{project}. 
The Skyrme field $U$ is then given by
Eq.\eqref{U}. Although this correspondence does not map solutions to 
solutions, the advantage now is that the 
baryon number $\Q[U]$ is exactly equal to the 
Hopf charge $\Q[\nn]$. This follows from the fact that 
${\bf A}^3_\mu=-{\mcal A}_\mu$ and  
${\mcal F}_{\mu\nu}=2({\bf A}^1_\mu{\bf A}^2_\nu-{\bf A}^1_\nu{\bf A}^2_\mu)$ 
due to Eq.\eqref{puregauge}. 
With this, the expression  \eqref{baryon} for the baryon number exactly reduces to the
expression \eqref{Q} for the Hopf charge. 
For example, using the parametrization \eqref{axial-CP} for the $CP^1$ 
scalars $\phi,\sigma$ with the 
Hopf charge $\Q[\nn]=nm$ gives with \eqref{U}
the Skyrme field $U$ with $\Q[U]=nm$.  

The Hopf charge $\Q[\nn]$ will determine in this case also the winding 
number of the pure gauge field
${\bf A}_k=iU^\dagger\partial_k U$. If $\nn$ is an energy minimizer
then the pure gauge ${\bf A}_k$  
is {maximally abelian} 
\cite{vanBaal:2001jm}.

The correspondence can also be used in the opposite direction to construct 
$\nn$ with a given Hopf charge out of $U$ with a given baryon number. 
If $U$ is spherically symmetric, then $\nn$ is axially symmetric,
as can be seen comparing Eq.\eqref{SU} and Eq.\eqref{W-Hopf1}.  
Using approximations for $U$ gives in this way approximate solutions
for $\nn$ \cite{Ward2}.

\subsubsection{Spinning skyrmions}

Let us consider the  $\Q=1$ static, spherically symmetric skyrmion  
described by  Eq.\eqref{SU}. Its spinning analog  
in the
 rigid body approximation is obtained by simply replacing 
in Eq.\eqref{SU} 
\be
\varphi\to\varphi+\omega t\,,
\ee 
which 
gives a non-zero value to the angular momentum. It is in fact 
precisely this type of approximation 
that  is often implicitly assumed in the literature when talking
about spinning solitons. 
In this approximation  spinning solitons do not change their shape 
and the radiation effects are neglected, which are natural assumptions  
if the rotation is slow. However, if $\omega$ is not small, then 
this description does not approximate the true spinning 
solutions any more but simply 
gives field configurations with  $J\neq 0$.  
If one uses them as initial data, then their temporal evolutions will be certainly 
accompanied by radiation carrying $J$ away, and it is not clear
if the system will finally relax to a state with $J\neq 0$.

In order to construct truly spinning skyrmions with $\Q=1$, 
Battye, Krusch and Sutcliffe (BKS) \cite{Battye:2005nx}
generalize the spherically symmetric ansatz 
\eqref{U},\eqref{SU} to the axially symmetric, 
non-manifestly stationary one, which can be parametrized as  
\be                                        \label{SU1}
\phi=\cos\frac{\Theta}{2}\,e^{i\psi},~~~~~
\sigma=\sin\frac{\Theta}{2}\,e^{i(\varphi+\omega t)}.
\ee 
It is instructive to compare this with the ansatz \eqref{axial-CP} for the 
axially symmetric hopfions with $m=n=1$.  
They also add to the Lagrangian \eqref{Skyrme} the mass term 
\be                         \label{Skyrmmass}
M^2\,{\rm tr}(U-1). 
\ee   
Instead of solving the field
equations for $\Theta(r,\vartheta),\psi(r,\vartheta)$ they 
minimize the energy with fixed $J$, whose expression is  
similar to the one in Eq.\eqref{EJ}. 
The frequency $\omega$ for 
their solutions is a parameter in the interval 
\be                          \label{limit}               
0\leq \omega^2 <M^2,
\ee
both the energy and angular momentum increasing with $\omega$. 
The latter feature is very interesting -- for spinning skyrmions $J$ is a 
continuous parameter that can be arbitrarily small, so that they can 
rotate slowly. 
To compare, $Q$-balls cannot rotate slowly. 
Although BKS do not emphasize  this, it seems that both $E$ and $J$ should blow up 
as $\omega^2\to M^2$ since the fields localized by the factor 
$\exp(-\sqrt{M^2-\omega^2}\, r)$
become then long range.

Trying to adjust $M^2$ and the energy scale 
to reproduce the pion and hadron masses and their spins, 
BKS discover that this requires moving to the parameter region where deviations
of spinning skyrmions from spherical symmetry are large. 
They conclude that 
the rigid body approximation does  not provide an adequate 
description of spinning skyrmions. 

The results of BKS have been confirmed  
by solving the field equations by Ioannidou,
Kleihaus and Kunz \cite{Ioannidou:2006nn} (who studied in fact gravity-coupled
skyrmions, but considered  the Minkowski space limit as well).  

The spinning skyrmions have been obtained by BKS by minimizing
the energy within the axially symmetric ansatz \eqref{SU1}.
Strictly speaking, this does not 
exclude the possibility of non-axially symmetric instabilities. However,
since the non-spinning skyrmions are stable, one can expect that at least for 
small $J$ their spinning analogs should be stable as well. 

Spinning solitons have also been studied  
in the context of the baby Skyrme model \cite{Piette},\cite{Betz}. This is simply
the Faddeev-Skyrme model restricted to $2+1$ spacetime dimensions and 
written in the form \eqref{vers} with an additional  mass term 
$M^2(\mbox{\boldmath$\phi$}^3-1)$. Spinning fields are then chosen to be 
\be                                           \label{baby}
\mbox{\boldmath$\phi$}^1+i\mbox{\boldmath$\phi$}^2=\sin\Theta(\rho)e^{i(\omega t+\varphi)},~~~~~
\mbox{\boldmath$\phi$}^3=\cos\Theta(\rho).
\ee
The field equations reduce in this case to an ODE for $\Theta(\rho)$ whose solutions 
are easy to study. Solutions exist for $\omega< M$ and  both $E$ and $J$
blow up as $\omega\to M$.   

Trying to further increase $\omega$, solutions become oscillatory, since  one has 
$\Theta\sim \exp(-\sqrt{M^2-\omega^2}\, \rho)$ for large $\rho$, 
both $E$ and $J$ being then infinite.  However, one can consider the initial data 
of the form \eqref{baby}, where $\omega>M$ and $\Theta(\rho)$ 
vanishes identically for large $\rho$.
$E,J$ will then be finite.   Evolving these data in time, 
the system radiates away a fraction of its energy and angular momentum and relaxes 
to a stationary, rotating configuration with $\omega<M$ \cite{Piette}.  

\subsubsection{Spinning gauged skyrmions}

Spinning solutions have also been constructed within the gauged version of 
the Skyrme model \cite{Callan:1983nx} 
by Radu and Tchrakian \cite{Radu:2005jp}. 
These solutions are 
 manifestly stationary and manifestly axisymmetric. 
Parametrizing the Skyrme field $U$ in terms of two complex scalars $\phi,\sigma$ 
according to Eq.\eqref{U}, 
the gauged Skyrme model is obtained from the Skyrme model as
\be                                                               \label{Skyrme2}
{\mcal L}_{\rm gS}[A_\mu,\phi,\sigma]=
-\frac14\,F_{\mu\nu}F^{\mu\nu}+
{\mcal L}_{\rm S}[\phi,\sigma].
\ee
Here ${\mcal L}_{\rm S}$ is the Skyrme Lagrangian \eqref{Skyrme} with the mass term 
\eqref{Skyrmmass} included, 
$F_{\mu\nu}=\partial_\mu A_\nu-\partial_\nu A_\mu$, and  the derivatives 
of the field $\sigma$ in ${\mcal L}_{\rm S}[\phi,\sigma]$ 
are replaced by the covariant derivatives,
$\partial_\mu\sigma\to(\partial_\mu-igA_\mu)\sigma$,  
while the derivatives  $\phi$ do not change, so that the theory is invariant 
under local U(1) gauge transformations
\be
A_\mu\to A_\mu+\partial_\mu\alpha,~~~~\sigma\to e^{i\alpha}\sigma,~~~~
\phi\to\phi.
\ee
Radu and Tchrakian \cite{Radu:2005jp} generalize the ansatz \eqref{SU1} as 
\be                          \label{SU2}        
A_\mu dx^\mu=A_0dt+A_\varphi\sin\vartheta\, d\varphi,~~~
\phi=\cos\frac{\Theta}{2}\,e^{i\psi},~~~~~
\sigma=\sin\frac{\Theta}{2}\,e^{i(m\varphi+\omega t)}, 
\ee
where $A_0,A_\varphi,\Theta,\psi$ depend on $r,\vartheta$. 
This ansatz 
is manifestly stationary and manifestly axially symmetric,
since its $t,\varphi$-dependence can be removed by a gauge 
transformation.  The topological charge \eqref{baryon} is $\Q=m$. 
Integrating the field equations gives
globally regular solutions with finite energy and a longrange gauge field, 
which is at large $r$
\be
A_0=\frac{Q_{\rm el}}{r}+\ldots,~~~~
A_\varphi=\frac{\bf{m}}{r^2}\,\sin\vartheta+\ldots ,
\ee
where $Q_{\rm el},{\bf m}$ are the electric charge and dipole moment. 
This feature is similar to that for the gauged $Q$-balls in Eq.\eqref{Ainf}. 
The angular momentum can be calculated in the same way as in Eq.\eqref{JJJ1},
with a similar result, 
\be                \label{Jmm}
J=\frac{4\pi Q_{\rm el}m}{g}\,.
\ee
However, unlike for $Q$-balls, $J$ can be arbitrarily small if $\omega$ is small.
The dependence of solutions on $\omega$ is 
 qualitatively similar to that     
for the global skyrmions: they exist only for a finite frequency range,
both $E$ and $J$ growing with $\omega$.
For $\omega=0$ the solutions reduce to the non-spinning gauged skyrmions 
found in Ref.\cite{Piette:1997ny}.

The stability of spinning gauged skyrmions has not been studied. They
probably have less chance to be stable than their
global counterparts -- since gauging introduces additional degrees of freedom
that can produce instabilities.

\subsection{Rotating monopole-antimonopole pairs  \label{mon-pairs}}

\begin{figure}[h]
\hbox to\linewidth{\hss%
  \resizebox{7cm}{5cm}{\includegraphics{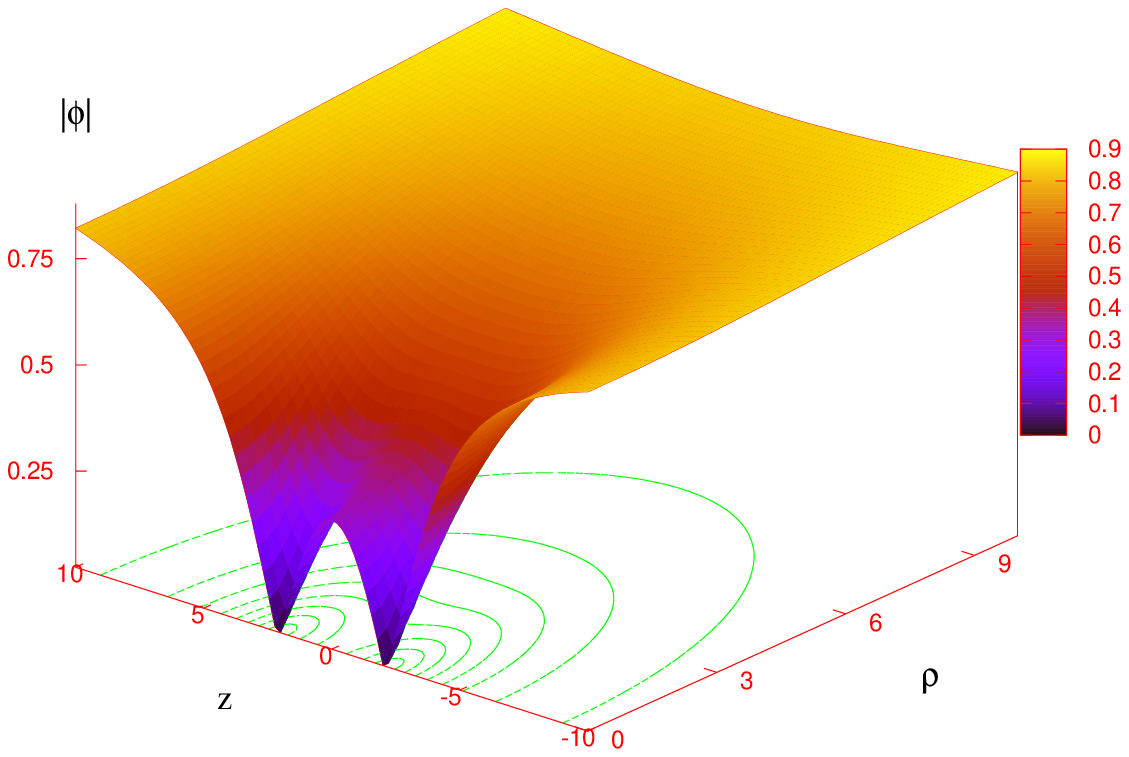}}%
\hspace{5mm}  \resizebox{6cm}{5cm}{\includegraphics{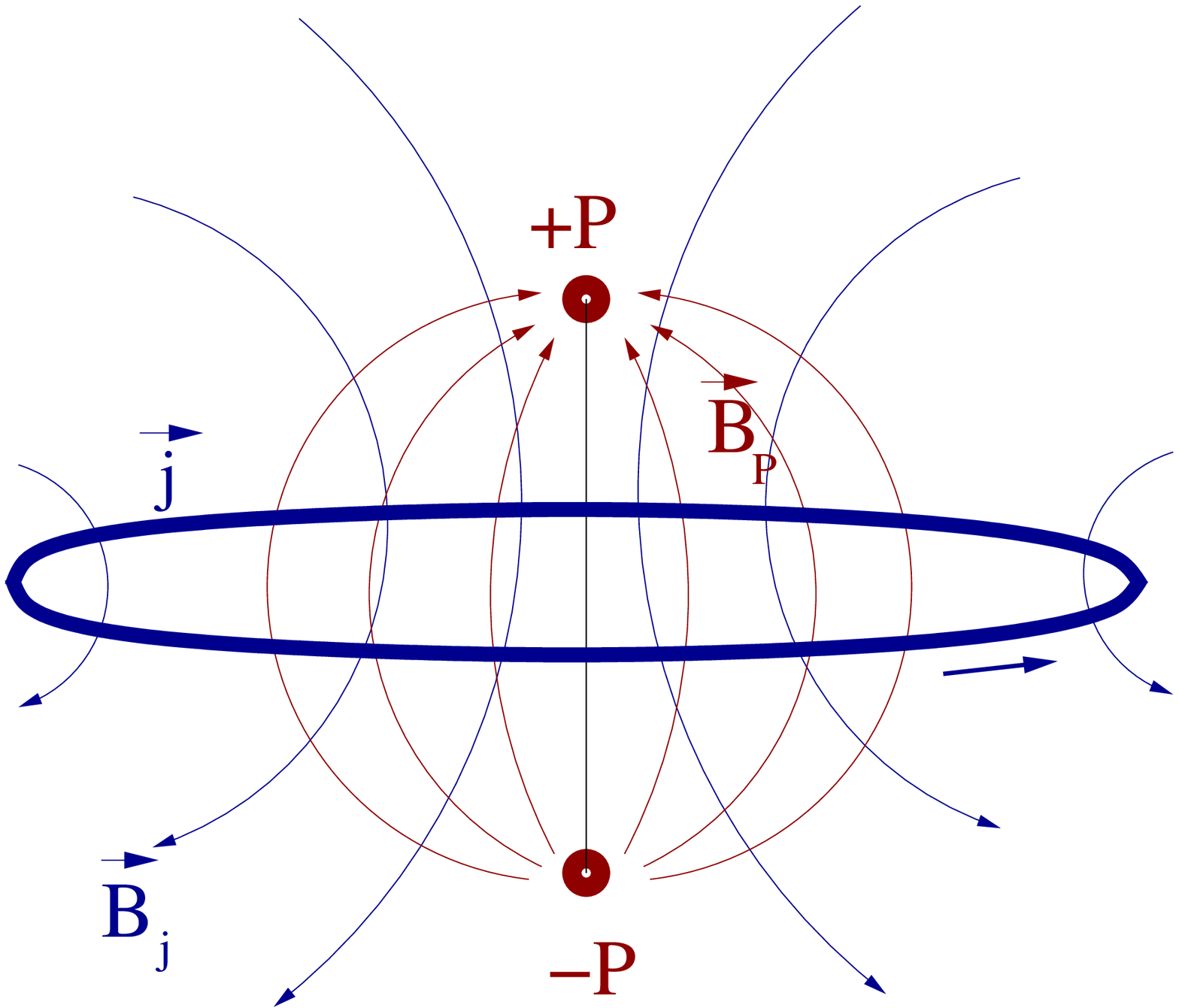}}%
\hss}
\caption{\small Left: the Higgs field amplitude for the monopole-antimonopole solution
in the limit where the Higgs field potential vanishes.  
It is shown only for a limited region around the origin, while at infinity 
it approaches the unit value. Right: the charge and current distributions 
for this solution exhibits essentially the same structure as for the monopole rings,
shown in Fig.\ref{Fig2}. 
}
\label{FigPqa}
\end{figure}

There are other known  solutions with non-zero angular momentum. They  
are manifestly stationary and manifestly axisymmetric, but they  
describe rotations in  multisoliton systems and not 
spinning of a single soliton. 
Let us consider the already mentioned above monopole-antimonopole solutions in the 
 Yang-Mills-Higgs theory \eqref{YMH}. Their existence was 
demonstrated by Taubes \cite{Taubes:1982ie} and they were explicitly constructed by 
Kleihaus and Kunz \cite{Kleihaus:1999sx} within the ansatz \eqref{anz0}
with $k=2$, $m=1$. 
These solutions  are static, purely magnetic and manifestly axisymmetric 
(since their $\varphi$-dependence can be gauged away) and  
with vanishing magnetic charge. The Higgs field $\Phi$ shows two simple zeros at
two spatial points separated by a finite distance (see Fig.\ref{FigPqa}). 
These two points correspond to 
the positions of the monopole and antimonopole. The
analysis of the charge and current distributions 
similar to that described in Sec.\ref{rings} shows  that the system also 
contains a circular electric current, as shown in Fig.\ref{FigPqa}. The magnetic 
field created by this current acts against the Coulombian attraction between  
the monopole and antimonopole, which presumably stabilizes the system. 

Since there is no electric field for these solutions, their angular momentum
\eqref{JJJ} is  zero. However, in the  
limit where the Higgs potential
vanishes, the Lagrangian \eqref{YMH} admits the global symmetry 
\cite{Bogomolnyi},\cite{Coleman:1976uk}
\be                               \label{hyper} 
A_0\to A_0\cosh\gamma+\Phi\sinh\gamma,~~~~~~~~
\Phi\to \Phi\cosh\gamma+ A_0\sinh\gamma,~~~~~~~A_k\to A_k,
\ee 
and this allows one to produce solutions with an electric field starting from 
purely magnetic solutions. As was noticed in \cite{Heusler:1998ec}, this 
transformation may also generate a non-zero angular momentum
\be
J\sim\sinh\gamma,
\ee
provided 
that the original  purely magnetic solution is not spherically symmetric and 
does not satisfy the first 
order Bogomol'nyi equations \cite{Bogomolnyi}. 
The latter condition excludes from consideration
the (multi)monopole solutions of these equations.
However, applying the symmetry \eqref{hyper}
to the monopole-antimonopole solution, 
which does not fulfill the Bogomol'nyi equations and 
whose Higgs field is long-range in the limit of vanishing Higgs potential, 
gives an electrically charged system, since both monopole and antimonopole
then receive an electric charge of the 
same sign \cite{Hartmann:2000ja}. The total 
electric charge calculated with the gauge-invariant definition \eqref{charge-current} is 
$Q_{\rm el}\sim\sinh\gamma$, and applying the surface integral formula \eqref{JJJ} 
gives the angular momentum \cite{vanderBij:2002sq}
\be                                           \label{JQQ} 
J=\frac{4\pi Q_{\rm el}}{g}
\ee
directed along the symmetry axis passing through the monopole 
and antimonopole. 

\begin{figure}[h]
\hbox to\linewidth{\hss%
\resizebox{7cm}{4cm}{\includegraphics{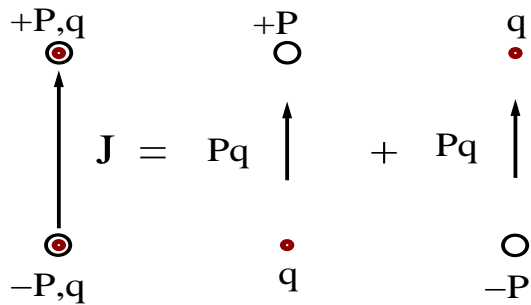}}%
\hss}
\caption{\small 
Schematic visualisation of the rotating monopole-antimonopole pair as superposition
of a monopole-electric charge pair  $(+P,q)$ and an antimonopole-electric charge 
pair $(-P,q)$, both of which making the same contribution to the angular momentum. 
}
\label{FigPq}
\end{figure}

It is interesting to understand the origin of this angular momentum. 
If it was an orbital angular momentum, then it would be orthogonal 
to the monopole-antimonopole symmetry axis, but it is parallel to it instead. 
Let us remember that a static system containing an electric charge 
$q$ and a magnetic charge $P$ has an angular momentum $J=Pq$ independent 
of the distance between the charges and 
directed from $q$ to $P$ \cite{Coleman:1982cx}.
This angular 
momentum can be obtained by integrating the 
Poynting vector as in Eq.\eqref{Poynt}. 
We know that a rotating monopole-antimonopole pair contains two magnetic charges
of the opposite sign and two electric charges of the same sign. It can therefore be
visualized as a superposition of two pairs, $(+P,q)$ and $(-P,q)$, both of which making 
{\it the same} contribution to the angular momentum, as shown in Fig.\ref{FigPq}.

A similar construction of rotating solutions 
can be carried  out using 
other known static solutions of the non-Abelian theory \eqref{YMH},
as for example the monopole rings described in Sec.\ref{rings}, 
also for the generic Higgs field potential \cite{Paturyan:2004ps},\cite{Kleihaus:2005fs}. 
It turns out that the following general relation holds
\be
J=\frac{2\pi m\, Q_{\rm el}}{g}\,[1+(-1)^k]
\ee 
where $m,k$ are the winding
numbers in the ansatz \eqref{mn}
 \cite{Paturyan:2004ps}, \cite{vanderBij:2002sq}, \cite{Kleihaus:2005fs}. 
Therefore, using Eq.\eqref{Qmagn}, the 
angular momentum is zero in the presence of a 
nonvanishing magnetic charge, even though  
$T_\varphi^0\neq 0$ \cite{Kleihaus:2005fs},
while the relation 
\eqref{Jmm} is still valid for solutions with zero magnetic charge. 

Since they are related to the monopole-antimonopole solution known to be 
unstable  \cite{Taubes:1982ie}, rotating monopole-antimonopole pairs are very 
probably unstable as well. However, the mechanism of their rotation is interesting
and suggests, in particular, a 
possibility to have stationary, non-radiating spinning solitons
{\it without} axial symmetry.  

Specifically, in theories without radiation, as in classical mechanics, 
steadily  rotating bodies can be totally asymmetric.  However, if they carry 
an electric charge distribution, say, then their electric dipole momentum will depend on time, 
hence  
producing radiation, unless they are axially symmetric. This suggests that 
non-radiating spinning field systems should be axially symmetric -- the assumption 
usually made when studying spinning solitons.   

An interesting illustration  of this has been found by Hen and 
Karliner \cite{Hen}, who analyze situations where rapidly spinning systems can loose 
axial symmetry. A classical example of this phenomenon
is provided by the rotating Jacobi ellipsoids in Newtonian gravity --
a theory without radiation. Hen and Karliner study spinning baby skyrmions on a 
{\it compact} (sphere or disk) 2D space, in which case there is no radiation,
and find that the system looses axial symmetry for large enough $J$. 
However, as was discussed above, in the limit of infinite $\mathbb{R}^2$,
when radiation can exist, spinning baby skyrmions remain 
axially symmetric for any value of $J$ \cite{Piette} --
since otherwise they would radiate.

Now, the rotating monopole-antimonopole pairs suggest a possibility to have
non-radiating, non-axially symmetric rotation even in theories with radiation. 
This can be achieved by simply taking three or more dyons not aligned 
along one direction.  Of course, this should be realized by constructing a 
smooth, static, non-axially symmetric, finite energy solutions  
in a non-Abelian gauge field theory. If they exist, they may 
have a non-zero $J$ due to the same mechanism as 
for the rotating monopole-antimonopole pairs.

\section{Vortons \label{sec-vortons}}
We are now ready to explicitly construct vortons 
as localized,
finite energy  
 solutions of the elliptic field equations. By construction, they are  
stationary and non-radiating. 
Our results can be viewed as complementary 
to those obtained by 
Battye, Cooper and Sutcliffe 
\cite{Battye:2001ec} (although in a completely different context) 
and by Lemperier and Shellard \cite{Lemperiere:2003yt}.  

\subsection{The Witten model}

Vortons were originally  suggested 
\cite{Davis:1988jp},
\cite{Davis:1988ij},
\cite{Davis:1988ip},
\cite{Davis:1988jq} 
in the context of Witten's model of
superconducting cosmic strings \cite{Witten:1984eb}. 
This model contains two Abelian vectors $A^{(a)}_\mu$ $(a=1,2)$ interacting
with two complex scalars $\phi$ and $\sigma$ with the Lagrangian density 
\be  
\label{W}
{\mathcal L}_{W}=-\frac14\sum_{a=1,2} F^{(a)}_{\mu\nu}F^{(a)\mu\nu}  
+D_\mu \phi^\ast D^\mu \phi + 
D_\mu \sigma^\ast D^\mu \sigma -U.
\ee 
Here $F^{(a)}_{\mu\nu}=\partial_\mu A^{(a)}_\nu-\partial_\nu A^{(a)}_\mu$
are the Abelian field strengths, the gauge covariant derivatives of the 
scalars are
$D_\mu \phi=(\partial_\mu-ig_{1}A^{(1)}_\mu)\phi$ and 
$D_\mu \sigma=(\partial_\mu-ig_{2}A^{(2)}_\mu)\sigma$
where $g_{1}$ and $g_{2}$ are the gauge coupling constants. 
The scalar field potential is 
\be                                                          \label{potential}
U=
\frac{1}{4}\lambda_\phi (|\phi|^2-\eta_\phi^2)^2
+\frac{1}{4}\lambda_\sigma |\sigma|^2(|\sigma|^2-2\eta_\sigma^2) 
+\gamma |\phi|^2 |\sigma|^2,
\ee
where $\lambda_\phi,\lambda_\sigma~,\eta_\phi,\eta_\sigma$ and $\gamma$ 
are positive constants. This theory is invariant under local 
U(1)$\times$U(1) gauge transformations, 
so that there are two conserved Noether 
currents.  

The theory admits stationary, cylindrically symmetric solutions of the vortex type,
supporting a constant non-zero value of one of the Noether currents
\cite{Witten:1984eb},
\cite{Babul:1987me},
\cite{Haws:1988ax},
\cite{Hill:1987qx},
\cite{Davis:1988jp},
\cite{Amsterdamski:1988zp},
\cite{Peter}.
These 
are the superconducting cosmic strings.
Vortons are supposed to be loops
made of these strings.

Within the full gauged model \eqref{W} explicit vorton constructions 
have never  been attempted -- due to the complexity of the problem.  
However, the problem simplifies in the global 
limit of this model, for $g_1=g_2=0$. The gauge fields then decouple 
and the theory reduces to   
\be                         \label{lag}
{\mathcal L}=  
\partial_\mu \phi^\ast \partial^\mu \phi + 
\partial_\mu \sigma^\ast \partial^\mu \sigma -U
\ee 
so that the U(1)$\times$U(1) internal symmetry becomes global. 
This global theory still keeps some essential 
features of the original local model. 
In particular, it still admits superconducting vortex solutions, even though
the corresponding current is now global and not local. One can therefore
study global vortons made of these vortices, which was in fact 
the subject of Refs.\cite{Battye:2001ec},\cite{Lemperiere:2003yt}. 
Below we shall 
construct the global vortons as stationary solutions of the field equations
in the model \eqref{lag}, 
which has not been done before. The question of whether
these solutions can be generalized within the full gauged model \eqref{W}
remains open.

One can  absorb $\eta_\phi$  in the definition
of $\phi,\sigma$ in \eqref{lag} to achieve $\eta_\phi=1$. 
Since the overall 
normalization of the potential could be changed by 
rescaling the spacetime coordinates, $x^\mu\to\Lambda x^\mu$,
one can impose one more condition on the remaining four parameters
$\lambda_\phi,\lambda_\sigma,\eta_\sigma,\gamma$, although we 
do not use this option.

A minimal value of the potential is achieved for
$|\phi|=\eta_\phi =1$ and $|\sigma|=0$, in which case $U=0$. 
This minimum is global if $4\gamma^2>\lambda_\sigma\lambda_\phi$.
The perturbative spectrum 
of field excitations around this vacuum consists of two massless 
Goldstone particles, corresponding to excitations of the phases of the fields,
and of two Higgs bosons with the masses
\be                                                        \label{MASSES}
M_\phi=\sqrt{\lambda_\phi},~~~~~
M_\sigma=\sqrt{\gamma-\frac12\,\lambda_\sigma\eta_\sigma^2}.
\ee

The global U(1)$\times$U(1) symmetry of the theory,
$\phi\to\phi e^{i\alpha_1}$, $\sigma\to\sigma e^{i\alpha_2}$, 
leads to the conserved currents
\be
\label{curr}  
j^\mu_{(\phi)}=2\Re(i\phi^\ast\partial^\mu \phi),~~~~~~~~~~
j^\mu_{(\sigma)}=2\Re(i\sigma^\ast\partial^\mu \sigma),
\ee 
with $\partial_\mu J^\mu_{(\phi)}=\partial_\mu J^\mu_{(\sigma)}=0$.
The energy-momentum tensor is
\be
\label{Tik}  
T_{\mu\nu}=
\partial_\mu \phi^\ast \partial_\nu \phi+\partial_\nu \phi^\ast \partial_\mu \phi
+\partial_\mu \sigma^\ast \partial_\nu \sigma
+\partial_\nu \sigma^\ast \partial_\mu \sigma
-g_{\mu\nu}{\mathcal L}~,
\ee
where $g_{\mu\nu}$ is the spacetime metric. 

The Lagrangian field equations in the theory \eqref{lag} read 
\be                             \label{eW}
\partial_\mu\partial^\mu\phi+
\frac{\partial U}{\partial |\phi|^2}\,\phi=0,~~~~~~~~~
\partial_\mu\partial^\mu\sigma+
\frac{\partial U}{\partial |\sigma|^2}\,\sigma=0.~
\ee
Expressing the complex scalar fields in terms of their amplitudes and phases as
\be
\phi=f_1\,e^{i\psi_1},~~~~~
\sigma=f_2\,e^{i\psi_2},
\ee
equations \eqref{eW} assume the form ($a=1,2$)
\be                                           \label{eqs}
\partial_\mu\partial^\mu f_a=(\partial_\mu\psi_a\partial^\mu\psi_2)f_a
-\frac12\,\frac{\partial U}{\partial f_a} \, ,~~~~~~
\partial_\mu(f_a^2\partial^\mu\psi_a)=0.
\ee
These equations admit cylindrically symmetric vortex solutions discussed 
by several authors (see $e.g.$
 \cite{Davis:1988jp},
\cite{Lemperiere:2003yt},
\cite{Hartmann:2008yr}).
 For these solutions one has  
\be
\label{vortices}  
\phi=f_1(\rho)e^{i n \varphi},~~~~~~~~
\sigma =f_2(\rho)e^{i(p z+\omega t)}~, 
\ee
where $n\in\mathbb{Z}$ and $p,\omega\in\mathbb{R}$, with 
$0\leftarrow f_1(\rho)\to 1$ and $f_2(0) \leftarrow f_2(\rho)\to 0$
as $0\leftarrow\rho\to\infty$, respectively. 
The field $\phi$ thus vanishes in the vortex core and approaches the finite vacuum 
value at infinity, while $\sigma$ develops a non-zero condensate value in the core
and vanishes at infinity. The fields $\phi,\sigma$ are sometimes 
referred to as vortex field and condensate field, respectively. 
The phase of $\phi$ changes by $2\pi n$ after one revolution around the vortex,
while the phase of $\sigma$ increases along the vortex. 
The $z$-dependence of the condensate field gives rise to a non-zero momentum 
along the vortex, $P=\int T^0_z d^3 x\sim p$. 

The qualitative vorton construction usually discussed in the literature 
suggests taking a piece of length $L$ of the vortex and identifying its 
extremities 
to form a loop. The momentum along the vortex will then circulate along the loop,
thus becoming an angular momentum, and the centrifugal force which arises 
is supposed to be able to compensate the tension of the loop, 
thereby  producing an equilibrium configuration. 
The coordinate $z$ along the vortex then becomes periodic and can be replaced
by the azimuthal angle $\varphi$, so that 
$p$ will have to assume discrete values, $p=2\pi m/L$ with $m\in\mathbb{Z}$.
The central axis of the vortex where the field $\phi$ vanishes 
then becomes a circle of radius $\RRR=L/2\pi$.  

\subsection{Vorton topology and boundary conditions}

\begin{figure}[ht]
\hbox to\linewidth{\hss%
	\resizebox{3.7cm}{7cm}{\includegraphics{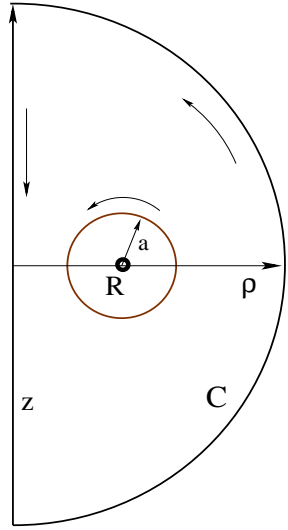}}
	\hss}
	\caption{Vorton topology. 
The phase of $\phi$ increases by $2\pi n$ after
one revolution around the boundary $C$ of the half-plane, 
$\phi$ then vanishing $n$ times somewhere inside $C$. 
In the simplest $n=1$ case the zero is located at a point $(\RRR,0)$
corresponding to the vortex center.
The vortex has a characteristic thickness $\ccc$.}
\label{Fig:top}
\end{figure}

The above considerations 
suggest describing the vortons by the 
ansatz
\be                              \label{anz}
\phi=f_1(\rho,z)e^{i\psi_1(\rho,z)},~~~~~~
\sigma=f_2(\rho,z)e^{im\varphi+i\omega t}.~
\ee
The phase of $\sigma$ 
increases by $2\pi m$ after one revolution around the $z$-axis, so that  
$f_2(\rho,z)$ should vanish at the axis for the field to be regular there. 
The phase $\psi_1$ 
increases by $2\pi n$ after one revolution around 
the boundary $C$ of the $(\rho,z)$ half-plane, which is the $z$-axis plus the
infinite semi-circle (see Fig.\ref{Fig:top}). 
The regularity condition then implies that $\phi(\rho,z)$ must have $n$ zeros
somewhere inside $C$.   
Therefore any vorton configuration can be characterized by a 
pair of integers ($n,m$), which we shall call, respectively,
vortex winding number and azimuthal winding number. 

It is instructive to compare these boundary conditions to those 
 for the Faddeev-Hopf knots in Eq.\eqref{axial-CP}.
They are the same --
the phases of $\phi,\sigma$ wind, respectively, along  the orthogonal directions 
shown in Fig.\ref{FigS}. The vortons and knots have therefore similar topology. 
This suggests introducing the `topological charge' 
analogues to the Hopf charge \eqref{Hopf-axial}
\be                                 \label{topol}
\Q=nm.
\ee
Strictly speaking, this quantity will be topologically invariant, that is
unchanged under arbitrary continues field deformations, only in the 
sigma model limit defined by the condition \eqref{constraint} below.
However, we shall call it topological charge in all cases and shall  
require  the vortons to have $\Q\neq 0$.

Let us  pass in \eqref{anz} to spherical coordinates,
\be                        \label{ansatz} 
\phi = f_1(r,\vartheta) e^{i\psi_1(r,\vartheta)},
~~~~~~~
\sigma = f_2(r,\vartheta) \exp[ i ( m \varphi+\omega t)].
\ee
Inserting this to Eqs.\eqref{eqs} gives 
\bea
\Delta f_1-\left((\nabla\psi_1)^2+\frac{\lambda_\phi}{2}(f_1^2-1)
+\gamma f_2^2\right)f_1&=&0\,, \notag \\
\Delta f_2-\left(\frac{m^2}{r^2\sin^2\vartheta}-\omega^2
+\frac{\lambda_\sigma}{2}(f_2^2-\eta_\sigma^2)
+\gamma f_1^2\right)f_2&=&0\,, \notag \\
\nabla(f_1^2\nabla\psi_1)&=&0.
\eea
At infinity the fields are required to approach the vacuum values, $f_1\to 1$,
$f_2\to 0$. One finds for large $r$ 
\bea                              \label{tail}
\psi_1&=&\frac{A\cos\vartheta}{r^2}+\ldots,~~~~~~~~~~
f_2=\frac{B}{r}\,\exp(-\sqrt{M_\sigma^2-\omega^2}~ r)+\ldots \,, \nonumber \\
f_1&=&1+\frac{A^2}{r^6}\,(1+3\cos^2\vartheta)+\ldots
+\frac{C}{r}\exp(-M_\phi~ r)+\ldots\,,
\eea
where $A,B,C$ are integration constants and the dots denote the subleading terms. 
These asymptotic expansions  show the presence of 
two massive Higgs modes and a long range massless Goldstone mode.
The second Goldstone field is not excited within the ansatz \eqref{ansatz}. 
Introducing the notation 
\be                           \label{ansatz:1}
X=f_1\cos\psi_1,~~~~~Y=f_1\sin\psi_1,~~~~~Z=f_2,
\ee
the equations assume the form 
\begin{eqnarray}
\nonumber 
&&\Delta X=\left(\frac{\lambda_\phi }{2}(X^2+Y^2-1)+\gamma Z^2\right)X,
\\
\label{eqs1} 
&&\Delta Y=\left(\frac{\lambda_\phi }{2}(X^2+Y^2-1)+\gamma Z^2\right)Y,\nonumber 
\\
\label{eqs2} 
&&\Delta Z=\left(
\frac{m^2}{r^2\sin^2\vartheta}-\omega^2+
\frac{\lambda_\sigma }{2}(Z^2-\eta_\sigma^2)+\gamma (X^2+Y^2)\right)Z. \label{eqs3} 
\end{eqnarray} 
Using \eqref{tail}, one has at large $r$
\be
X=1+O(r^{-4}),~~~~Y=O(r^{-2}),~~~~
Z\sim \frac{1}{r}\,\exp(-\sqrt{M_\sigma^2-\omega^2}~ r).
\ee
Introducing 
\be
\label{T} 
E_2=\int(|\nabla\phi|^2+|\nabla\sigma|^2)d^3x\,,~~~
E_0=\int U d^3x, ~~~
\ssigma=\int Z^2 d^3x\,, 
\ee 
the energy is 
\be
\label{Energy}  
E=\int T_0^0 d^3x= \omega^2\ssigma+E_2+E_0~,
\ee 
while the Lagrangian
\be
L=\int{\mathcal L}\,d^3x=\omega^2\ssigma-E_2-E_0\,.
\ee
One of the two Noether charges is non-vanishing,  
\be
\label{charge}
Q=\int d^3 xJ^0_{(\sigma)}=2\omega\ssigma,
\ee
and this gives rise to the angular momentum  
\be
\label{J}  
J=\int T_\varphi^0 d^3x =mQ.
\ee
Rescaling the spatial coordinates as $\bx\to\Lambda \bx$, the 
Lagrangian  $L$ should be stationary, which implies 
 the virial relation
\be
\label{virial} 
 E_2=3(\omega^2\ssigma-E_0).
\ee
The presence of the term $({m^2Z^2}/{r^2\sin^2\vartheta})$ in  
$|\nabla\sigma|^2$ requires that for finite energy fields one should have $Z=0$
at the $z$-axis.   
Let us now remember that the phase $\psi_1$ of $\phi=X+iY=f_1e^{i\psi_1}$ 
should increase by 
$2\pi n$ after one revolution around the boundary of the $(\rho,z)$ half-plane.  
Since one has $X=1$, $Y=0$ at $r\to\infty$, it follows that 
the phase is constant at infinity,  so that it 
can only change along the $z$-axis (see Fig.\ref{Fig:top}). Therefore,
\be                                            \label{topology}
\psi_1(\rho=0,z=\infty)-\psi_1(\rho=0,z=-\infty)=-2\pi n,
\ee
from where it follows that the functions $X$ and $Y$ have (at least) 
$2n$  and $2n+1$ zeros at the $z$-axis, respectively. Such a behavior
is compatible with the assumption that $X$ is symmetric and $Y$ is 
antisymmetric under $z\to -z$. Assuming that $Z$ is symmetric,
one arrives  at the following parity conditions, 
\be
X(r,\vartheta)=X(r,\pi-\vartheta),~~~~~
Y(r,\vartheta)=-Y(r,\pi-\vartheta),~~~~~
Z(r,\vartheta)=Z(r,\pi-\vartheta),
\ee
which allow one to restrict in the analysis the range of $\vartheta$ to the 
interval $[0,\pi/2]$. Summarizing everything together,
solutions of Eqs.\eqref{eqs3} should satisfy the following boundary conditions. 

{At the symmetry axis, $\vartheta=0$},
functions $X,Y$ should 
have, respectively, $n$ and $n-1$ zeros for $0<r<\infty$,
and one should also have 
\be                                      \label{1}
\partial_\vartheta X=\partial_\vartheta Y=Z=0.
\ee 
{At the origin, $r=0$},
\be                                \label{2}
\partial_r X=Y=Z=0.  
\ee
{At infinity, $r=\infty$}, 
\be                                 \label{3}
X=1,~~~~ Y=Z=0.  
\ee
{In the equatorial plane, $\vartheta=\pi/2$},
\be                                   \label{4}
\partial_\vartheta X=\partial_\vartheta Z=Y=0.
\ee


\subsection{The sigma model limit \label{o}}
We have succeeded for the first time 
to solve the equations  in the special limit 
of the theory obtained by choosing  
the parameters in the potential \eqref{potential} as
\be                                  \label{sm-limit}
 \lambda_\sigma=\lambda_\phi=\beta,~~\eta_\sigma=1,~~
 \gamma=\frac{1}{2}\,\beta+\gamma_0,
\ee
such that the potential becomes
\be                                \label{pot1} 
U=\frac{\beta}{4}(|\phi|^2+|\sigma|^2-1)^2+\gamma_0|\phi|^2|\sigma|^2.
\ee
Taking 
$\beta\to\infty$
enforces the constraint 
\be                                       \label{constraint}
|\phi|^2+|\sigma|^2=1\,
\ee
and the theory \eqref{lag} becomes a sigma model,
\be 
\label{sig}
{\mathcal L}=  
\partial_\mu \phi^\ast \partial^\mu \phi + 
\partial_\mu \sigma^\ast \partial^\mu \sigma -\gamma_0|\phi|^2|\sigma|^2.
\ee
%
This model is simpler than the full theory \eqref{lag}.  
It has less degrees of freedom
and no free parameters, since the value of $\gamma_0$ 
can be changed by rescaling the spacetime
coordinates. The field mass $M_\sigma$ determined by  
\eqref{MASSES} reduces now 
to $\sqrt{\gamma_0}$, while the mass $M_\phi$ 
becomes infinite, which means that field 
degrees of freedom violating the constraint  
\eqref{constraint} are excluded from the dynamics. 
The theory therefore contains only one massive 
particle and two Goldstone bosons in the spectrum. 
It is worth noting that $\Q=mn$ now becomes genuinely
topological charge, invariant under arbitrary smooth field deformations.

 It is convenient to use the Lagrange multiplier method by  
adding to the Lagrangian \eqref{sig} 
a term $\mu (1-|\phi|^2-|\sigma|^2)$ 
and varying with respect to $\phi,\sigma$ and $\mu$.
Using the same ansatz \eqref{ansatz},\eqref{ansatz:1}
as before gives  
\begin{align}                  \label{eqsa}
\nonumber 
&\Delta X=\left(\gamma_0 Z^2+\mu\right)X,     \\
&\Delta Y=\left(\gamma_0 Z^2+\mu\right)Y,\nonumber   \\
&\Delta Z=\left(
\frac{m^2}{r^2\sin^2\vartheta}-\omega^2+
\gamma_0 (X^2+Y^2)+\mu\right)Z. 
\end{align} 
and also 
\be                   \label{c}
X^2+Y^2+Z^2-1=0.
\ee
Multiplying the three equations \eqref{eqsa}, respectively, by 
$X,Y,Z$ and taking their sum and using the constraint \eqref{c}
one finds the expression for the Lagrange multiplier,  
\bea                                    \label{lambda}
 \mu &=& X\Delta X+
Y\Delta Y
+
Z\Delta Z
-2\gamma_0 (X^2+Y^2)Z^2
-(\frac{m^2}{r^2\sin^2 \theta}-\omega^2)Z^2 \nonumber \\
&-&\frac12\, \Delta(X^2+Y^2+Z^2),
\eea
where the second line has been added in order to 
cancel the second derivatives appearing in the first line. 
Inserting this to \eqref{eqsa}, 
the constraint \eqref{c} will be imposed automatically
on the solutions, so that it can be excluded from considerations 
from now on.

The boundary conditions for Eqs.\eqref{eqsa} are given by 
Eqs.\eqref{1}--\eqref{4}. 
The regularity condition 
$Z|_{\theta=0,\pi}=0$ imposes the 
constraint $X^2+Y^2=1$ on the $z$-axis.
In addition, since at the origin one has $Y=0$, 
the constraint requires that 
$X^2=1$, and so one can choose 
\be                                 \label{n}
X|_{r=0}=-1\,.
\ee
Since one has $X|_{r=\infty}=+1$, this will guarantee a non-zero 
value of $n$, which is the principal advantage of the 
sigma model theory \eqref{sig}. 
 
In fact, making sure that the phase of $\phi$ winds $n\neq 0$ times around 
the contour $C$ is not simple. 
This condition can be naturally implemented  in 
toroidal coordinates \eqref{toroidal}, but these coordinates 
are somewhat singular at infinity. 
Spherical or cylindrical coordinates are  better suited for numerics, but 
the field topology is then determined by 
zeros of the $X,Y$ amplitudes 
(see Eq.\eqref{topology}), whose number 
 cannot be generically prescribed,
since they  are able to disappear  during numerical iterations. 
 Now,
the condition \eqref{n} enforces
a non-trivial field topology at least in the $n=1$ case. 

We can now solve the problem.
Starting from a field configuration satisfying the 
 boundary conditions \eqref{1}--\eqref{4}, \eqref{n}
we numerically iterate it 
until the convergence is achieved. 
 The resulting solutions   
are qualitatively very similar to the generic vortons described
in detail below, and they also 
agree with the `skyrmions' 
in the theory of Bose-Einstein condensation previously 
obtained in a completely different way  by Battye, Cooper and Sutcliffe 
\cite{Battye:2001ec}. The latter issue will be discussed below in more detail.

\subsection{Explicit vorton solutions}

Having obtained vortons in the sigma model limit, 
we can use them as the starting profiles in order to iteratively 
descend from  infinite to finite values of $\beta$ in 
the potential \eqref{pot1}, thus relaxing the sigma model condition.
 After this, we can relax also 
the conditions \eqref{sm-limit}, thereby  recovering vorton
solutions within the full original theory \eqref{lag}.

In our numerics we use the program FIDISOL (written in Fortran),
based on the iterative Newton-Raphson method.
 A detailed presentation of the FIDISOL code is given in \cite{FIDISOL},\cite{FIDISOL1}.
In this approach  the field equations are discretized on a ($r,~\vartheta$) 
grid with $N_r\times N_{\vartheta}$ points,
and the resulting system is iterated until convergence is achieved. 
The grid spacing in the $r$-direction is non-uniform, while the 
values of the grid points
in the angular direction are given by $\vartheta_k=(k-1)\pi/(2(N_{\vartheta}-1))$.
Instead of $r$, 
a new radial variable $x$ is introduced which maps the 
semi-infinite region $[0,\infty)$ to the finite interval $[0,1]$.
 There
are various possibilities for such a mapping, 
a flexible enough choice being $ x=r/(c+r)$, where $c$
is a properly chosen constant.
Typical grids for the $n=1$ solutions have around $200 \times 30$ points. 
The typical  numerical error 
for the solutions is estimated to be of order $10^{-3}$.
In addition, the virial relation (\ref{virial}) was also monitored.
The deviation from unity of the  ratio  $E_2/3(\omega^2\ssigma-E_0)$
is less than $10^{-3}$ for most of the solutions we have considered.

\begin{figure}[ht]
\hbox to\linewidth{\hss%
        \resizebox{7cm}{7cm}{\includegraphics{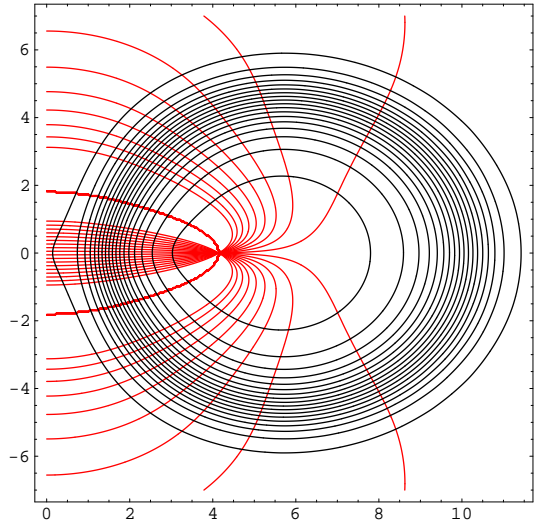}}	
\hss}
	\caption{Levels of constant amplitude $|\phi|$ (closed lines)
and phase $\psi_1$ (radial lines, red online) on the $\rho$ 
(horizontal direction)--$z$ (vertical direction) plane 
for the vortex field $\phi=|\phi|\exp(i\psi_1)$ 
corresponding to the
vorton solution with $n=1$, $m=2$, $\omega=0.85$, 
$\lambda_\phi=41.12$,
$\lambda_\sigma=40$,
$\eta_\sigma=1$, 
$\gamma=22.3$.}
\label{Fig:phase}
\end{figure}

As a result, for given values of the parameters  
$\lambda_\phi$, $\lambda_\sigma$, $\eta_\sigma$ and $\gamma$ 
in the potential, we can specify the azimuthal winding number $m$
and the frequency $\omega$ and obtain a vorton solution. 
The value of $n$, the charge $Q$ and   
energy $E$ are computed from the numerical output.
A complete analysis of the parameter space of solutions  
is a time consuming  task that we did not aim at.
Instead,  
we analyzed in detail a few particular classes of solutions, 
which hopefully reflects the essential properties 
of the general pattern.


We mainly considered the case where  the function $X$ vanishes
once at the positive $z$ semi-axis, which corresponds to the vortex winding
number  $n=1$.
The amplitude of the   scalar field $\phi$
has in this case one zero 
located on a circle in the $\vartheta=\pi/2$ plane.
The corresponding vorton topology  
can be illustrated
by the diagram in Fig.\ref{Fig:phase}, where the level lines
of the complex vortex function $\phi=|\phi|\exp(i\psi_1)$ are shown. 
The levels of constant amplitude, $|\phi|$, are closed lines encircling
the center of the vortex where $\phi$ vanishes. 
Emanating from this center there are lines of constant phase. 
The phase, $\psi_1$, 
increases  by $2\pi$ after one revolution around the center.  

\begin{figure}[ht]
\hbox to\linewidth{\hss%
	\resizebox{7cm}{6cm}{\includegraphics{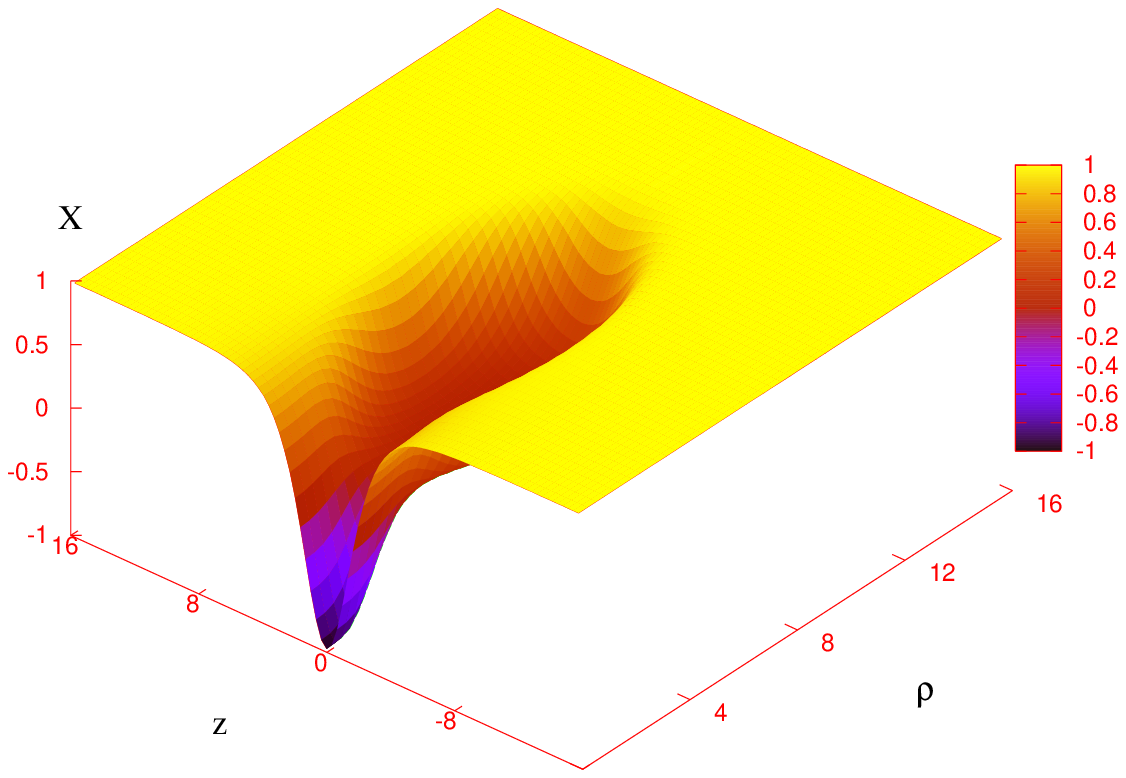}}
\hspace{5mm}%
        \resizebox{7cm}{6cm}{\includegraphics{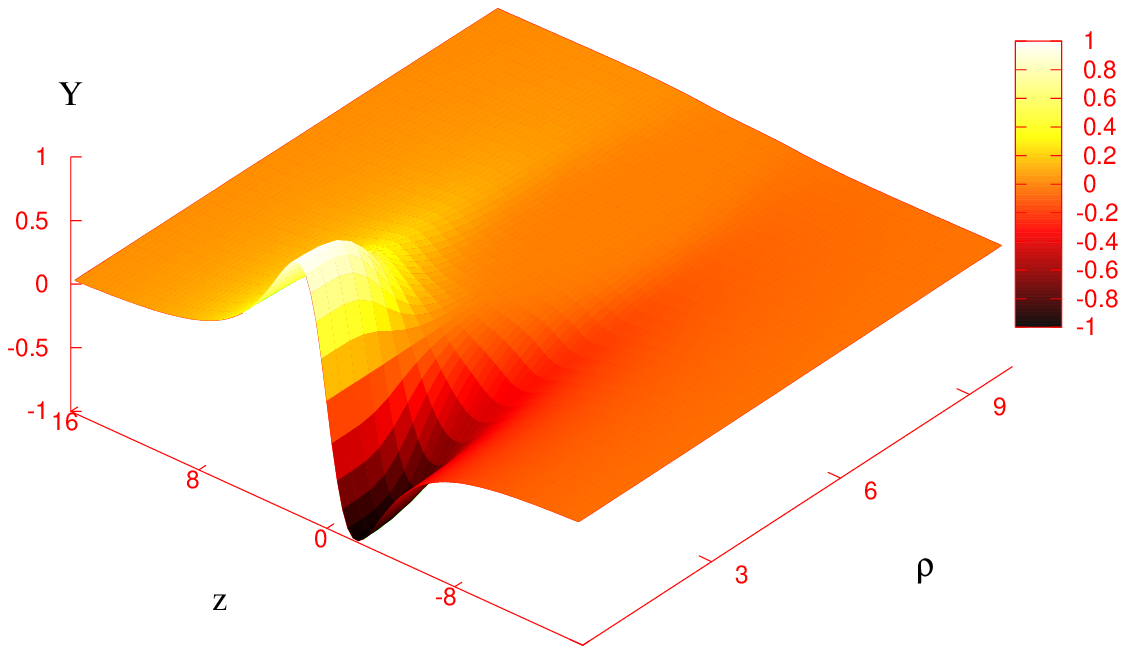}}	
\hss}
	\caption{The amplitudes $X(\rho,z),Y(\rho,z)$ 
for the typical vorton solution; here 
$n=1$, $m=2$, $\omega=0.85$,
   $\lambda_\phi=41.12$,
$\lambda_\sigma=40$, $\eta_\sigma=1$, $\gamma=22.3$.}
\label{Fig3}
\end{figure}

\begin{figure}[ht]
\hbox to\linewidth{\hss%
	\resizebox{7cm}{6cm}{\includegraphics{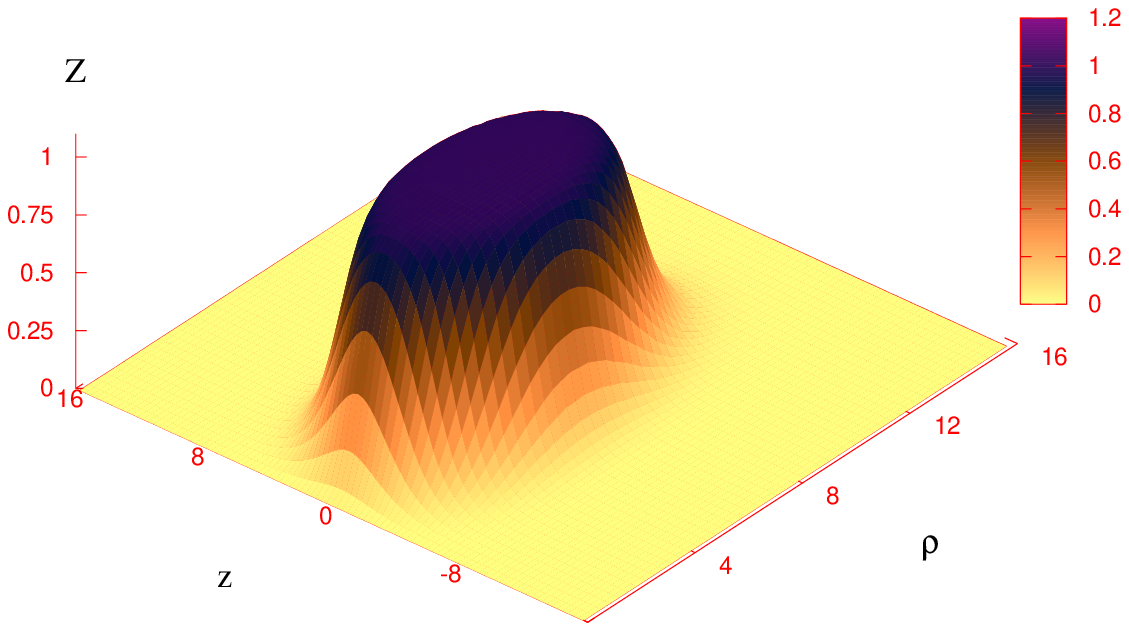}}
\hspace{5mm}%
        \resizebox{7cm}{6cm}{\includegraphics{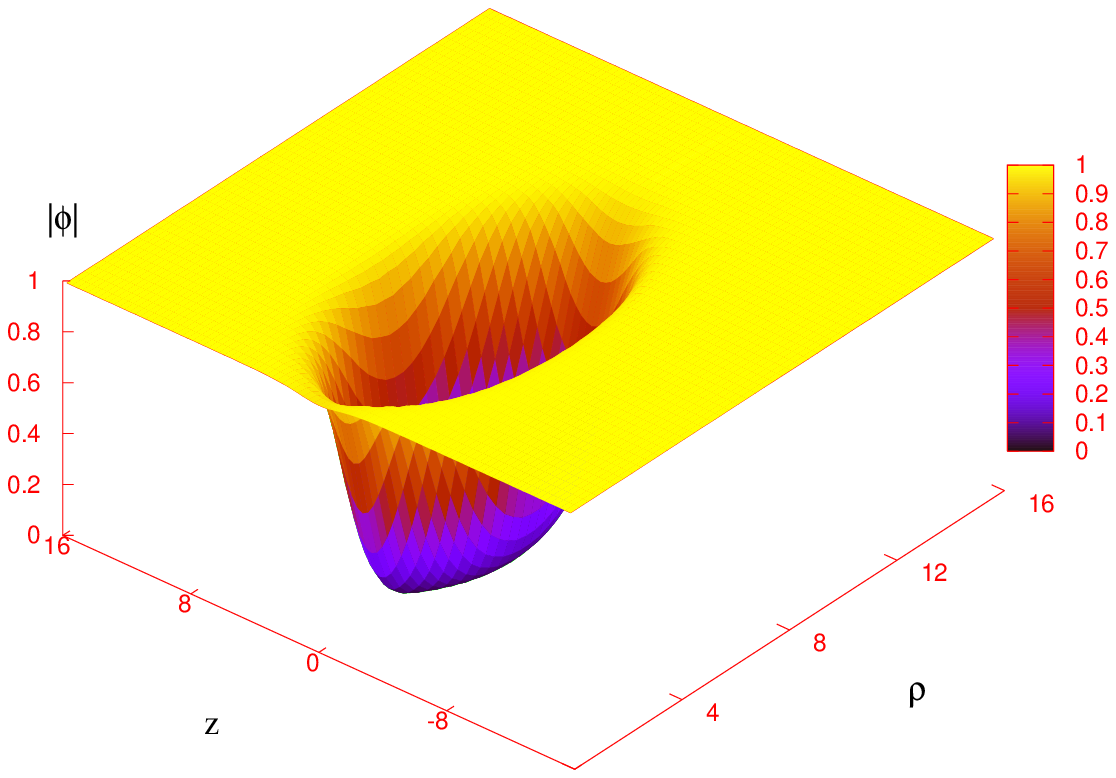}}	
\hss}
	\caption{The amplitudes $Z(\rho,z)$ and $|\phi(\rho,z)|$ for the same
solution as in Fig.\ref{Fig3}.}
\label{Fig3a}
\end{figure}

All vortons have toroidal distributions of the 
energy density and charge.
It seems that they  do not 
exist for arbitrary values of the parameters of the model but only for 
some regions in the parameter space.  
In Figs. \ref{Fig3}, \ref{Fig3a} the  3D plots of $X,Y,Z$ and also 
of $|\phi|=\sqrt{X^2+Y^2}$
for a typical solution with  $m=2$ are presented. 
One can see that these functions exhibit a strong  dependence
both on $\rho$ and $z$, with the zero of $|\phi|$ 
located in the $z=0$ plane at $\RRR\simeq 0.82$.
\begin{figure}[ht]
\hbox to\linewidth{\hss%
	\resizebox{7cm}{6cm}{\includegraphics{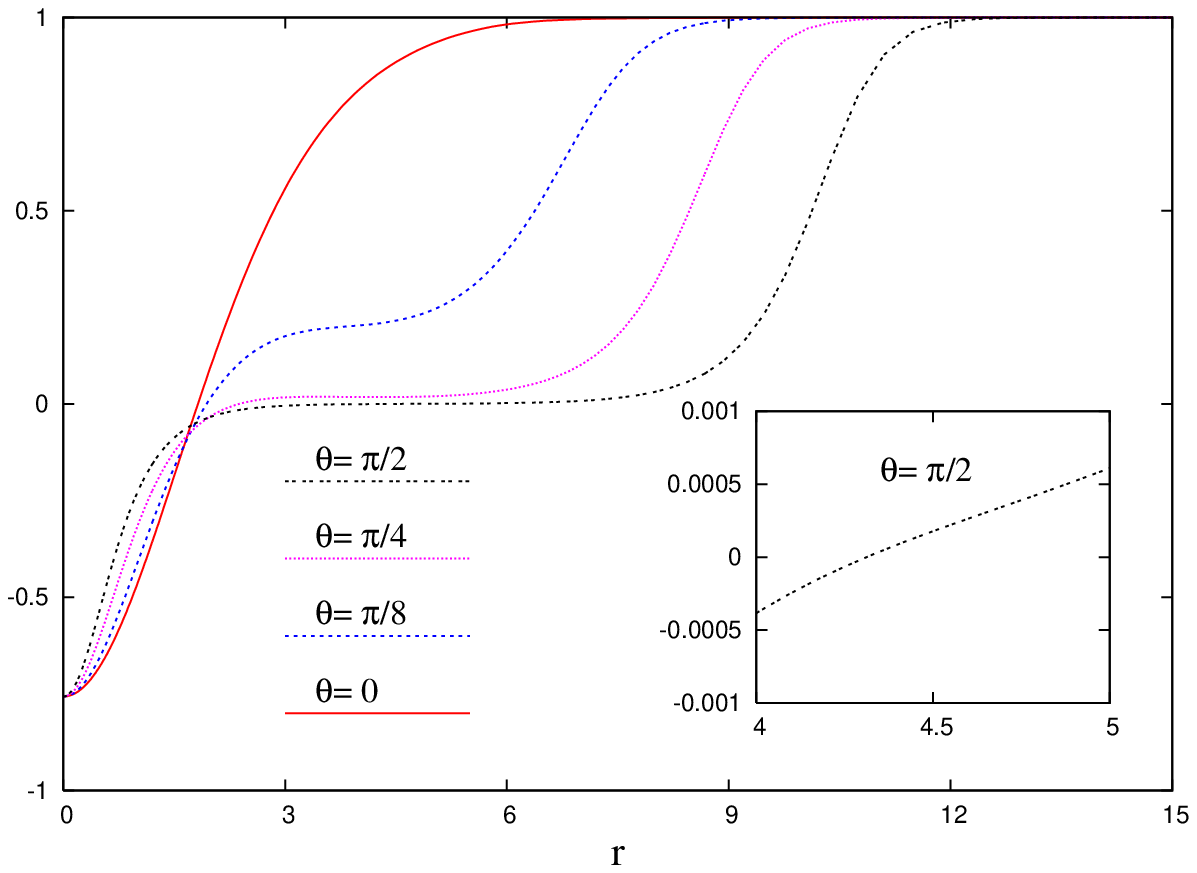}}
\hspace{5mm}%
        \resizebox{7cm}{6cm}{\includegraphics{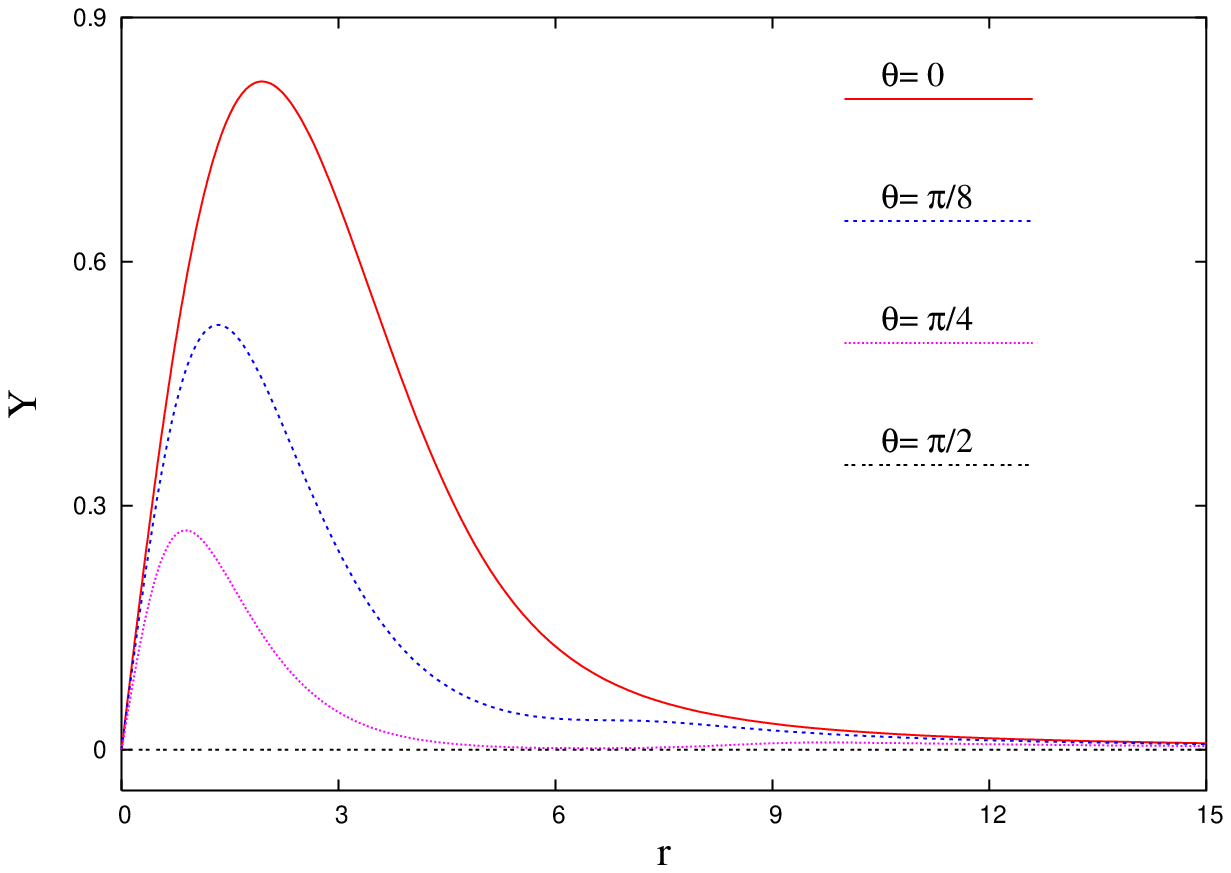}}	
\hss}
	\caption{Profiles of  $X(r,\vartheta), Y(r,\vartheta)$ for 
fixed $\vartheta$
	for the $n=m=1$ vorton solution with   $\omega=0.8$,   $\lambda_\phi=33.26$,
$\lambda_\sigma=32.4$, $\eta_\sigma=1$, $\gamma=18.4$.}
\label{Fig4}
\end{figure}
\begin{figure}[ht]
\hbox to\linewidth{\hss%
	\resizebox{7cm}{6cm}{\includegraphics{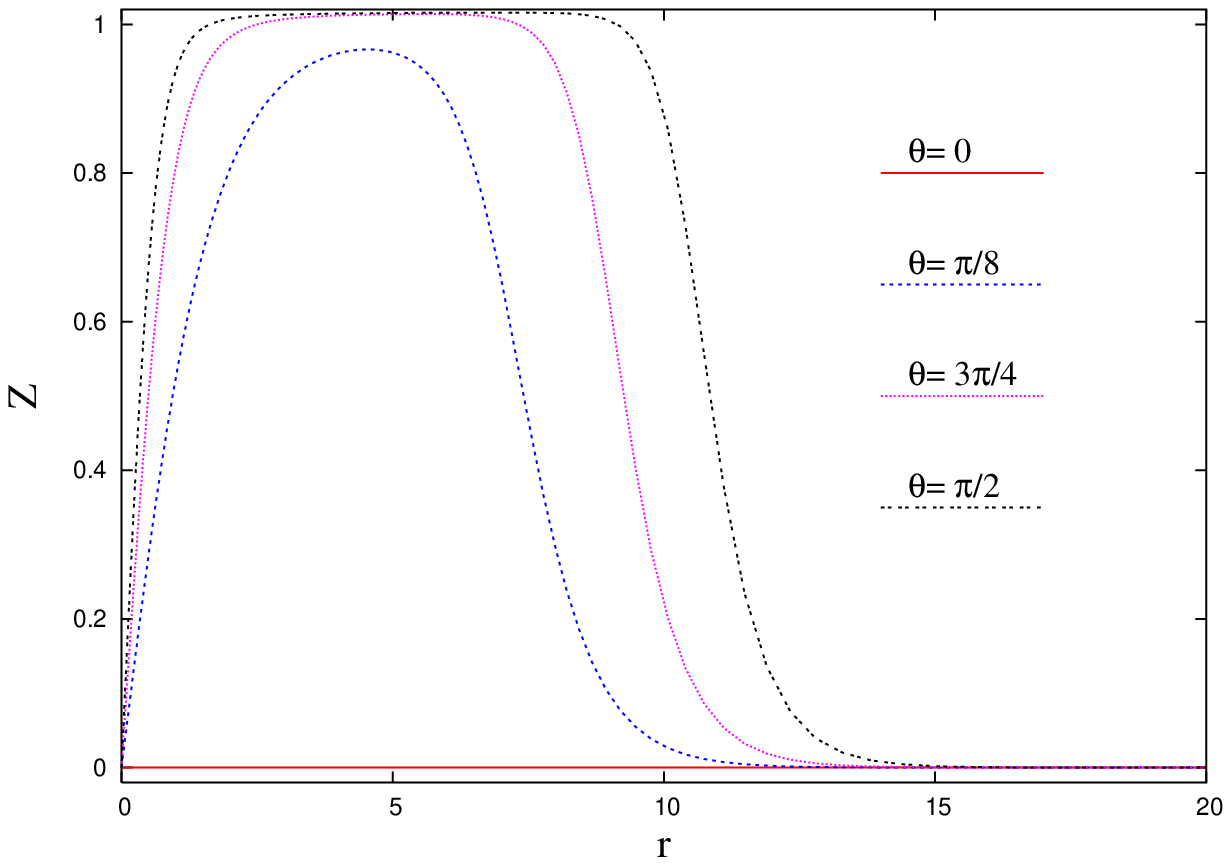}}
\hspace{5mm}%
        \resizebox{7cm}{6cm}{\includegraphics{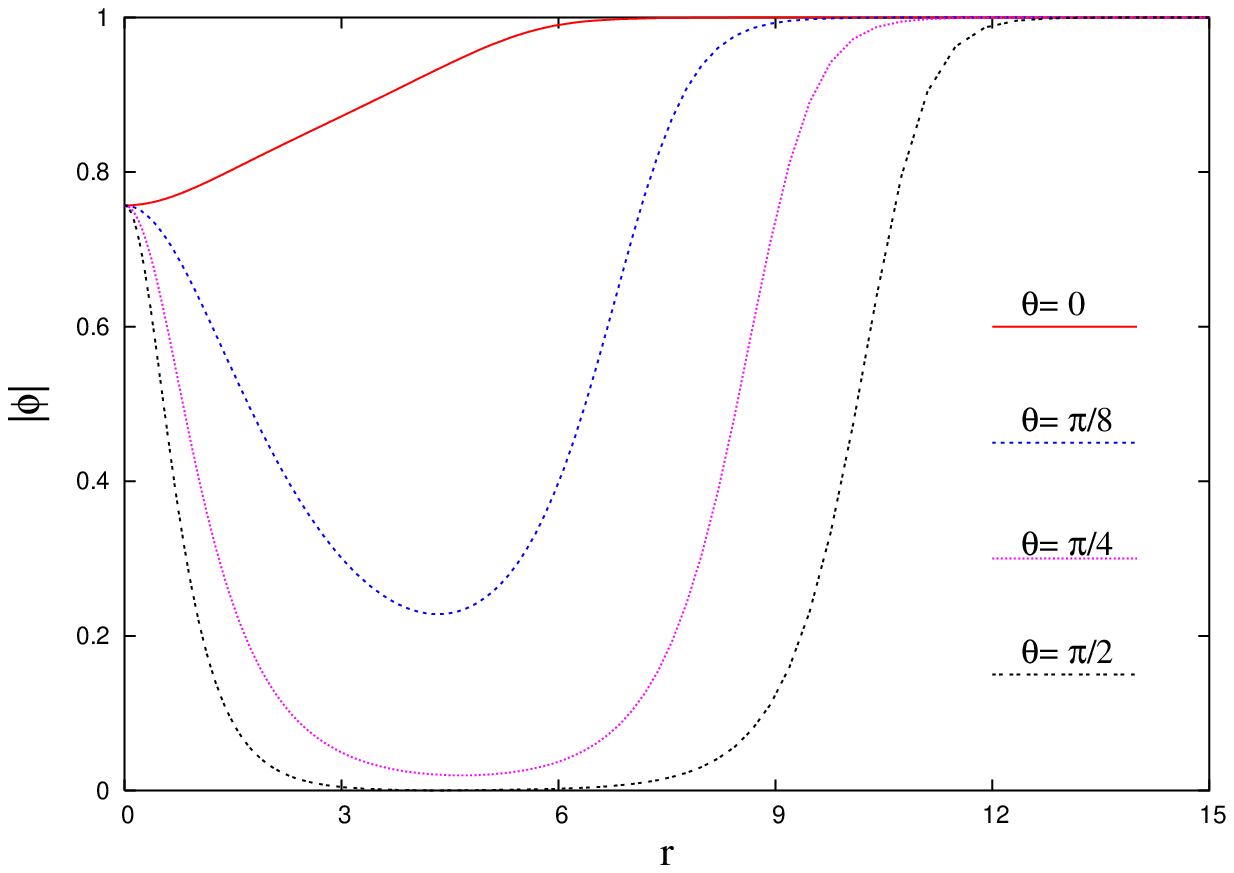}}	
\hss}
	\caption{Profiles of $Z(r,\vartheta),|\phi(r,\vartheta)|$ 
for fixed values of $\vartheta$
	for the same solution as in Fig.\ref{Fig4}. 
}
\label{Fig4a}
\end{figure}

In Fig. \ref{Fig4}, \ref{Fig4a} the 2D profiles of $X,Y$,$Z$ and
$|\phi|$ for a typical  $m=1$ solution for several values of $\vartheta$ are shown; 
these pictures remain 
qualitatively the same for higher values of $m$.
As seen in these plots,
the space can be partitioned into three regions. 
In the first region, located near the origin, the functions  $X,Y,Z$ present  a
strong variation, with $Y$ approaching its extremum.
The second region is located inside the vortex, where $X$
stays very close to zero, while the field $Z$ is almost constant 
and close to its maximal value.
As seen in the insertion in the first plot, $X$ crosses zero value with
a non-zero first derivative, as it should, since the degree of zero of $\phi$
at the vortex center should be one. 
In the third region, outside the vortex ring, 
the fields approach the vacuum values.
Both $X$ and $Z$ show strong variations in a transition domain between second 
and third regions. We notice also that the value of $X$ at the origin, which was  
$-1$ in the sigma model limit, is now larger, although 
it stays always negative and smaller that $-0.5$ for all solutions we have found.

\begin{figure}[ht]
\hbox to\linewidth{\hss%
	\resizebox{7cm}{6cm}{\includegraphics{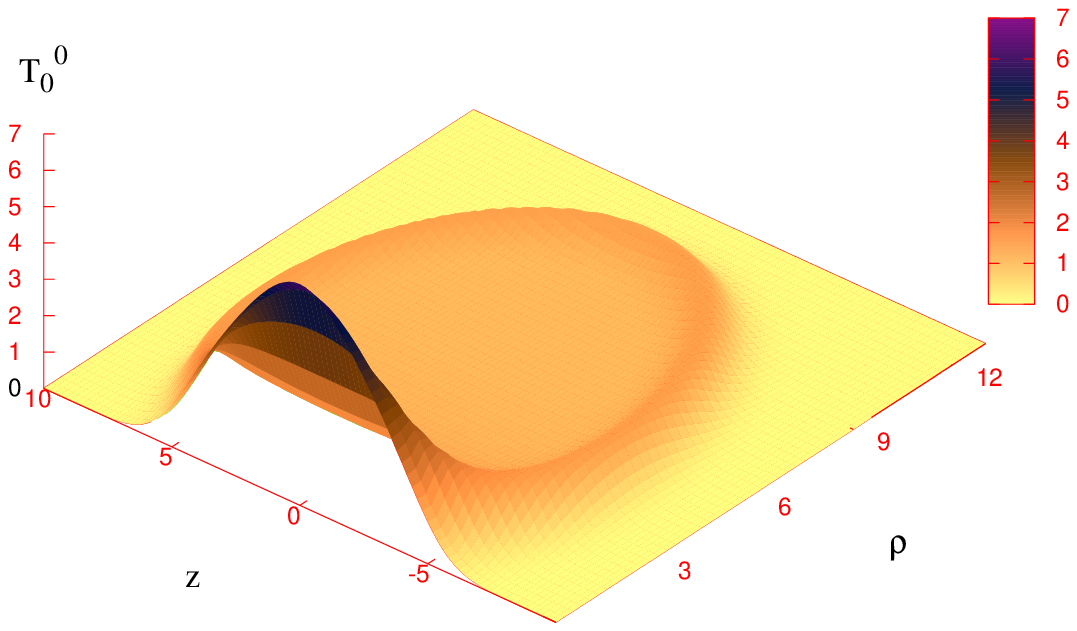}}
\hspace{5mm}%
        \resizebox{7cm}{6cm}{\includegraphics{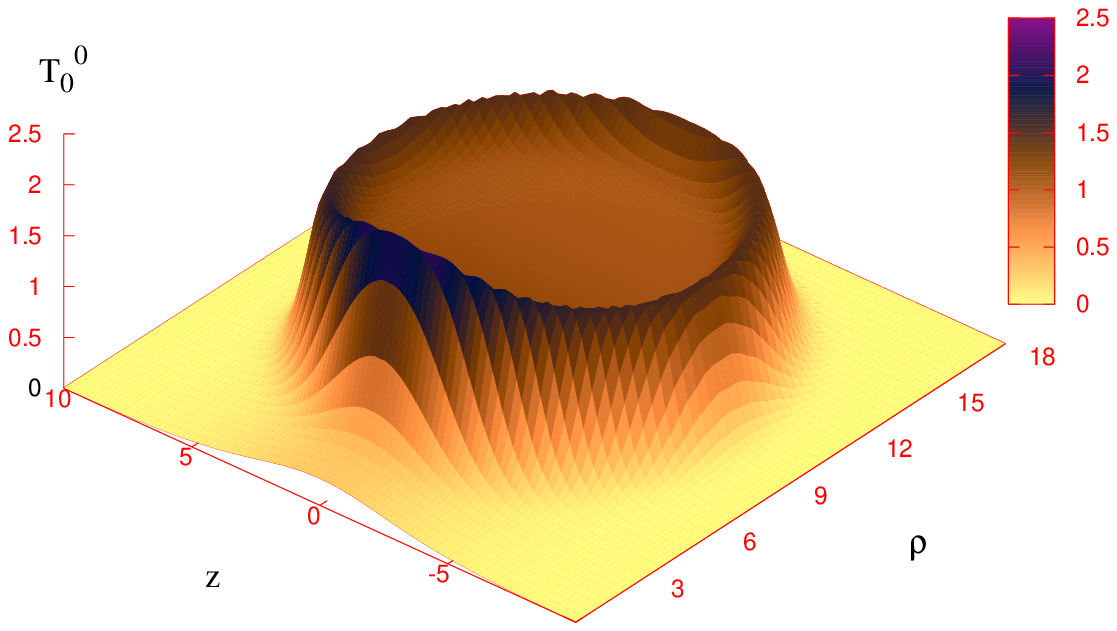}}	
\hss}
\caption{The energy density $T_0^0$ for 
the $n=m=1$ (left) and  $n=1$, $m=3$ (right) vortons with 
$\omega=0.8$,   $\lambda_\phi=39.25$,
$\lambda_\sigma=38.4$, $\eta_\sigma=1$, $\gamma=21.39$.  }
\label{Fig6}
\end{figure}
\begin{figure}[ht]
\hbox to\linewidth{\hss%
	\resizebox{7cm}{6cm}{\includegraphics{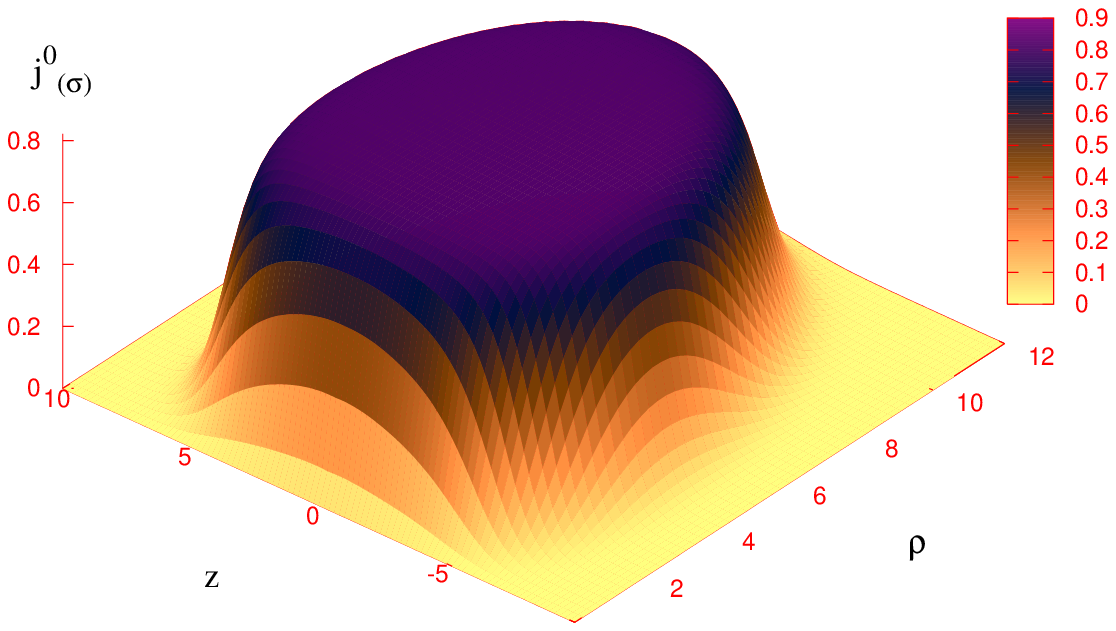}}
\hspace{5mm}%
        \resizebox{7cm}{6cm}{\includegraphics{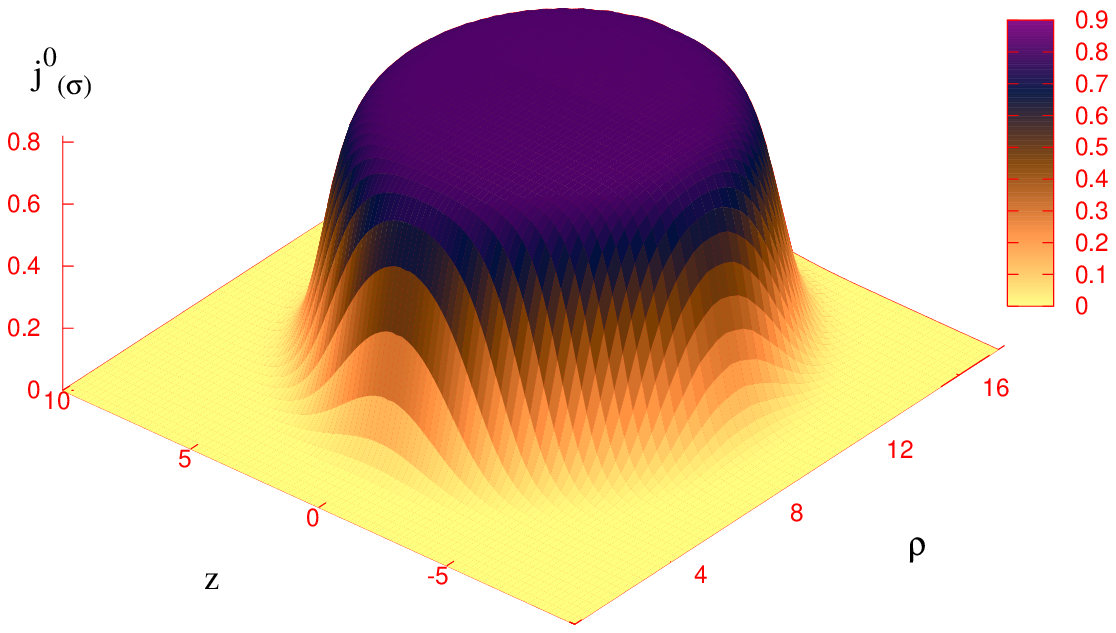}}	
\hss}
	\caption{The charge density for the same 
solutions as in Fig.\ref{Fig6}.}
\label{Fig6a}
\end{figure}
\begin{figure}[ht]
\hbox to\linewidth{\hss%
	\resizebox{7cm}{6cm}{\includegraphics{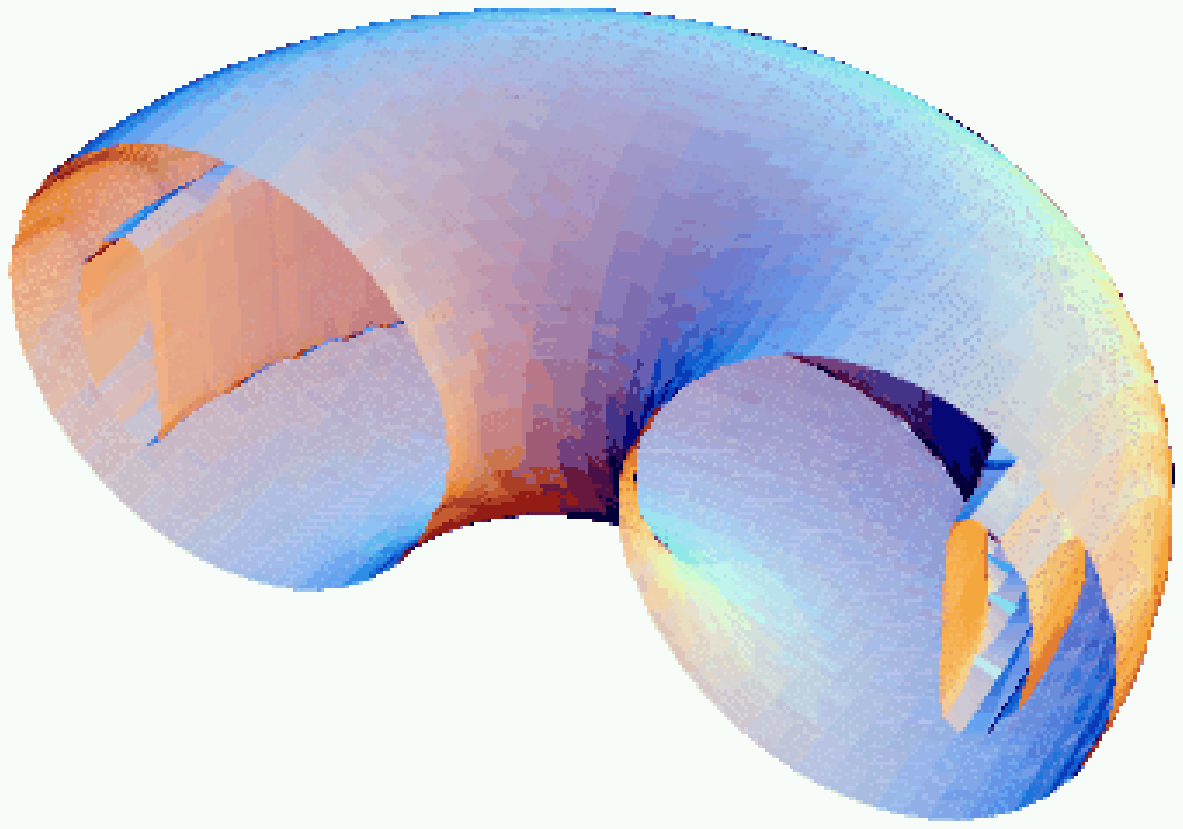}}
\hspace{5mm}%
        \resizebox{7cm}{6cm}{\includegraphics{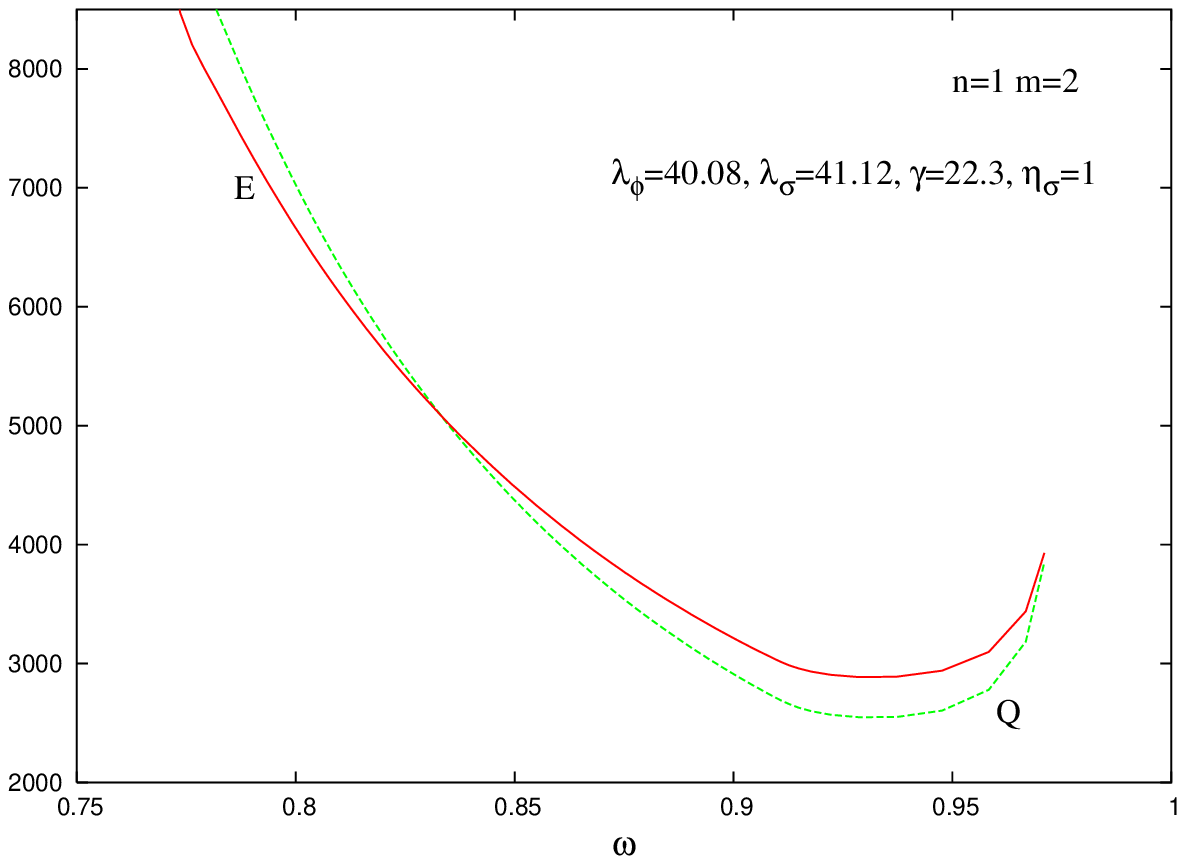}}	
\hss}
	\caption{Left: the surface of constant $(T^0_0=1.05)$ energy density 
for the same $n=1$, $m=3$ vorton solution as in Fig.\ref{Fig6}. 
It has the structure of a torus containing
inside another toroidal surface. Right: 
the energy $E(\omega)$ and charge $Q(\omega)$ for the 
$n=1, m=2$ vorton solutions.}
\label{Fig6aa}
\end{figure}

 For the $m=1$ solutions the 
energy density is maximal at the origin.
Emanating from this central maximum there is 
a toroidal shell of energy containing  the ring of radius $\RRR$ 
where $\phi$ vanishes 
(see Fig.\ref{Fig6}). When $m$ increases
$R$ grows, 
whereas the height of the maximum at the origin 
decreases (see Fig.\ref{Fig6}), so that  
the shape of the  energy density resembles   a hollow tube 
(see Fig.\ref{Fig6aa}). 
There is, however, an almost  constant  energy density  
inside the tube, whose value decreases with $m$. 
As seen in Fig.\ref{Fig6a},
the charge of the solutions, although also localized in a compact region,
does not present the tube structure of the energy density.

Nontrivial vorton solutions are likely to exist for all values of $m$.
We have studied solutions up to $m=5$, but beyond this value 
the numerical accuracy of our procedure decreases drastically.
The energy of the solutions increases with $m$, and 
for solutions with the same $\omega$ one has 
$
E(m)<mE(m=1).
$ 
All solutions we have found  have the ratio
$\RRR/\ccc$ between 1 and 1.5, so that the radius of the vortex ring
is almost the same as the thickness of the vortex. 
The construction of vortons in the thin ring limit, for $\RRR\gg \ccc$, 
remains a numerical challenge.

Fig.\ref{Fig6aa} (right) shows the energy $E$ and charge $Q$ 
as functions of the frequency 
$\omega$ for $m=2$ vortons.  
These functions are actually quite similar to those
found in the $Q$-ball case. 
As one can see,  
solutions exist for a limited range of frequencies, 
$\omega_{-}<\omega<\omega_{+}$.
Unfortunately, the numerical accuracy decreases considerably near the limits
of this interval. 
However, it seems that
 both $Q(\omega)$ and $E(\omega)$ diverge
in these limits, as in the $Q$-ball case.

In analogy with $Q$-balls, the vortons become large as $\omega\to\omega_{\pm}$. 
However, it is not the vorton radius $\RRR$ that becomes large, since it appears to be
not very sensitive to $\omega$ 
and remains finite and nonzero when $\omega$ approaches 
its limiting values. It appears that the torus thickness increases for 
$\omega\to\omega_{\pm}$ so that its volume grows, thereby increasing the 
energy and charge. Both $E$ and $Q$
assume their minimal values at
$\omega=\omega_{\rm crit}$.
 The dependence $E(Q)$ 
appears to be double-valued, very similar to the one shown in Fig.\ref{FigQ3} for $Q$-balls, 
again exhibiting two branches with a cusp.  Solutions with $\omega<\omega_{\rm crit}$
are less energetic than those for $\omega>\omega_{\rm crit}$ and with the
same $Q$. 

Using again the analogy with $Q$-balls, it seems likely that only solutions 
from the lower branch can be stable, probably those for large enough charge. 
Stable vortons certainly exist, perhaps not for all values of the parameters
of the potentials, but at least for large enough values, when the theory 
approaches the sigma model limit. Vortons in the sigma model limit 
are stable, since, as will be discussed below, they can be obtained  via 3D 
energy minimization \cite{Battye:2001ec}. This suggests 
that they should remain stable at least
when they are close to this limit.

Although we did not study  in detail the dependence of the solutions 
on the parameters of the potential, 
one should mention that for $\eta_\sigma\simeq 1$ and $\omega\simeq 0.85$
(the values we mainly considered)
we could find solutions only for $\lambda_\phi\gtrsim 22$,
while
the ratio $\lambda_\phi/\lambda_\sigma$ was always
 close to one. 
In other words, our vortons are not that far from the sigma model limit,
which suggests that they could be stable.

Vortons with $n>1$ also exist, and  we were able to construct them 
for $n=2$. 
For these solutions there are two concentric rings 
in the $\vartheta=\pi/2$ plane where $|\phi|$ vanishes.
All $n=2$ solutions we have found have very large values of
$\lambda_\phi$, $\lambda_\sigma$, $\gamma$. 

Finally, an  interesting class of solutions of Eqs.(\ref{eqs1})
is obtained by setting $Y=0$, which means that the phase of $\phi$ 
is trivial and so $n=0$. Such solutions 
have not been discussed in the literature. 
Although $m\neq 0$ in this case, 
one has $\Q=nm=0$, 
and so according to our definition these solutions are not vortons,
since they are `topologically trivial'. 
However, they  also have a ring structure, resembling  somewhat 
spinning $Q$-balls. One can view them 
as `$\phi$-dressed' $Q$-balls made of the complex $\sigma$ field, 
the non-renormalizable
$|\sigma|^6$ term in the $Q$-balls potential 
being replaced by the interaction with a real scalar field $\phi$,
as was first suggested in Ref.\cite{Friedberg}.

\section{Ring solitons in non-relativistic systems}

So far we have been considering relativistic ring solitons that could be
relevant in models of high energy physics. At the same time, 
similar objects also exist 
in non-relativistic physics, and below we shall consider some of them which 
are most closely related to our previous discussion. 

A famous example of soliton-type 
toroidal systems is provided by the magnetically confined plasma in 
the TOKAMAK's. However, since considering this subject would lead us 
far beyond the scope of our present 
discussion, we shall restrict ourselves to simply giving a reference 
to a recent review \cite{plasma}. In addition, since 
the external field is necessary to confine the plasma in this case, 
this example does not quite fit in with our discussion, because we are
considering closed, self-interacting systems. Another example of plasma confinement,
although somewhat controversial, but
which could perhaps be explained by some solitonic structures, 
is provided by the phenomenon of ball lightning 
(see \cite{Ball} for a bibliography). 
It has in fact been conjectured that some analogs of Faddeev-Skyrme solitons 
in plasma  might be responsible for their existence 
\cite{Faddeev:2000rp},\cite{Faddeev:2000qw}, although we are unaware
of any confirmations of this conjecture via constructing explicit solutions. 
 
Field theory models of condensed matter physics are more closely related to our
discussion.  As in the relativistic case, they  can be either 
theories with local gauge invariance,
as for example the Ginzburg-Landau theory of superconductivity \cite{Ginzburg}, 
or models with 
global internal symmetries,  as in the Gross-Pitaevskii  
\cite{Pitaevskii},
\cite{Pitaevskii1},
\cite{Gross}
theory of Bose-Einstein 
condensation. It has been conjectured that some analogs of the 
Faddeev-Skyrme knot solitons
could exist in the multicomponent Ginzburg-Landau models \cite{Babaev:2001jt}.
However, we are again
unaware of any explicit solutions, and moreover, as was discussed above, 
their existence is not very plausible, at least in the purely
magnetic case. 
On the other hand, global models of condensed matter physics do admit 
ring type solitons.

 \subsection{Vortons versus `skyrmions' in Bose-Einstein condensates  \label{VS}}

Surprisingly, it turns out that the described above vortons
in the global limit of Witten's model can also be considered  as solutions of the 
Gross-Pitaevskii (GP) equation, 
\be                                    \label{GP}
i\frac{\partial\Psi_a}{\partial t}=
\left(-\frac12\,\Delta + V(\bx)+\frac12\sum_{b}\kappa_{ab}|\Psi_b|^2\right)\Psi_a\,.
\ee
This equation, also called nonlinear Schr\"odinger equation, 
provides an effective description of the Bose-Einstein condensates. 
Here $\Psi_a$ are the condensate order parameters, $V(\bx)$ is the external
trapping potential and $\kappa_{ab}$ describe interactions between different condensate
components. In the case of a two-component condensate the indices  $a,b$ assume 
values $1,2$.  

 Let 
\be                              \label{an}
\phi=\phi(\bx),~~~~~~
\sigma=\sigma(\bx)e^{i\omega t}
\ee
be a solution of the equations \eqref{eW} of the global model of Witten \eqref{lag} 
in the case where the parameters in the potential \eqref{potential} are related as
\be                                \label{rel}
\lambda_\phi=\lambda_\sigma\eta_\sigma^2.
\ee
Then setting 
\be
\Psi_1=\,e^{-i\frac{\lambda_\phi}{4}\,t}\,\phi(\bx),
~~~~\Psi_2= e^{-i\left(\frac{\lambda_\phi}{4}+\frac{\omega^2}{2}\right) t}\,\sigma(\bx)
\ee
gives a solution of the GP  equation \eqref{GP} in the case where 
\be
V(\bx)=0,~~~\kappa_{11}=\frac{\lambda_\phi}{2},~~~
\kappa_{22}=\frac{\lambda_\sigma}{2},~~~
\kappa_{12}=\kappa_{21}=\gamma.
\ee
It follows that any vorton solution satisfying the condition \eqref{rel} (this condition
is typically not difficult to achieve numerically) provides a solutions to the  
Gross-Pitaevskii equation and hence can be interpreted in terms of condensed
matter physics. 

In particular, the condition \eqref{rel} is achieved in the 
sigma model limit \eqref{sm-limit},\eqref{constraint},
in which case one obtains solutions of the GP equation with 
\be
V(\bx)=\kappa_{11}=\kappa_{22}=0,~~~~~~~~~
\kappa_{12}=\kappa_{21}=\gamma_0
\ee
by setting $\Psi_1=\phi(\bx)$, $\Psi_2= e^{-i\frac{\omega^2}{2} t}\,\sigma(\bx)$. 
Remarkably, solutions of the GP equation in this case have been independently 
studied by Battye, Cooper and Sutcliffe (BCS) 
\cite{Battye:2001ec}, so that we can compare our results in Sec.\ref{o} with theirs.  
They 
parametrize the fields as 
\be                              \label{anz:3}
\phi(\bx)=\cos\frac{\Theta}{2}\,e^{in\psi}\,,~~~~~~
\sigma(\bx)=\sin\frac{\Theta}{2}\,e^{im\varphi}\,,~
\ee
which exactly coincides with the parametrization \eqref{axial-CP} of the 
Faddeev-Hopf field.  
Instead of directly solving the GP equation, BCS minimize the energy
\be                                    
E_2+E_0=\int(|\nabla\phi|^2+|\nabla\sigma|^2+\gamma_0|\phi|^2|\sigma|^2)d^3\bx
\ee
in a given $\Q$ sector 
by keeping fixed the particle number
\be
{\mcal N}=\int |\sigma|^2 d^3\bx\,.
\ee
They call the resulting configurations $\phi_\ssigma(\bx),\sigma_\ssigma(\bx)$ 
`skyrmions' -- because Eq.\eqref{anz:3} agrees with the 
skyrmion parametrization \eqref{SU1}.  
At first view it is not completely obvious how these 
`skyrmions' are
related 
to our vortons $\phi_\omega(\bx),\sigma_\omega(\bx)$ obtained by solving 
Eqs.\eqref{eqsa} for a given $\omega$. 
However, these are actually the same solutions.
In particular,  the profiles of the vortons described in the previous sections,
their energy density distributions, 
are similar to those for the `skyrmions' 
given in Ref.\cite{Battye:2001ec}.

The two different ways to obtain the solutions, 
either by minimizing the energy or via
solving the equations, are actually equivalent\footnote{We thank 
Richard Battye for explaining this point to us.}. 
This issue has in fact already been discussed above in the $Q$-ball context. 
$Q$-balls  can be obtained either by solving Eq.\eqref{Qe}, which gives 
$\phi_\omega(\bx)$, or via 
minimizing the truncated energy functional $E_0+E_2$ \eqref{NE} 
with the particle number $\ssigma$ \eqref{NN} fixed, which gives 
$\phi_\ssigma(\bx)$. Once one knows 
the value  $\omega=\omega(\ssigma)$, obtained either  
with the formula \eqref{omega2} or from the virial relation \eqref{Qvirial}, the 
two results can be related to each other: 
$\phi_\omega(\bx)=\phi_{\omega(\ssigma)}(\bx)=\phi_\ssigma(\bx)$.

Similarly, the BCS skyrmions  $\phi_\ssigma(\bx),\sigma_\ssigma(\bx)$  
can be related to our vortons $\phi_\omega(\bx),\sigma_\omega(\bx)$ 
by simply establishing the relation $\omega(\ssigma)$. 
The knowledge of $\omega$ is important to make sure that the 
no radiation condition
\be
\omega^2\leq M_\sigma^2=\gamma_0\,
\ee
is fulfilled, which is the case for our solutions,
while their profiles agree  with the features described by BCS 
in \cite{Battye:2001ec}.
This provides an independent confirmation of our numerical results 
described in Sec.\ref{o}.  Since they can be obtained
by the 3D energy minimization, these solutions are stable. 

Apart from the work of BCS there have been other studies of `skyrmions'
in Bose-Einstein condensates 
\cite{Savage:2003hh},
\cite{Ruostekoski:2003qx},
\cite{Ruostekoski:2004pj},
\cite{Ueda},
\cite{Metlitski:2003gj}. Such solutions  
have been  constructed by numerically resolving the GP equation  
for more general choices of 
$V(\bx)$ and $\kappa_{ab}$ and also for more 
than two condensate components. 
Possibilities for an experimental creation
and observation of such objects have also been discussed 
\cite{Ruostekoski:2001fc},
\cite{Ruostekoski:2005zd}.  
In all studies the solutions are typically
presented for the winding numbers $n=1$, with the  
exception of Ref.\cite{Ruostekoski:2004pj},
where results for $n=2$ are reported, although without giving many
details. In our language this corresponds to the double vortons. 

\subsection{Spinning rings in non-linear optics -- 
$Q$-balls as light bullets \label{light}}

Yet another application of the considered above ring solitons arises in the 
theory of light propagating in media whose polarization vector $\vec{\mcal P}$
depends non-linearly
on the electric field $\vec{\mcal E}$. The sourceless Maxwell equations become in this case 
non-linear and in a number of important cases they can be reduced to a 
non-linear Schr\"odinger (NLS) equation  similar to the Gross-Pitaevskii equation \eqref{GP}.
For a Kerr medium, when the non-linear part of  $\vec{\mcal P}$ 
is cubic in $\vec{\mcal E}$, the NLS equation has exactly the same structure as Eq.\eqref{GP}, 
but for more general media it can have different non-linearities. Soliton solutions 
of this NLS equation describe non-linear light pulses, 
sometimes called light bullets, which are
very interesting from the purely theoretical viewpoint and which can 
actually be observed,
as for example in the optical fibers.  
Unfortunately, discussing these issues  in more detail would lead us away
from our subjects,  and so we
simply refer to a monograph \cite{akhmediev} on optical solitons.  

Most of the known solitons of the NLS equation describe plane waves or 
cylindrical beams of light \cite{akhmediev}. However, solutions
describing solitons localized in all three dimensions are also known, 
they have been studied by 
Mihalache $et~ al$ \cite{mihalache}. These solutions describe spinning rings, 
which is
very interesting in the context of our discussion. 

Mihalache $et~ al$ consider light pulses 
travelling in the $z$ direction with a group velocity $V$ 
in a medium with cubic and quintic non-linearities. 
After a suitable rescaling of the coordinates and fields, 
the envelope of the pulse, $\Psi$, satisfies the equation \cite{mihalache}
\be                                       \label{bullet}
i\frac{\partial \Psi}{\partial z}=\left(
-\frac{\partial^2}{\partial x^2}
-\frac{\partial^2}{\partial y^2}
-\frac{\partial^2}{\partial \zeta^2}
+
|\Psi|^4-|\Psi|^2\right)\Psi
\ee
 where $\zeta=z-Vt$. Solutions of this equation, numerically constructed in 
Ref.\cite{mihalache}, describe `spinning light bullets'. 
Interestingly, it turns out that 
these solitons correspond to the spinning
$Q$-balls already described above.  

Specifically, let us rename the variables as 
$z\to t$, $\zeta\to z$
and set
$\Psi=\exp\{i(\omega^2-M^2)t\}\phi$. 
Eq.\eqref{bullet} then assumes the form 
\be                                       \label{bullet1}
i\dot{\phi}=(-\Delta+|\phi|^4-|\phi|^2+M^2-\omega^2)\phi\,.
\ee
Let us now consider Eq.\eqref{Qe00} describing $Q$-balls with the 
potential \eqref{pQ}. We notice that rescaling the spacetime 
coordinates and the field as $x^\mu\to \Lambda_1 x^\mu$, $\Phi\to \Lambda_2\Phi$
is equivalent to changing the values of the parameters $\lambda,a$ in the potential. 
Therefore, as long as they do not vanish, one can choose without any loss of generality
$\lambda=1/3$, $a=3/2$. 
Setting then  $\Phi=e^{i\omega t}\phi$, Eq.\eqref{Qe00} assumes the form 
\be                                       \label{bullet2}
-\ddot{\phi}-2i\omega\dot{\phi}   
=(-\Delta+|\phi|^4-|\phi|^2+M^2-\omega^2)\phi\,,
\ee
with $M^2=\lambda b$.  
We see that if 
$\dot{\phi}=0$ then Eq.\eqref{bullet1} and  Eq.\eqref{bullet2} become identical.
They have therefore the same solutions. 
We thus conclude that the $Q$-balls solutions discussed above in Sec.\ref{Qballs}
can describe light pulses in non-linear media.  
The solutions found in Ref.\cite{mihalache} correspond to the 
simplest $m^{+}$, that is non-twisted, even parity $Q$-balls. Let us remind that for 
a given $m$ these solutions can be parametrized by the value of $\omega$ varying 
within the range 
$\omega^2_{-}(m) < \omega^2< \omega^2_{+}=M^2$. 

Mihalache $et~ al$ also study the dynamical stability of 
the solutions by analysing their small perturbations \cite{mihalache}.  
Perturbing in 
Eq.\eqref{bullet1} the field as $\phi\to\phi+\delta\phi$ and 
linearizing with respect to $\delta\phi$
gives
\be                                       \label{bullet3}
i\dot{\delta\phi}=\hat{{\mcal D}}[\phi]\delta\phi\,,
\ee
where $\hat{{\mcal D}}[\phi]$ is a linear differential operator. 
If the spectrum of this operator
is known, 
\be                                       \label{bullet4}
\hat{{\mcal D}}[\phi]\xi_\lambda=\lambda\xi_\lambda\,,
\ee
then Eq.\eqref{bullet3} is solved by setting 
$\delta\phi=\exp\{i\lambda t\}\xi_\lambda$. 
It follows that  if there are complex 
eigenvalues $\lambda$ with a negative imaginary part, 
then the perturbations grow for $t\to\infty$ and so the 
background is unstable. 
Mihalache $et~ al$ study the spectral  problem \eqref{bullet4} 
and find complex eigenvalues for all 
 spinning solitons with $m=2$. They  also  
integrate Eq.\eqref{bullet} to trace 
the full non-linear perturbation dynamics, and they observe that the spinning
toroidal  soliton splits into {\it three} individual non-spinning spheroidal pieces
\cite{mihalache}. However, they find that the $m=0,1$ solitons can be stable,
both at the linear and non-linear levels, if  
$\omega$ is less than a certain value $\omega_\ast(m)<M$
but still within the domain of the solution existence, 
$\omega_{-}(m) < \omega<\omega_\ast(m)$. 
According to their results, for $m=0$ one has 
$\omega_\ast=\omega_{\rm crit}$, which is the value where 
$E(\omega),Q(\omega)$ pass through the minimum (see Fig.\ref{FigQ3}).  
For $m=1$ the stability region shrinks, since 
$\omega_\ast<\omega_{\rm crit}$, so that 
not all parts of the less energetic solution branch 
(see Fig.\ref{FigQ3}) correspond to stable
solutions. 

These results are very interesting, 
since they allow us to draw  certain
 conclusions about stability of the relativistic $Q$-balls. 
Linearizing  Eq.\eqref{bullet2} gives 
\be                                       \label{bullet5}
-\ddot{\delta\phi}-2i\omega\dot{\delta\phi}=\hat{{\mcal D}}[\phi]\delta\phi\,,
\ee
whose solution is $\delta\phi=\exp\{i\gamma t\}\xi_\lambda$ with 
$\gamma=-\omega\pm\sqrt{\lambda+\omega^2}$. All eigenmodes with complex 
$\lambda$ therefore correspond to complex $\gamma$ and so to unstable modes. 
In addition, there could be unstable modes with 
real $\lambda<-\omega^2$, since $\gamma$ would then be complex. 
As a result, the relativistic $Q$-balls 
have the same unstable modes as their non-relativistic optical counterparts
(although with different instability growth rates) 
and perhaps also some additional instabilities.   

It follows that 
the $2^{+}$ spinning $Q$-balls are unstable, 
while the $1^{+}$ ones, if they belong to the lower solution branch 
and their charge is large enough, should be stable   
(unless they have negative modes with  
$\lambda<-\omega^2$). The latter conclusion agrees with the 
results shown in Fig.\ref{FigQ3}.

As was said above, 
$Q$-balls corresponding to the lower solution branch in Fig.\ref{FigQ3}
cannot decay into free particles. As we now see,
this property does not yet guarantee their stability,
since there can be other decay channels. Their non-relativistic optical counterparts 
 can decay into several solitonic constituents. 
In order to see if the same could be true
for the relativistic $Q$-balls, one should integrate 
the full time evolutions equation Eq.\eqref{bullet2}. 
The existence of negative modes for spinning $Q$-balls is also suggested 
by the results of Ref.\cite{Axenides:2001pi},\cite{BattyeQ}.

The results of Ref.\cite{mihalache} have been generalized in 
Refs.\cite{mihalache1},\cite{mihalache2} to 
describe the case of light pulses with two 
independent polarizations and also for 
more general non-linearities of the medium.

\subsection{Moving vortex rings }

For the sake of completeness, we would also like to consider  
another well known type of ring solitons: moving vortex rings in continuous media. 
Although they are stabilized by the 
interaction with the medium and not by intrinsic forces, 
their structure can be quite similar to that for the 
other solitons considered above.

\begin{figure}[h]
\hbox to\linewidth{\hss%
  \resizebox{12cm}{4cm}{\includegraphics{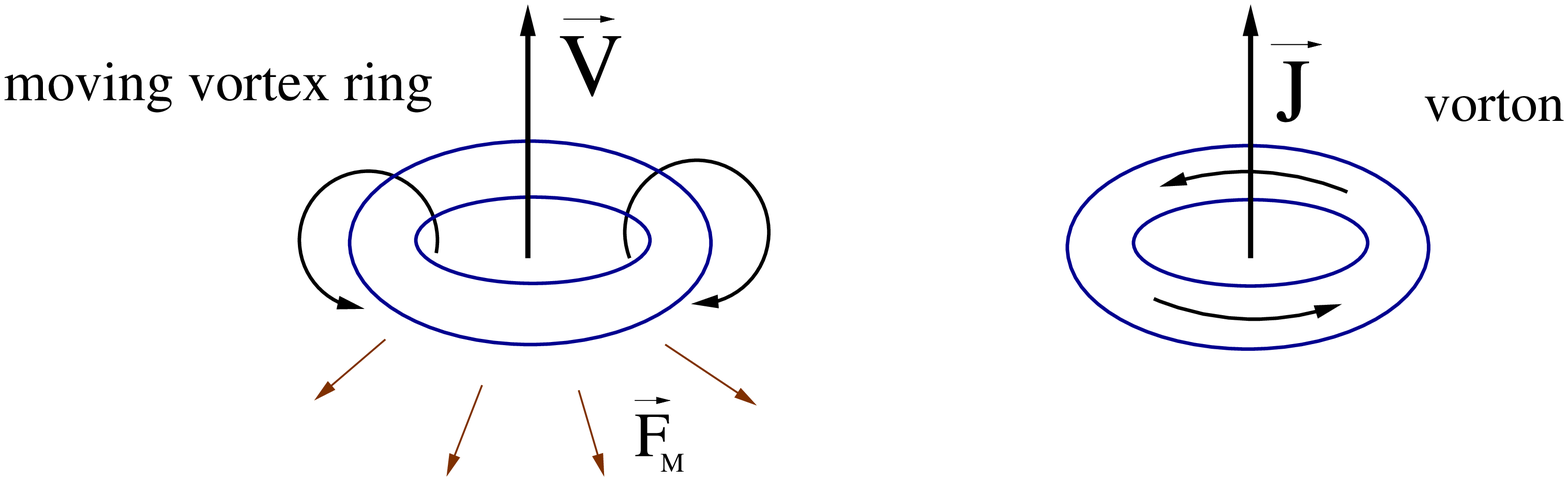}}%
\hss}
\caption{\small 
Moving vortex rings are stabilized by an analog $\vec{F}_M$ of the Magnus force
produced by the phase rotation {\it around} the vortex core.
Vortons are stabilized by the centrifugal force
produced by the phase rotation {\it along} the vortex. 
}
\label{FigVV}
\end{figure}

A vortex ring moving in a medium encounters the `wind'  that 
produces  a circulation around the vortex core, giving  rise to the 
Magnus force -- the same force that acts on a spinning ping-pong ball in 
the direction orthogonal to its velocity. 
This force is directed outwards orthogonally to the ring velocity, 
stabilizing the ring against shrinking (see Fig.\ref{FigVV}). As a result, 
the ring travels with a constant speed (if there is no dissipation),
dynamically keeping a constant radius, so that in the comoving reference 
frame its configuration is stationary. 

The  moving vortex rings are best known in fluid dynamics, as for example 
smoke rings. The vortices themselves correspond in this case to their traditional 
definition -- these are line type objects containing a non-zero 
rotation of the fluid in their core, as for example swirls in water or tornadoes. 
Vortices and 
vortex rings can be created in the wake of rigid bodies travelling  through the fluid. 
Their hydrodynamical theory was created in the 19-th 
century by Helmholtz and Kelvin and has been developing ever since, 
finding numerous practical applications, as for example in the aircraft engineering. 
We do not intend  to discuss
this vast subject here and simply refer to the standard monograph on vortices
\cite{Saffman}. We also give the classical 
formulas of Kelvin relating the vortex ring velocity $V$ and its energy $E$  
and momentum $P$
to the ring radius, $\RRR$, and the vortex core radius, $\ccc$, 
\be                                           \label{Kelvin}
E=\frac12\,\eta^2\Gamma^2\left(\ln\frac{8\RRR}{\ccc}-\frac{7}{4}\right),~~~~
P=\pi\eta\Gamma\RRR^2,~~~~
V=\frac{\Gamma}{4\pi \RRR}\left(\ln\frac{8\RRR}{\ccc}-\frac{1}{4}\right),
\ee
which are valid in the thin ring limit, $\RRR\gg \ccc$.
Here $\eta$ is the fluid mass density
and $\Gamma$ is 
the circulation of the fluid velocity around the vortex.
These formulas show that large rings have large energy and momentum
but move slowly. 
 Very large and slowly moving rings, which could 
perhaps be described by some version of hydrodynamics of magnetized plasma, 
are observed in Sun's atmosphere.  

\subsubsection{Moving vortex rings in the superfluid helium} 

Moving vortex rings exist also in the 
superfluid helium, where they are created by moving impurities 
 \cite{Rayfield} (see \cite{Volovik} for a survey of defects in helium).
It turns out that large rings in helium
can be  well described by the classical hydrodynamics. However, 
in the generic case 
their adequate description should be quantum. 
The superfluid helium can be approximately modeled by a weakly interacting  
Bose-Einstein condensate (although in reality the interactions 
in helium are not weak, see $e.g.$ \cite{Berloff4}),
in which case vortices can be described by solutions of 
the Gross-Pitaevskii equation \eqref{GP}. 
In the simplest case 
one can consider the one-component version of this equation, 
which, upon setting $V(\bx)=0$, 
$\kappa_{11}=\kappa$ 
and $\Psi_1\equiv e^{-i\frac{\kappa}{2} t}\,\Psi$ reduces to
\be                                    \label{GP1}
2i\frac{\partial\Psi}{\partial t}=
\left(-\Delta +\kappa(|\Psi|^2-1)\right)\Psi\,.
\ee
Rescaling coordinates and time one can set $\kappa=1$, if it is positive. 
Choosing $\Psi=e^{in\varphi}f(\rho)$ gives
\be                                    \label{GP2}
f^{\prime\prime}+\frac{1}{\rho}\,f^\prime-\frac{n^2}{\rho^2}\,f
=(f^2-1)f\,,
\ee
whose solution with the boundary conditions $f(0)=0$ and $f(\infty)=1$
describes a global vortex with $n$ quanta of circulation
\cite{Pitaevskii}. 

Solutions of the GP equation \eqref{GP1} describing moving vortex rings 
and their various properties 
have been comprehensively analyzed by P.H.~Roberts and collaborators in a series of articles 
written over the last 25 years,
of which we only mention 
Refs.\cite{Grant},
\cite{Jones},
\cite{Berloff},
\cite{Berloff1},
\cite{Berloff3}.
The fundamental vortex ring solutions 
have been numerically obtained by Jones and Roberts 
\cite{Jones} by setting 
\be
\Psi=\Psi(\rho,z-Vt),
\ee
such that $\Psi(\rho,z)$ satisfies the equation 
\be                                    \label{GP3}
-i2V\partial_z\Psi=
\left(-\Delta +(|\Psi|^2-1)\right)\Psi\,.
\ee
They imposed the reflection condition $\Psi(\rho,z)=\Psi(\rho,-z)^\ast$
and also the asymptotic conditions for large $r$,
\be
\Re(\Psi)=1+O(r^{-3}),~~~~\Im(\Psi)=O(r^{-2}).
\ee
Numerically integrating Eq.\eqref{GP3}, they obtained
a family of solutions labeled by $V\in(0,V_{\rm max})$,
where $V_{\rm max}=1/\sqrt{2}$ 
is the speed of sound in the dimensionless units chosen. 
For small $V$ the solutions have ring profiles: the function $\Psi$ 
vanishes at a point $(\RRR,0)$ in the $(\rho,z)$ plane while its phase
increases by $2\pi$ after one revolution around this point,
so that $\Psi$ has the `vorton topology'  (see
Fig.\ref{Fig:top}). 
When $V$ is small, the radius of the ring, $\RRR$,
as well as its energy and momentum, 
\begin{align}
E&=\frac12\int\left(|\nabla\Psi|^2+\frac12(|\Psi|^2-1)^2\right)d^3\bx\,,\notag \\
P_3&=\Im\int (\Psi^\ast-1)\partial_z\Psi\, d^3\bx,
\end{align}
are large, and in the leading order  they 
agree with the classical Kelvin  formulas 
\eqref{Kelvin} \cite{Grant}.  
As $V$ increases, $E$, $P_3$ and $\RRR$ all decrease, 
and the following relation holds,
\be
V=\frac{\partial E}{\partial P_3}\,.
\ee
The ring radius 
$\RRR$ vanishes for 
$V=V_0\approx 0.62$ and  
for $V_0<V<V_{\rm max}$ solutions change their character and describe 
acoustic excitations in helium, called in  \cite{Jones}  rarefaction
pulses, in which case $\Psi$ 
no longer has zeroes . 
The energy and momentum  diverge when $V$ approaches $0,V_{\rm max}$, 
 and they both  assume their minimal values for some 
$V=V_{\rm crit}\approx 0.657$.
The functions $E(V)$ and $P_3(V)$ exhibit therefore qualitatively the same 
behaviour as $E(\omega)$ and $Q(\omega)$ for the $Q$-balls (see Fig.\ref{FigQ3}). 
The function $E(P_3)$ shows the two-branch structure with a cusp, 
similarly to $E(Q)$ in Fig.\ref{FigQ3}. The moving rings belong to the 
lower branch, which suggests that they could be stable.
Their perturbative stability has been demonstrated by Berloff and Roberts
\cite{Berloff}, who also studied the 
ring creation by solving 
the GP equation for a superfluid flow around a moving solid sphere \cite{Berloff1}
and showed that rings are created in its wake,
provided that its velocity exceeds some critical value.  

The moving ring solutions have been generalized for a non-zero trapping
potential $V(\bx)$ in the GP equation (see \cite{Komineas} for a review)
 and for the two-component
condensate case \cite{Berloff2}. An interesting solution of the 
GP equation describing  a non-linear superposition of a straight vortex and
a vortex ring encircling it and moving along it, 
somewhat similar to the `hoop' solution shown in Fig.\ref{hoop}, 
 has been presented by Berloff in
 \cite{Berloff3}.  

One may also wonder whether the vortex ring solutions 
could be generalized within a
relativistic field theory (as is the case for the vortex 
solution of Eq.\eqref{GP2}). 
In fact, the answer to this question is affirmative, since taking
 a solution $\Psi(\rho,z)$ of Eq.\eqref{GP3} and  setting
\be
\phi=e^{iV(z\pm t)}\Psi(\rho,z)
\ee
and also $\sigma=0$ solves
Eqs.\eqref{eW} of Witten's model  
for $\lambda_\phi=2$.
 However, since $\Psi\to 1$ at infinity,
it follows that $\phi\to e^{iV(z\pm t)}$ in this limit, which 
corresponds to a massless
Goldstone radiation.
The relativistic
analogs of the moving vortex rings are therefore 
radiative solutions and their total 
energy is infinite.

\subsubsection{Moving magnetic rings \label{magnet}}

An interesting example of moving ring solitons, 
found by Cooper \cite{Cooper} and  
studied also by 
Sutcliffe  \cite{Sutcliffe:2007vm}, 
arises in ferromagnetic systems
whose dynamics is described in the continuum 
limit by the Landau-Lifshitz equation
\be                                                               \label{LL1}
\frac{\partial{n^a}}{\partial t}=\epsilon_{abc}n^b\Delta n^c.
\ee  
Here $n^a$ is a unit three-vector, $n^a n^a=1$, giving the local orientation
of the magnetization. The temporal evolution defined by the 
Landau-Lifshitz equation
preserves the value of the energy 
\be
E= \frac{1}{4}\int (\partial_k n^a)^2d^3\bx\,.
\ee
In addition, it does not change the 
momentum,
\be                       \label{P}
P_i=\frac12 \int \epsilon_{ijk}(x_j {\mcal B}_k) d^3\bx,
\ee
where ${\mcal B}_i=\frac{1}{2}\epsilon_{ijk}{\mcal F}_{jk}$ and ${\mcal F}_{jk}$ is defined
as in the Faddeev-Skyrme model  by Eq.\eqref{F}. 
The number of spin flips,
\be
{\mcal N}=\frac12\int(1-n^3)d^3\bx,
\ee
is also preserved by the evolution; 
for this number to be finite one should have $n^3=1$ at infinity.
Since the theory contains a unit vector field,
it can be characterized by the topological index of Hopf defined by the same 
equation \eqref{Q} as in the Faddeev-Skyrme model. 
An  axially symmetric field configuration 
with a non-zero Hopf charge can be parametrized 
exactly as in Eq.\eqref{Hopf:axial} (with $n=1$),
\be                        \label{nn}
n^1+in^2=\sin\Theta(\rho,z) \exp\{i[m\varphi-\psi(\rho,z)]\},~~~~
n^3=\cos\Theta(\rho,z).
\ee  
The energy density then has a ring structure and the 
Hopf charge is 
$\Q=m.$
Using this to calculate
the momentum \eqref{P} shows that its $z$-component
does not vanish,
\be
P_z=\frac14\int \rho(\partial_z \Theta\partial_\rho\psi
-\partial_\rho \Theta\partial_z\psi) d^3\bx\,.
\ee 
This means that if one uses \eqref{nn} as the initial data for the 
Landau-Lifshitz equation, then this data set will have a 
non-zero initial momentum.
To verify this assertion, 
Sutcliffe \cite{Sutcliffe:2007vm} directly integrates the Landau-Lifshitz equation 
to reconstruct the temporal dynamics of the  ring.  
For largely arbitrary functions $\Theta(\rho,z)$, $\psi(\rho,z)$ in \eqref{nn}
he then sees the whole ring configuration drift  with an almost constant 
velocity along the $z$-axis. He also observes 
that the phase of $n^1+in^2$ increases as a linear function of the 
distance travelled by  the ring,
so that there are internal rotations in the ring as it moves. 

These empirical features can be understood by making the following ansatz 
for the dynamical fields \cite{Cooper},
\be                        \label{nn1}
n^1+in^2=\sin \Theta(\rho,z-Vt)\exp\{i[m\varphi+\omega t-\psi(\rho,z-Vt)]\},~~~~
n^3=\cos\Theta(\rho,z-Vt),
\ee  
with constant $V,\omega$. Inserting this ansatz to the Landau-Lifshitz equation
\eqref{LL1}, the $t,\varphi$ variables separate and the problem reduces 
to the two dimensional elliptic problem for $\Theta(\rho,z),\psi(\rho,z)$,
\begin{align}                    \label{eq-magn}
\Delta \Theta&=(\omega+V\partial_z\psi)\sin\Theta
+\left[(\partial_\rho\psi)^2+(\partial_z\psi)^2)
+\frac{m^2}{\rho^2}\right] \sin\Theta\cos\Theta, \notag \\
\sin\Theta\Delta\psi&=-V\partial_z \Theta
-2\cos\Theta[\partial_\rho \Theta\partial_\rho\psi+\partial_z \Theta\partial_z\psi].
\end{align}
This problem is in fact very much similar to that in the 
vorton case.  This can be seen by applying the same change
of variables which was used to rewrite the Faddeev--Skyrme model \eqref{action0}
in the $CP^1$ form \eqref{action1}, 
$
n^a=\Phi^\dagger \tau^a\Phi\,,
$
with $\Phi^\dagger\Phi=1$.  This parametrization of the system allows one to  
write down a simple Lagrangian for the  Landau-Lifshitz equation \eqref{LL1},
\be                               \label{LagLL}
{\mcal L}=i\Phi^\dagger\partial_t\Phi
-\frac14(\partial_k n^a)^2
+\mu(\Phi^\dagger\Phi-1),
\ee
where $\mu$ is the Lagrange multiplier. One can parametrize
$\Phi$ exactly in the same way as in the vorton case,
$$
\Phi=
\left(\begin{array}{c}
X+iY\\
Z e^{i\omega t+i m\varphi}\\
\end{array}
\right).
$$
Inserting this  to the Lagrangian \eqref{LagLL} and 
varying with respect to $X,Y,Z,\mu$ gives equations similar to 
Eqs.\eqref{eqsa} for vortons in the sigma model limit. 
These equations can be solved  with the same boundary conditions 
\eqref{1}--\eqref{4}
for  $X,Y,Z$ as for vortons. On the other hand, setting 
$X+iY=\cos(\Theta/2)e^{i\psi}$ and $Z=\sin(\Theta/2)$,
as in Eq.\eqref{anz:3}, gives 
again Eqs.\eqref{eq-magn}. 

Fixing $m$  
one can solve the equations for suitable  input values of $\omega,V$. 
Having obtained the solutions, one can calculate
 $E(\omega,V)$, ${\mcal N}(\omega,V)$ and $P_3(\omega,V)$.  
Alternatively, 
instead of solving the equations, one can 
minimize the energy $E$ with fixed 
${\mcal N},P_3$. Having found the 
minimal value $E({\mcal N},P_3)$, one can reconstruct  $\omega,V$ as 
\cite{Cooper},\cite{Sutcliffe:2007vm}
\be
\omega=\frac{\partial E}{\partial {\mcal N}},~~~~~
V=\frac{\partial E}{\partial P_3}.
\ee
Cooper \cite{Cooper} and
 Sutcliffe \cite{Sutcliffe:2007vm} choose the latter approach:
they minimize the energy with fixed $\ssigma,P_3$ 
for several values of $m=\Q$.
Their numerics converge to non-trivial  
configurations  $\Theta(\rho,z),\psi(\rho,z)$, inserting which 
to \eqref{nn1} 
gives dynamical rings travelling with a constant velocity.  
Fixing all other parameters, the ring velocity decreases with growing $m$. 

In fact, these rings also have an angular momentum,
$J=m\ssigma $,
but apparently this feature is  not so essential for their 
stabilization as it is for vortons, 
since solutions with $m=\Q=J=0$ also exist 
and do not exhibit any particular features  
as compared to the solutions with $m\neq 0$ \cite{Sutcliffe:2007vm}.  
Similarly to the other moving ring solitons discussed above,
magnetic rings are stabilized by a Magnus-type force produced 
by the phase circulation 
{\it around} the vortex core and directed outwards, 
orthogonally to the velocity (see Fig.\ref{FigVV}). 
The magnetic rings are  stable not only against shrinking but also
with respect to all other deformations -- since they can be obtained via the 3D energy 
minimization  \cite{Sutcliffe:2007vm}.

\section{Concluding remarks}

We have reviewed  the known field theory solutions 
describing stationary vortex loops.
From the physical point of view, loops of magnetic 
flux would perhaps be the most interesting. 
However, as we have seen, almost all explicit solutions  
describe vortex loops  in global field theories,
which are effective theories with a relatively limited range of applicability. 
Perspectives for generalizing these solutions in the context of gauge field theory 
do not look very promising at the moment, at least as far as  knot solitons are concerned.  
The best known static knot solitons, Faddeev-Skyrme knots,
do not seem to admit immediate gauge field theory generalizations, unless 
one makes additional physical assumptions to fix the charges in the 
Protogenov-Verbus formula. Such assumptions can be made, as shows the example
of Schmid and Shaposhnikov. However this requires a rather exotic physical 
environment.  

The non-Abelian monopole and sphaleron 
rings are very interesting 
theoretically, however,  they are presumably unstable. 
It is therefore unclear whether they can find important
physical applications. Summarizing, the existence of knots -- magnetic vortex loops 
stabilized against shrinking by internal stresses -- 
does not seem to be very plausible
in the context of physically interesting relativistic gauge field theories.

Perhaps gauged vortons have more chances 
to exist. 
As we have seen, their global counterparts  
do exist  as stationary, non-radiating field theory objects.
Already for these global vortons an additional analysis to study the structure 
of the parameter space is needed.  
A natural problem to address would be their analysis  
in the thin ring limit, for  large values of the azimuthal winding number $m$, 
when the ring radius is much larger than the vortex core thickness. 
One can expect that the field theory solutions in this limit should agree 
with the effective macroscopic description. This limit, however, is difficult to 
explore numerically, since the vorton fields are then almost everywhere 
constant, except for  a narrow ring region containing all the field gradients.
It is difficult to properly adjust the grid in this case, so that some  
other numerical methods are necessary, as perhaps 
the multi-domain spectral method \cite{Grandclement}. 
Is seems also that some other methods
are necessary to study solutions with $n>1$.

Yet another interesting related problem would be to find 
vortons with {\it three} 
winding
numbers. We know that the phase of $\phi$ winds around 
the vortex core, while
that of $\sigma$ winds along the ring, the 
corresponding winding numbers being $n$ and $m$. 
However, nothing forbids the phase of $\sigma$ to wind {\it both} 
around the core and
along the ring, as does the phase of the twisted $Q$-balls, 
in which case one
would need a third integer, $k$, counting the 
windings of $\sigma$ around the 
vortex core. Constructing vortons with $k\neq 0$ remains a challenge.   

A very important open problem related to all vortons is their 
dynamical stability analysis. This problem can be handled either by 
studying the spectrum of small field  fluctuations around the vorton background,
or by looking for vorton negative modes within the energy minimization method,
or by reconstructing the full temporal evolution of the perturbed
vorton configuration. Stable vortons definitely exist, at least in the 
sigma model limit, but in the 
generic case their stability should be studied.

Generalizing global vortons within gauge field theory does not seem
to be impossible. In theories where all the internal symmetries are gauged,
as in Witten's model \eqref{W}, all spinning phases of the fields can 
be  gauged away, so that if the fields are stationary and axisymmetric, these  
symmetries are manifest.
In this case the surface integral formula \eqref{JJJ} for the 
angular momentum applies, providing  higher chances for
$J$ to vanish due to 
the fast asymptotic falloff of the fields. Although   
the example of spinning gauged $Q$-balls shows that spinning is possible
in local theories, 
it seems that vortons have better chances to exist in models with both local
and global internal symmetries. 
Such models contain gauge fields, such that one can have magnetic vortices, 
but the gauge freedom is not enough to gauge away  all the phases of the scalars. 
The symmetries of the fields are therefore not manifest and the surface integral 
formula does not apply, in which case the 
angular momentum should normally be non-zero. 

For example, it would be natural to analyze
 the existence of stationary vortons in a half gauged
version of Witten's modes obtained by coupling the two scalars in Eq.\eqref{lag}
to a U(1) gauge field. 
Such solutions could then perhaps be generalized within the
fully gauged U(1)$\times$U(1) Witten's modes to confirm  the existence of 
stationary loops made of superconducting cosmic strings  -- the idea first 
put forward more than 20 years ago. 

However, a real challenge would be to construct stationary superconducting loops
in Standard Model. There are  indications that this might be possible.
The electroweak sector of Standard Model contains stationary 
current carrying vortices -- superconducting strings \cite{Volkov:2006ug}. 
Their stability 
has been studied so far only in the semilocal limit, where the SU(2) gauge field 
decouples and the current becomes global \cite{Forgacs:2006pm},\cite{Forgacs:2005sf}, 
and it has been found that sufficiently short pieces of strings 
are perturbatively stable \cite{Garaud:2007ti}.
This suggests that small loops 
made of these strings may also be stable. If such loops could be 
constructed for the physical value of the weak mixing angle, this would 
give stable solitons in Standard Model. 

For the sake of completeness, having in mind possible applications of vortex loops
in astrophysics and cosmology, it is also interesting to  mention the effects
of gravity on these solutions. On general grounds, one expects all known ring 
solitons to admit self-gravitating generalizations with essentially the same properties
as in the zero gravity limit,  
at least for small enough values of the gravitational coupling constant. However,
the soliton structure can change considerably
in the strong gravity regime \cite{VG}.
The solitons can then become gravitationally closed and may also contain 
a small black hole in their center. The spinning  of the latter can 
endow the whole configuration with an angular momentum, 
even if the soliton 
itself cannot spin -- as for example the monopole \cite{monBH}. 
However, if the soliton can spin on its own,
as for example a vorton, it would be  interesting to put a spinning black hole in its center. 
Since there are no asymptotically flat toroidal black holes in four dimensions 
\cite{Friedman}, the resulting configuration is expected to be a black hole with spherical 
horizon topology, surrounded by a vortex ring.

\vspace{1 cm}
{\bf Acknowledgements.}
The work of E.R. was supported by the ANR grant NT05-$1_{-}$42856 
 `Knots and Vortons'. We would like to thank Richard Battye, Brandon Carter, 
Maxim Chernodub, Ludwig Faddeev, Peter Forgacs, Philippe Grandclement, Betti Hartmann, 
Xavier Martin, Antti Niemi, 
Mikhail Shaposhnikov,
Matthias Schmid, Paul Shellard, Paul Sutcliffe 
and Tigran Tchrakian  
 for discussions during various stages of this work. 


\begin{thebibliography}{100}

\bibitem{Abrikosov}
A.A. Abrikosov.
\newblock {\ssl On the magnetic properties of superconductors of the second
  group}.
\newblock {\em {\XPEH Sov.Phys.JETP}}, {\bbf 5}, 1174--1182, 1957.

\bibitem{Achucarro:1999it}
A.~Achucarro and T.~Vachaspati.
\newblock {\ssl Semilocal and electroweak strings}.
\newblock {\em {\XPEH Phys.Rept.}}, {\bbf 327}, 347--426, 2000.

\bibitem{Adam:2006wg}
C.~Adam, J.~Sanchez-Guillen, R.A. Vazquez, and A.~Wereszczynski.
\newblock {\ssl Investigation of the Nicole model}.
\newblock {\em {\XPEH J.Math.Phys.}}, {\bbf 47}, 052302, 2006.

\bibitem{Adkins:1983hy}
G.~Adkins and C.R. Nappi.
\newblock {\ssl The Skyrme model with pion masses}.
\newblock {\em {\XPEH Nucl.Phys.}}, {\bbf B233}, 109--115, 1984.

\bibitem{Adkins:1983ya}
G.~Adkins, C.R. Nappi, and W.~Witten.
\newblock {\ssl Static properties of nucleons in the Skyrme model}.
\newblock {\em {\XPEH Nucl.Phys.}}, {\bbf B228}, 552--566, 1983.

\bibitem{akhmediev}
N.N. Akhmediev and A.~Ankiewicz.
\newblock {\em {\ssl Solitons, nonlinear pulses and beams}}.
\newblock {\XPEH Chapman and Hall, London}, 1997.
\newblock 299 p.

\bibitem{Amsterdamski:1988zp}
P.~Amsterdamski and P.~Laguna-Castillo.
\newblock {\ssl Internal structure and the spacetime of superconducting bosonic
  strings}.
\newblock {\em {\XPEH Phys.Rev.}}, {\bbf D37}, 877--884, 1988.

\bibitem{Aratyn:1999cf}
H.~Aratyn, L.A. Ferreira, and A.H. Zimerman.
\newblock {\ssl Exact static soliton solutions of 3+1 dimensional integrable
  theory with nonzero Hopf numbers}.
\newblock {\em {\XPEH Phys.Rev.Lett.}}, {\bbf 83}, 1723--1726, 1999.

\bibitem{Aratyn}
H.~Aratyn, L.A. Ferreira, and A.H. Zimerman.
\newblock {\ssl Toroidal solitons in 3+1 dimensional integrable theories}.
\newblock {\em {\XPEH Phys.Lett.}}, {\bbf B456}, 162--170, 1999.

\bibitem{Axenides:2001pi}
M.~Axenides, E.~Floratos, S.~Komineas, and L.~Perivolaropoulos.
\newblock {\ssl Q-rings}.
\newblock {\em {\XPEH Phys.Rev.Lett.}}, {\bbf 86}, 4459--4462, 2001.

\bibitem{Babaev:2001jt}
E.~Babaev.
\newblock {\ssl Knotted solitons in triplet superconductors}.
\newblock {\em {\XPEH Phys.Rev.Lett.}}, {\bbf 88}, 177002, 2002.

\bibitem{Babaev:2001zy}
E.~Babaev, L.D. Faddeev, and A.J. Niemi.
\newblock {\ssl Hidden symmetry and duality in a charged two-condensate Bose
  system}.
\newblock {\em {\XPEH Phys.Rev.}}, {\bbf B65}, 100512, 2002.

\bibitem{Babul:1987me}
A.~Babul, T.~Piran, and D.M. Spergel.
\newblock {\ssl Bosonic superconducting cosmic strings. I. Classical field
  theory solutions}.
\newblock {\em {\XPEH Phys.Lett.}}, {\bbf B202}, 307--314, 1988.

\bibitem{Battye:1998pe}
R.~Battye and P.~Sutcliffe.
\newblock {\ssl Knots as stable soliton solutions in a three-dimensional
  classical field theory}.
\newblock {\em {\XPEH Phys.Rev.Lett.}}, {\bbf 81}, 4798--4801, 1998.

\bibitem{Battye:1998zn}
R.~Battye and P.~Sutcliffe.
\newblock {\ssl Solitons, links and knots}.
\newblock {\em {\XPEH Proc.Roy.Soc.Lond.}}, {\bbf A455}, 4305--4331, 1999.

\bibitem{BattyeQ}
R.~Battye and P.~Sutcliffe.
\newblock {\ssl $Q$-ball dynamics}.
\newblock {\em {\XPEH Nucl.Phys.}}, {\bbf B590}, 329--363, 2000.

\bibitem{Battye:2001ec}
R.A. Battye, N.R. Cooper, and P.M. Sutcliffe.
\newblock {\ssl Stable Skyrmions in two-component Bose-Einstein condensates}.
\newblock {\em {\XPEH Phys.Rev.Lett.}}, {\bbf 88}, 080401, 2002.

\bibitem{Battye:2005nx}
R.A. Battye, S.~Krusch, and P.M. Sutcliffe.
\newblock {\ssl Spinning Skyrmions and the Skyrme parameters}.
\newblock {\em {\XPEH Phys.Lett.}}, {\bbf B626}, 120--126, 2005.

\bibitem{Battye-kink}
R.A. Battye and P.~Sutcliffe.
\newblock {\ssl Kinky vortons}, 2008.
\newblock {\tt e-Print: arXiv:0806.2212 [hep-th]}.

\bibitem{Berloff2}
N.G. Berloff.
\newblock {\ssl Nucleation of solitary wave complexes in two-component mixture
  Bose-Einstein condensates}, 2005.
\newblock arXiv:cond-mat/0412743.

\bibitem{Berloff3}
N.G. Berloff.
\newblock {\ssl Solitary waves on vortex lines in Ginzburg-Landau models for
  the example of Bose-Einstein condensates}.
\newblock {\em {\XPEH Phys.Rev.Lett.}}, {\bbf 94}, 010403, 2005.

\bibitem{Berloff1}
N.G. Berloff and P.H. Roberts.
\newblock {\ssl Motion in a Bose condensate: VII. Boundary-layer separation}.
\newblock {\em {\XPEH J.Phys.}}, {\bbf A33}, 4025--4038, 2000.

\bibitem{Berloff4}
N.G. Berloff and P.H. Roberts.
\newblock {\ssl Nonlinear Schr\"odinger equation as a model of superfluid
  helium}.
\newblock In {\em {\XPEH Quantized Vortex Dynamics and Superfluid Turbulence}},
  volume~{\bbf 571} of {\em Lecture Notes in Physics}. Springer-Verlag, 2001.
\newblock Edited by C.F. Barenghi, R.J. Donnelly and W.F. Vinen.

\bibitem{Berloff}
N.G. Berloff and P.H. Roberts.
\newblock {\ssl Motion in a Bose condensate: X. New results on stability of
  axisymmetric solitary waves of the Gross-Pitaevskii equation}.
\newblock {\em {\XPEH J.Phys.}}, {\bbf A37}, 11333--11351, 2004.

\bibitem{Betz}
M.~Betz, H.B. Rodrigues, and T.~Kodama.
\newblock {\ssl Rotating skyrmion in (2+1)-dimensions}.
\newblock {\em {\XPEH Phys.Rev.}}, {\bbf D54}, 1010--1019, 1996.

\bibitem{Bevis:2006mj}
N.~Bevis, M.~Hindmarsh, M.~Kunz, and J.~Urrestilla.
\newblock {\ssl CMB power spectrum contribution from cosmic strings using
  field-evolution simulations of the Abelian Higgs model}.
\newblock {\em {\XPEH Phys.Rev.}}, {\bbf D75}, 065015, 2007.

\bibitem{Bogomolnyi}
E.B. Bogomol'nyi.
\newblock {\ssl Stability of classical solutions}.
\newblock {\em {\XPEH Sov.J.Nucl.Phys.}}, {\bbf 24}, 449--454, 1976.

\bibitem{plasma}
A.H. Boozer.
\newblock {\ssl Physics of magnetically confined plasmas}.
\newblock {\em {\XPEH Rev.Mod.Phys.}}, {\bbf 76}, 1071--1141, 2004.

\bibitem{Brandenberger:1996zp}
R.H. Brandenberger, B.~Carter, A.C. Davis, and M.~Trodden.
\newblock {\ssl Cosmic vortons and particle physics constraints}.
\newblock {\em {\XPEH Phys.Rev.}}, {\bbf D54}, 6059--6071, 1996.

\bibitem{Brihaye:2007tn}
Y.~Brihaye and B.~Hartmann.
\newblock {\ssl Interacting Q-balls}, 2007.
\newblock {\tt arXiv:0711.1969 [hep-th]}.

\bibitem{Callan:1983nx}
C.G. Callan and E.~Witten.
\newblock {\ssl Monopole catalysis of skyrmion decay}.
\newblock {\em {\XPEH Nucl.Phys.}}, {\bbf B239}, 161--176, 1984.

\bibitem{Carter:1989dp}
B.~Carter.
\newblock {\ssl Duality relation between charged elastic strings and
  superconducting cosmic strings}.
\newblock {\em {\XPEH Phys.Lett.}}, {\bbf B224}, 61--66, 1989.

\bibitem{Carter:1989xk}
B.~Carter.
\newblock {\ssl Stability and characteristic propagation speeds in
  superconducting cosmic and other string models}.
\newblock {\em {\XPEH Phys.Lett.}}, {\bbf B228}, 466--470, 1989.

\bibitem{Carter:1990sm}
B.~Carter.
\newblock {\ssl Mechanics of cosmic rings}.
\newblock {\em {\XPEH Phys.Lett.}}, {\bbf B238}, 166--171, 1990.

\bibitem{Carter:1994hn}
B.~Carter and P.~Peter.
\newblock {\ssl Supersonic string model for Witten vortices}.
\newblock {\em {\XPEH Phys.Rev.}}, {\bbf D52}, 1744--1748, 1995.

\bibitem{Cho:2001gc}
Y.M. Cho.
\newblock {\ssl Knot solitons in Weinberg-Salam model}, 2001.
\newblock arXiv:hep-th/0110076.

\bibitem{Cho}
Y.M. Cho.
\newblock {\ssl Monopoles and knots in Skyrme theory}.
\newblock {\em {\XPEH Phys.Rev.Lett.}}, {\bbf 87}, 252001, 2001.

\bibitem{Coleman:1982cx}
S.R. Coleman.
\newblock {\ssl The magnetic monopole fifty years later}.
\newblock Lectures given at Int. Sch. of Subnuclear Phys., Erice, Italy, Jul
  31-Aug 11, 1981, at 6th Brazilian Symp. on Theor. Phys., Jan 7-18, 1980, at
  Summer School in Theoretical Physics, Les Houches, France, and at Banff
  Summer Inst. on Particles $\&$ Fields, Aug 16-28, 1981.

\bibitem{Coleman:1985ki}
S.R. Coleman.
\newblock {\ssl Q balls}.
\newblock {\em {\XPEH Nucl.Phys.}}, {\bbf B262}, 263--283, 1985.

\bibitem{Coleman:1976uk}
S.R. Coleman, S.~Parke, A.~Neveu, and C.M. Sommerfield.
\newblock {\ssl Can one dent a dyon?}
\newblock {\em {\XPEH Phys.Rev.}}, {\bbf D15}, 544--545, 1977.

\bibitem{Cooper}
N.R. Cooper.
\newblock {\ssl Propagating magnetic vortex rings in ferromagnets}.
\newblock {\em {\XPEH Phys.Rev.Lett.}}, {\bbf 82}, 1554--1557, 2001.

\bibitem{Copeland:1987yv}
E.J. Copeland, D.~Haws, M.~Hindmarsh, and N.~Turok.
\newblock {\ssl Dynamics of and radiation from superconducting strings and
  springs}.
\newblock {\em {\XPEH Nucl.Phys.}}, {\bbf B306}, 908--930, 1988.

\bibitem{Copeland:1987th}
E.J. Copeland, M.~Hindmarsh, and N.~Turok.
\newblock {\ssl Dynamics of superconducting cosmic strings}.
\newblock {\em {\XPEH Phys.Rev.Lett.}}, {\bbf 58}, 1910--1913, 1987.

\bibitem{Correia}
F.~Paccetti Correia and M.G. Schmidt.
\newblock {\ssl Q balls: some analytical results}.
\newblock {\em {\XPEH Eur.Phys.Journ.}}, {\bbf C21}, 181--191, 2001.

\bibitem{Davis:1988ip}
R.L. Davis.
\newblock {\ssl Semitopological solitons}.
\newblock {\em {\XPEH Phys.Rev.}}, {\bbf D38}, 3722--3730, 1988.

\bibitem{Davis:1988jp}
R.L. Davis and E.P.S. Shellard.
\newblock {\ssl The physics of vortex superconductivity}.
\newblock {\em {\XPEH Phys.Lett.}}, {\bbf B207}, 404--410, 1988.

\bibitem{Davis:1988jq}
R.L. Davis and E.P.S. Shellard.
\newblock {\ssl The physics of vortex superconductivity. 2}.
\newblock {\em {\XPEH Phys.Lett.}}, {\bbf B209}, 485--490, 1988.

\bibitem{Davis:1988ij}
R.L. Davis and E.P.S. Shellard.
\newblock {\ssl Cosmic vortons}.
\newblock {\em {\XPEH Nucl.Phys.}}, {\bbf B323}, 209--224, 1989.

\bibitem{deVega:1977rk}
H.J. de~Vega.
\newblock {\ssl Closed vortices and the Hopf index in classical field theory}.
\newblock {\em {\XPEH Phys.Rev.}}, {\bbf D18}, 2945--2951, 1978.

\bibitem{VanderBij:2001nm}
J.J.~Van der Bij and E.~Radu.
\newblock {\ssl On rotating regular nonabelian solutions}.
\newblock {\em {\XPEH Int.J.Mod.Phys.}}, {\bbf A17}, 1477--1490, 2002.

\bibitem{vanderBij:2002sq}
J.J.~Van der Bij and E.~Radu.
\newblock {\ssl Magnetic charge, angular momentum and negative cosmological
  constant}.
\newblock {\em {\XPEH Int.J.Mod.Phys.}}, {\bbf A18}, 2379--2393, 2003.

\bibitem{Derrick:1964ww}
G.H. Derrick.
\newblock {\ssl Comments on nonlinear wave equations as models for elementary
  particles}.
\newblock {\em {\XPEH J.Math.Phys.}}, {\bbf 5}, 1252--1254, 1964.

\bibitem{Wipf}
L.~Dittmann, T.~Heinzl, and A.~Wipf.
\newblock {\ssl A lattice study of the Faddeev-Niemi action}.
\newblock {\em {\XPEH Nucl.Phys.Proc.Suppl.}}, {\bbf B106}, 649--651, 2002.

\bibitem{Donnelly}
R.J. Donnelly.
\newblock {\em {\ssl Quantized vortices in helium II}}.
\newblock {\XPEH Cambridge University Press}, 1991.
\newblock 346 p.

\bibitem{Doudoulakis:2007ti}
C.G. Doudoulakis.
\newblock {\ssl On vortices and rings in extended Abelian models}.
\newblock {\em {\XPEH Physica}}, {\bbf D234}, 1--10, 2007.

\bibitem{Doudoulakis:2007xz}
C.G. Doudoulakis.
\newblock {\ssl On vortices and solitons in Goldstone and Abelian-Higgs
  models}, 2007.
\newblock {\tt arXiv:0709.3709 [hep-ph]}.

\bibitem{Doudoulakis:2006iw}
C.G. Doudoulakis.
\newblock {\ssl Search of axially symmetric solitons}.
\newblock {\em {\XPEH Physica}}, {\bbf D228}, 159--165, 2007.

\bibitem{Emparan}
R.~Emparan and H.S. Reall.
\newblock {\ssl Black rings}.
\newblock {\em {\XPEH Class.Quant.Grav.}}, {\bbf 23}, R169, 2006.

\bibitem{movies}
J.~Hietarinta et~al.
\newblock {\tt http://users.utu.fi/hietarin/knots/}.

\bibitem{Yao}
W.M.~Yao {\it et al.}
\newblock {\ssl The review of particle physics}.
\newblock {\em {\XPEH Journ.Phys.}}, {\bbf G33}, 1--1232, 2006.

\bibitem{Faddeev:1975}
L.D. Faddeev.
\newblock {\ssl Quantization of solitons}, 1975.
\newblock Princeton preprint IAS-75-QS70; also {\ssl Einstein and several
  contemporary tendencies in the theory of elementary particles} in {
  Relativity, quanta, and cosmology, Vol.1}, M. Pantaleo and F. De Finis, eds.,
  pp. 247-266 (1979), reprinted in {\it L. Faddeev, 40 years in mathematical
  physics}, pp. 441-461 (World Scientific, 1995).

\bibitem{Faddeev:1976pg}
L.D. Faddeev.
\newblock {\ssl Some comments on the many-dimensional solitons}.
\newblock {\em {\XPEH Lett. Math. Phys.}}, {\bbf 1}, 289--293, 1976.

\bibitem{Faddeev:2000qw}
L.D. Faddeev, L.~Freyhult, A.J. Niemi, and P.~Rajan.
\newblock {\ssl Shafranov's virial theorem and magnetic plasma confinement}.
\newblock {\em {\XPEH J.Phys.}}, {\bbf A35}, L133--L140, 2002.

\bibitem{Faddeev:1996zj}
L.D. Faddeev and A.J. Niemi.
\newblock {\ssl Knots and particles}.
\newblock {\em {\XPEH Nature}}, {\bbf 387}, 58--61, 1997.

\bibitem{Faddeev:1997pf}
L.D. Faddeev and A.J. Niemi.
\newblock {\ssl Toroidal configurations as stable solitons}, 1997.
\newblock {\tt hep-th/9705176}.

\bibitem{Faddeev:1998eq}
L.D. Faddeev and A.J. Niemi.
\newblock {\ssl Partially dual variables in SU(2) Yang-Mills theory}.
\newblock {\em {\XPEH Phys.Rev.Lett.}}, {\bbf 82}, 1624--1627, 1999.

\bibitem{Faddeev:2000rp}
L.D. Faddeev and A.J. Niemi.
\newblock {\ssl Magnetic Geometry and the Confinement of Electrically
  Conducting Plasmas}.
\newblock {\em {\XPEH Phys.Rev.Lett.}}, {\bbf 85}, 3416--3419, 2000.

\bibitem{Faddeev:2006sw}
L.D. Faddeev and A.J. Niemi.
\newblock {\ssl Spin-charge separation, conformal covariance and the SU(2)
  Yang-Mills theory}.
\newblock {\em {\XPEH Nucl.Phys.}}, {\bbf B776}, 38--65, 2007.

\bibitem{Faddeev:2003aw}
L.D. Faddeev, A.J. Niemi, and U.~Wiedner.
\newblock {\ssl Glueballs, closed fluxtubes and eta(1440)}.
\newblock {\em {\XPEH Phys.Rev.}}, {\bbf D70}, 114033, 2004.

\bibitem{Fayzullaev:2004xa}
B.A. Fayzullaev, M.M. Musakhanov, D.G. Pak, and M.~Siddikov.
\newblock {\ssl Knot soliton in Weinberg-Salam model}.
\newblock {\em {\XPEH Phys.Lett.}}, {\bbf B609}, 442--448, 2005.

\bibitem{Forgacs:1979zs}
P.~Forgacs and N.~Manton.
\newblock {\ssl Space-time symmetries in gauge theories}.
\newblock {\em {\XPEH Commun.Math.Phys.}}, {\bbf 72}, 15--35, 1980.

\bibitem{Forgacs:2005sf}
P.~Forgacs, S.~Reuillon, and M.S. Volkov.
\newblock {\ssl Superconducting vortices in semilocal models}.
\newblock {\em {\XPEH Phys.Rev.Lett.}}, {\bbf 96}, 041601, 2006.

\bibitem{Forgacs:2006pm}
P.~Forgacs, S.~Reuillon, and M.S. Volkov.
\newblock {\ssl Twisted superconducting semilocal strings}.
\newblock {\em {\XPEH Nucl.Phys.}}, {\bbf B751}, 390--418, 2006.

\bibitem{Forgacs}
P.~Forgacs and M.S. Volkov.
\newblock {\ssl On the existence of knot solitons in gauge field theory}, 2002.
\newblock unpublished.

\bibitem{Friedberg}
R.~Friedberg, T.D. Lee, and A.~Sirlin.
\newblock {\ssl A class of scalar-field soliton solutions in three space
  dimensions}.
\newblock {\em {\XPEH Phys.Rev.}}, {\bbf D13}, 2739--2761, 1976.

\bibitem{Garaud:2007ti}
J.~Garaud and M.S. Volkov.
\newblock {\ssl Stability analysis of the twisted superconducting electroweak
  strings}.
\newblock {\em {\XPEH Nucl.Phys.}}, {\bbf B799}, 430--455, 2008.

\bibitem{Ginzburg}
V.L. Ginzburg and L.D. Landau.
\newblock {\ssl On the theory of superconductivity}.
\newblock {\em {\XPEH Zh.Eksp.Teor.Fiz.}}, {\bbf 20}, 1064, 1950.

\bibitem{Pitaevskii}
V.L. Ginzburg and L.P. Pitaevskii.
\newblock {\ssl On the theory of superfluidity}.
\newblock {\em {\XPEH Sov.Phys.JETP}}, {\bbf 34}, 858--861, 1958.

\bibitem{Gladikowski:1996mb}
J.~Gladikowski and M.~Hellmund.
\newblock {\ssl Static solitons with non-zero Hopf number}.
\newblock {\em {\XPEH Phys.Rev.}}, {\bbf D56}, 5194--5199, 1997.

\bibitem{Grandclement}
P.~Grandclement and J.~Novak.
\newblock {\ssl Spectral methods for numerical relativity}, 2007.
\newblock arXiv:0706.2286 [gr-qc].

\bibitem{Grant}
J.~Grant and P.H. Roberts.
\newblock {\ssl Motion in a Bose condensate III. The structure and effective
  masses of charged and uncharged impurities}.
\newblock {\em {\XPEH J.Phys.}}, {\bbf A7}, 260--279, 1974.

\bibitem{Ball}
B.~Greenwood.
\newblock {\ssl Ball lightning bibliography.} 
\newblock {\tt www.project1947.com/shg/${\tt bl_{-}db.html}$}.

\bibitem{Gross}
E.P. Gross.
\newblock {\ssl Structure of quantized vortex}.
\newblock {\em {\XPEH Nuovo Cimento}}, {\bbf 20}, 454--477, 1961.

\bibitem{Hartmann:2008yr}
B.~Hartmann and B.~Carter.
\newblock {\ssl The logarithmic equation of state for superconducting cosmic
  strings}, 2008.
\newblock {\tt arXiv:0803.0266 [hep-th]}.

\bibitem{Hartmann:2000ja}
B.~Hartmann, B.~Kleihaus, and J.~Kunz.
\newblock {\ssl Dyons with axial symmetry}.
\newblock {\em {\XPEH Mod.Phys.Lett.}}, {\bbf A15}, 1003--1012, 2000.

\bibitem{Haws:1988ax}
D.~Haws, M.~Hindmarsh, and N.~Turok.
\newblock {\ssl Superconducting strings or springs ?}
\newblock {\em {\XPEH Phys.Lett.}}, {\bbf B209}, 255--261, 1988.

\bibitem{Hen}
I.~Hen and M.~Karliner.
\newblock {\ssl Spontaneous breaking of rotational symmetry in rotating
  solitons: A toy model of excited nucleons with high angular momentum}.
\newblock {\em {\XPEH Phys.Rev.}}, {\bbf D77}, 116002, 2008.

\bibitem{Heusler:1998ec}
M.~Heusler, N.~Straumann, and M.~Volkov.
\newblock {\ssl On rotational excitations and axial deformations of BPS
  monopoles and Julia-Zee dyons}.
\newblock {\em {\XPEH Phys.Rev.}}, {\bbf D58}, 105021, 1998.

\bibitem{Hietarinta:1998kt}
J.~Hietarinta and P.~Salo.
\newblock {\ssl Faddeev-Hopf knots: Dynamics of linked un-knots}.
\newblock {\em {\XPEH Phys.Lett.}}, {\bbf B451}, 60--67, 1999.

\bibitem{Hietarinta:2000ci}
J.~Hietarinta and P.~Salo.
\newblock {\ssl Ground state in the Faddeev-Skyrme model}.
\newblock {\em {\XPEH Phys.Rev.}}, {\bbf D62}, 081701, 2000.

\bibitem{Hill:1987qx}
C.T Hill, H.M. Hodges, and M.S. Turner.
\newblock {\ssl Bosonic superconducting cosmic strings}.
\newblock {\em {\XPEH Phys.Rev.}}, {\bbf D37}, 263--282, 1988.

\bibitem{Hobart}
R.H. Hobart.
\newblock {\ssl On the instability of a class of unitary field model}.
\newblock {\em {\XPEH Proc.Phys.Soc.}}, {\bbf 82}, 201--203, 1963.

\bibitem{Huang:1980bz}
K.~Huang and R.~Tipton.
\newblock {\ssl Vortex excitations in the Weinberg-Salam theory}.
\newblock {\em {\XPEH Phys.Rev.}}, {\bbf D23}, 3050--3057, 1981.

\bibitem{Ioannidou:2006nn}
T.~Ioannidou, B.~Kleihaus, and J.~Kunz.
\newblock {\ssl Spinning gravitating skyrmions}.
\newblock {\em {\XPEH Phys.Lett.}}, {\bbf B643}, 213--220, 2006.

\bibitem{Jaykka:2006gf}
J.~Jaykka, J.~Hietarinta, and P.~Salo.
\newblock {\ssl Investigation of the stability of Hopfions in the two-component
  Ginzburg-Landau model}.
\newblock {\em {\XPEH Phys.Rev.}}, {\bbf B77}, 094509, 2008.

\bibitem{Friedman}
J.L.Friedman, K.~Schliech, and D.M. Witt.
\newblock {\ssl Topological censorship}.
\newblock {\em {\XPEH Phys.Rev.Lett.}}, {\bbf 71}, 1486--1489, 1993.

\bibitem{Jones}
C.A. Jones and P.H. Roberts.
\newblock {\ssl Motion in a Bose condensate: IV. Axisymmetric solitary waves}.
\newblock {\em {\XPEH J.Phys.}}, {\bbf A15}, 2599--2619, 1982.

\bibitem{Julia:1975ff}
B.~Julia and A.~Zee.
\newblock {\ssl Poles with both magnetic and electric charges in nonabelian
  gauge theory}.
\newblock {\em {\XPEH Phys.Rev.}}, {\bbf D11}, 2227--2232, 1975.

\bibitem{Ueda}
Y.~Kawaguchi, M.~Nitta, and M.~Ueda.
\newblock {\ssl Knots in a spinor Bose-Einstein condensate}, 2008.
\newblock {\tt arXiv:0802.1968 [cond-mat.other]}.

\bibitem{Kleihaus:1999sx}
B.~Kleihaus and J.~Kunz.
\newblock {\ssl A monopole antimonopole solution of the SU(2) Yang-Mills- Higgs
  model}.
\newblock {\em {\XPEH Phys.Rev.}}, {\bbf D61}, 025003, 2000.

\bibitem{Kleihaus:2008gn}
B.~Kleihaus, J.~Kunz, and M.~Leissner.
\newblock {\ssl Sphalerons, antisphalerons and vortex rings}, 2008.
\newblock arXiv:0802.3275 [hep-th].

\bibitem{Kleihaus:2005me}
B.~Kleihaus, J.~Kunz, and M.~List.
\newblock {\ssl Rotating boson stars and Q-balls}.
\newblock {\em {\XPEH Phys.Rev.}}, {\bbf D72}, 064002, 2005.

\bibitem{Kleihaus:2007vk}
B.~Kleihaus, J.~Kunz, M.~List, and I.~Schaffer.
\newblock {\ssl Rotating boson stars and Q-balls II: negative parity and
  ergoregions}, 2007.
\newblock {\tt arXiv:0712.3742 [gr-qc]}.

\bibitem{monBH}
B.~Kleihaus, J.~Kunz, and F.~Navarro-Lerida.
\newblock {\ssl Rotating black holes with monopole hair}.
\newblock {\em {\XPEH Phys.Lett.}}, {\bbf B599}, 294--300, 2004.

\bibitem{Kleihaus:2005fs}
B.~Kleihaus, J.~Kunz, and U.~Neemann.
\newblock {\ssl Gravitating stationary dyons and rotating vortex rings}.
\newblock {\em {\XPEH Phys.Lett.}}, {\bbf B623}, 171--178, 2005.

\bibitem{Kleihaus:2003xz}
B.~Kleihaus, J.~Kunz, and Y.~Shnir.
\newblock {\ssl Monopoles, antimonopoles and vortex rings}.
\newblock {\em {\XPEH Phys.Rev.}}, {\bbf D68}, 101701, 2003.

\bibitem{Kleihaus:2004is}
B.~Kleihaus, J.~Kunz, and Y.~Shnir.
\newblock {\ssl Monopole-antimonopole chains and vortex rings}.
\newblock {\em {\XPEH Phys.Rev.}}, {\bbf D70}, 065010, 2004.

\bibitem{Klinkhamer:1984di}
F.R. Klinkhamer and N.S. Manton.
\newblock {\ssl A saddle point solution in the Weinberg-Salam theory}.
\newblock {\em {\XPEH Phys.Rev.}}, {\bbf D30}, 2212--2220, 1984.

\bibitem{Koma:1999sm}
Y.~Koma, H.~Suganuma, and H.~Toki.
\newblock {\ssl Flux-tube ring and glueball properties in the dual
  Ginzburg-Landau theory}.
\newblock {\em {\XPEH Phys.Rev.}}, {\bbf D60}, 074024, 1999.

\bibitem{Komineas}
S.~Komineas.
\newblock {\ssl Vortex rings and solitary waves in trapped Bose-Einstein
  condensates}.
\newblock {\em {\XPEH Eur.Phys.J. Special Topics}}, {\bbf 147}, 133--152, 2007.

\bibitem{Kundu}
A.~Kundu and Yu.P. Rybakov.
\newblock {\ssl Closed-vortex-type solitons with Hopf index}.
\newblock {\em {\XPEH J.Phys.}}, {\bbf A15}, 269--275, 1982.

\bibitem{Kusenko}
A.~Kusenko.
\newblock {\ssl Small Q-balls}.
\newblock {\em {\XPEH Phys.Lett.}}, {\bbf B404}, 285--290, 1997.

\bibitem{Kusenko:1997zq}
A.~Kusenko.
\newblock {\ssl Solitons in the supersymmetric extensions of the standard
  model}.
\newblock {\em {\XPEH Phys.Lett.}}, {\bbf B405}, 108--113, 1997.

\bibitem{Kusenko:1997si}
A.~Kusenko and M.~Shaposhnikov.
\newblock {\ssl Supersymmetric Q-balls as dark matter}.
\newblock {\em {\XPEH Phys.Lett.}}, {\bbf B418}, 46--54, 1998.

\bibitem{Lee:1988ag}
K.~Lee, J.A. Stein-Schabes, R.~Watkins, and L.M. Widrow.
\newblock {\ssl Gauged Q-balls}.
\newblock {\em {\XPEH Phys.Rev.}}, {\bbf D39}, 1665--1673, 1989.

\bibitem{Lee:1991ax}
T.~D. Lee and Y.~Pang.
\newblock {\ssl Nontopological solitons}.
\newblock {\em {\XPEH Phys.Rept.}}, {\bbf 221}, 251--350, 1992.

\bibitem{Leggett}
A.J. Leggett.
\newblock {\ssl Bose-Einstein condensation in the alkali gazes: Some
  fundamental concepts}.
\newblock {\em {\XPEH Rev.Mod.Phys.}}, {\bbf 73}, 307--356, 2001.

\bibitem{Lemperiere:2003yt}
Y.~Lemperiere and E.~P.~S. Shellard.
\newblock {\ssl Vorton existence and stability}.
\newblock {\em {\XPEH Phys.Rev.Lett.}}, {\bbf 91}, 141601, 2003.

\bibitem{Lin}
F.~Lin and Y.~Yang.
\newblock {\ssl Existence of energy minimizers as stable knotted solitons in
  the Faddeev model}.
\newblock {\em {\XPEH Comm.Math.Phys.}}, {\bbf 249}, 273--303, 2004.

\bibitem{Makhankov}
V.G. Makhankov, Y.P. Rybakov, and V.I. Sanyuk.
\newblock {\em {\ssl The Skyrme model, Fundamentals, Methods, Applications}}.
\newblock Springer-Verlag, Berlin, 1993.

\bibitem{Manton:2004tk}
N.~Manton and P.~Sutcliffe.
\newblock {\em {\ssl Topological Solitons}}.
\newblock Cambridge University Press, 2004.
\newblock 493 p.

\bibitem{Martins:1998gb}
C.J.A.P. Martins and E.P.S. Shellard.
\newblock {\ssl Vorton formation}.
\newblock {\em {\XPEH Phys.Rev.}}, {\bbf D57}, 7155--7176, 1998.

\bibitem{Meissner}
U.G. Meissner.
\newblock {\ssl Toroidal solitons with unit Hopf charge}.
\newblock {\em {\XPEH Phys.Lett.}}, {\bbf B154}, 190--192, 1985.

\bibitem{Metlitski:2003gj}
M.~Metlitski and A.R. Zhitnitsky.
\newblock {\ssl Vortex Rings in two Component Bose-Einstein Condensates}.
\newblock {\em {\XPEH JHEP}}, {\bbf 06}, 017, 2004.

\bibitem{mihalache}
D.~Mihalache, D.~Mazilu, L.C. Crasovan, I.~Towers, A.V. Buryak, B.A. Malomed,
  L.~Torner, J.P. Torres, and F.~Lederer.
\newblock {\ssl Stable spinning optical solitons in three dimensions}.
\newblock {\em {\XPEH Phys.Rev.Lett.}}, {\bbf 88}, 073902, 2002.

\bibitem{mihalache1}
D.~Mihalache, D.~Mazilu, L.C. Crasovan, I.~Towers, B.A. Malomed, A.V. Buryak,
  L.~Torner, and F.~Lederer.
\newblock {\ssl Stable three-dimensional spinning optical solitons supported by
  competing quadratic and cubic nonlinearities}.
\newblock {\em {\XPEH Phys.Rev.}}, {\bbf E66}, 016613, 2002.

\bibitem{mihalache2}
D.~Mihalache, D.~Mazilu, I.~Towers, B.A. Malomed, and F.~Lederer.
\newblock {\ssl Stable spatiotemporal spinning solitons in a bimodal
  cubic-quintic medium}.
\newblock {\em {\XPEH Phys.Rev.}}, {\bbf E67}, 056608, 2003.

\bibitem{Nicole}
A.D. Nicole.
\newblock {\ssl Solitons with non-vanishing Hopf index}.
\newblock {\em {\XPEH J.Phys.}}, {\bbf 4}, 1363--1369, 1978.

\bibitem{NO}
H.B. Nielsen and P.~Olesen.
\newblock {\ssl Vortex line models for dual strings}.
\newblock {\em {\XPEH Nucl.Phys.}}, {\bbf B61}, 45--61, 1973.

\bibitem{Niemi:2000ny}
A.J. Niemi, K.~Palo, and S.~Virtanen.
\newblock {\ssl (Meta)stable closed vortices in (3+1)+dimensional gauge
  theories with an extended Higgs sector}.
\newblock {\em {\XPEH Phys.Rev.}}, {\bbf D61}, 085020, 2000.

\bibitem{Paturyan:2004ps}
V.~Paturyan, E.~Radu, and D.H. Tchrakian.
\newblock {\ssl Rotating regular solutions in Einstein-Yang-Mills-Higgs
  theory}.
\newblock {\em {\XPEH Phys.Lett.}}, {\bbf B609}, 360--366, 2005.

\bibitem{Peter}
P.~Peter.
\newblock {\ssl Superconducting cosmic string: Equation of state for spacelike
  and timelike current in the neutral limit}.
\newblock {\em {\XPEH Phys.Rev.}}, {\bbf D45}, 1091--1102, 1992.

\bibitem{Piette}
B.M.A.G. Piette, B.J. Schroers, and W.~Zakrzewski.
\newblock {\ssl Dynamics of baby skyrmions}.
\newblock {\em {\XPEH Nucl.Phys.}}, {\bbf B439}, 205--238, 1995.

\bibitem{Piette:1997ny}
B.M.A.G. Piette and D.H. Tchrakian.
\newblock {\ssl Static solutions in the U(1) gauged Skyrme model}.
\newblock {\em {\XPEH Phys.Rev.}}, {\bf D62}, 025020, 2000.

\bibitem{Pitaevskii1}
L.P. Pitaevskii.
\newblock {\ssl Vortex lines in an imperfect Bose gas}.
\newblock {\em {\XPEH Sov.Phys.JETP}}, {\bbf 13}, 451--454, 1961.

\bibitem{Polyakov:1974ek}
A.M. Polyakov.
\newblock {\ssl Particle spectrum in quantum field theory}.
\newblock {\em {\XPEH JETP Lett.}}, {\bbf 20}, 194--195, 1974.

\bibitem{Protogenov:2002bt}
A.P. Protogenov and V.A. Verbus.
\newblock {\ssl Energy bounds of linked vortex states}.
\newblock {\em {\XPEH JETP Lett.}}, {\bbf 76}, 53--55, 2002.

\bibitem{Radu:2005jp}
E.~Radu and D.H. Tchrakian.
\newblock {\ssl Spinning U(1) gauged skyrmions}.
\newblock {\em {\XPEH Phys.Lett.}}, {\bbf B632}, 109--113, 2006.

\bibitem{Rajaraman}
R.~Rajaraman.
\newblock {\em {\ssl Solitons and Instantons}}.
\newblock North Holland, Amsterdam, 1982.
\newblock 418 p.

\bibitem{Rayfield}
G.W. Rayfield and F.~Reif.
\newblock {\ssl Quantized vortex rings in superfluid helium}.
\newblock {\em {\XPEH Phys.Rev.}}, {\bbf A136}, 1194--1208, 1964.

\bibitem{Ren}
J.~Ren, R.~Li, and Y.~Duan.
\newblock {\ssl Inner topological structure of Hopf invariant}.
\newblock {\em {\XPEH Journ.Math.Phys.}}, {\bbf 48}, 073502, 2007.

\bibitem{Rubakov:1985nk}
V.A. Rubakov.
\newblock {\ssl On the electroweak theory at high fermion density}.
\newblock {\em {\XPEH Prog.Theor.Phys.}}, {\bbf 75}, 366--385, 1986.

\bibitem{Rubakov:1986am}
V.A. Rubakov and A.N. Tavkhelidze.
\newblock {\ssl Stable anomalous states of superdense matter in gauge
  theories}.
\newblock {\em {\XPEH Phys.Lett.}}, {\bbf B165}, 109--112, 1985.

\bibitem{Ruostekoski:2004pj}
J.~Ruostekoski.
\newblock {\ssl Stable particlelike solitons with multiply-quantized vortex
  lines in Bose-Einstein condensates}.
\newblock {\em {\XPEH Phys.Rev.}}, {\bbf A70}, 041601, 2004.

\bibitem{Ruostekoski:2005zd}
J.~Ruostekoski and Z.~Dutton.
\newblock {\ssl Engineering vortex rings and systems for controlled studies of
  vortex interactions in Bose-Einstein condensates}.
\newblock {\em {\XPEH Phys.Rev.}}, {\bbf A72}, 063626, 2005.

\bibitem{Ruostekoski:2001fc}
J.~Ruostekoski and J.~R. J.R.~Anglin.
\newblock {\ssl Creating vortex rings and three-dimensional skyrmions in
  Bose-Einstein condensates}.
\newblock {\em {\XPEH Phys.Rev.Lett.}}, {\bbf 86}, 3934--3937, 2001.

\bibitem{Ruostekoski:2003qx}
J.~Ruostekoski and J.~R. J.R.~Anglin.
\newblock {\ssl Monopole core instability and Alice rings in spinor Bose-
  Einstein condensates}.
\newblock {\em {\XPEH Phys. Rev. Lett.}}, {\bbf 91}, 190402, 2003.

\bibitem{Saffman}
P.G. Saffman.
\newblock {\em {\ssl Vortex dynamics}}.
\newblock {\XPEH Cambridge University Press}, 1992.
\newblock 311 p.

\bibitem{Savage:2003hh}
C.M. Savage and J.~Ruostekoski.
\newblock {\ssl Energetically stable particle-like Skyrmions in a trapped
  Bose-Einstein condensate}.
\newblock {\em {\XPEH Phys.Rev.Lett.}}, {\bbf 91}, 010403, 2003.

\bibitem{FIDISOL1}
M.~Schauder, R.~Wei\ss\, and W.~Sch\"onauer.
\newblock {\ssl The CADSOL Program Package, Universit\"at Karlsruhe Interner
  Bericht Nr. 46/92 }, 1992.

\bibitem{Schmid:2007dm}
M.~Schmid and M.~Shaposhnikov.
\newblock {\ssl Anomalous Abelian solitons}.
\newblock {\em {\XPEH Nucl.Phys.}}, {\bbf B775}, 365--389, 2007.

\bibitem{FIDISOL}
W.~Sch\"onauer and R.~Wei\ss.
\newblock {\ssl The Fidisol blackbox solver}.
\newblock {\em {\XPEH J. Comput. Appl. Math.}}, {\bbf 27}, 279--297, 1989.

\bibitem{Shabanov:1999xy}
S.V. Shabanov.
\newblock {\ssl An effective action for monopoles and knot solitons in
  Yang-Mills theory}.
\newblock {\em {\XPEH Phys.Lett.}}, {\bbf B458}, 322--330, 1999.

\bibitem{Shabanov:1999uv}
S.V. Shabanov.
\newblock {\ssl Yang-Mills theory as an Abelian theory without gauge fixing}.
\newblock {\em {\XPEH Phys.Lett.}}, {\bbf B463}, 263--272, 1999.

\bibitem{Shnir:2005te}
Y.~Shnir.
\newblock {\ssl Electromagnetic interaction in the system of multimonopoles and
  vortex rings}.
\newblock {\em {\XPEH Phys.Rev.}}, {\bbf D72}, 055016, 2005.

\bibitem{Shnir}
Ya.M. Shnir.
\newblock {\em {\ssl Magnetic monopoles}}.
\newblock {\XPEH Springer-Verlag, Berlin-Heidelberg}, 2005.
\newblock 532 p.

\bibitem{Skyrme:1961vq}
T.H.R. Skyrme.
\newblock {\ssl A nonlinear field theory}.
\newblock {\em {\XPEH Proc.Roy.Soc.Lond.}}, {\bbf A260}, 127--138, 1961.

\bibitem{Sutcliffe:2007ui}
P.~Sutcliffe.
\newblock {\ssl Knots in the Skyrme-Faddeev model}, 2007.
\newblock {\tt arXiv:0705.1468 [hep-th]}.

\bibitem{Sutcliffe:2007vm}
P.~Sutcliffe.
\newblock {\ssl Vortex rings in ferromagnets}, 2007.
\newblock {\tt arXiv:0707.1383 [cond-mat.mes-hall]}.

\bibitem{'tHooft:1974qc}
G.~'t~Hooft.
\newblock {\ssl Magnetic monopoles in unified gauge theories}.
\newblock {\em {\XPEH Nucl.Phys.}}, {\bbf B79}, 276--284, 1974.

\bibitem{Taubes:1982ie}
C.H. Taubes.
\newblock {\ssl The existence of a non-minimal solution to the SU(2)
  Yang-Mills-Higgs equations on $\mathbb{R}^3$}.
\newblock {\em {\XPEH Commun.Math.Phys.}}, {\bbf 86}, 257--298, 1982.

\bibitem{Kelvin}
W.H. Thomson.
\newblock {\ssl On vortex motion}.
\newblock {\em {\XPEH Trans.R.Soc.Edin.}}, {\bbf 25}, 217--260, 1867.

\bibitem{Vakulenko:1979uw}
A.F. Vakulenko and L.V. Kapitansky.
\newblock {\ssl Stability of solitons in $S^2$ in the nonlinear
  $\sigma$-model.}
\newblock {\em {\XPEH Sov.Phys.Dokl.}}, {\bbf 24}, 433--434, 1979.

\bibitem{vanBaal:2001jm}
P.~van Baal and A.~Wipf.
\newblock {\ssl Classical gauge vacua as knots}.
\newblock {\em {\XPEH Phys.Lett.}}, {\bbf B515}, 181--184, 2001.

\bibitem{Vilenkin}
A.~Vilenkin and E.P.S. Shellard.
\newblock {\em {\ssl Cosmic Strings and Other Topological Defects}}.
\newblock Cambridge University Press, 1994.
\newblock 517 p.

\bibitem{Volkov:2006ug}
M.S. Volkov.
\newblock {\ssl Superconducting electroweak strings}.
\newblock {\em {\XPEH Phys.Lett.}}, {\bbf B644}, 203--207, 2007.

\bibitem{VG}
M.S. Volkov and D.V. Gal'tsov.
\newblock {\ssl Gravitating non-Abelian solitons and black holes with
  Yang-Mills fields}.
\newblock {\em {\XPEH Phys.Rep.}}, {\bbf 319}, 1--83, 1999.

\bibitem{Volkov:2002aj}
M.S. Volkov and E.~Wohnert.
\newblock {\ssl Spinning Q-balls}.
\newblock {\em {\XPEH Phys.Rev.}}, {\bbf D66}, 085003, 2002.

\bibitem{Volkov:2003ew}
M.S. Volkov and E.~Wohnert.
\newblock {\ssl On the existence of spinning solitons in gauge field theory}.
\newblock {\em {\XPEH Phys.Rev.}}, {\bbf D67}, 105006, 2003.

\bibitem{Volovik}
G.E. Volovik.
\newblock {\em {\ssl The Universe in a helium droplet}}.
\newblock {\XPEH Oxford University Press}, 2003.
\newblock 507 p.

\bibitem{Ward:1998pj}
R.S. Ward.
\newblock {\ssl Hopf solitons on $S^3$ and $R^3$}, 1998.
\newblock {\tt hep-th/9811176}.

\bibitem{Ward:2000qj}
R.S. Ward.
\newblock {\ssl The interaction of two Hopf solitons}.
\newblock {\em {\XPEH Phys.Lett.}}, {\bbf B473}, 291--296, 2000.

\bibitem{Ward2}
R.S. Ward.
\newblock {\ssl Hopf solitons from instanton holonomy}, 2001.
\newblock {\tt hep-th/0108082}.

\bibitem{Ward:2002vq}
R.S. Ward.
\newblock {\ssl Stabilizing textures with magnetic fields}.
\newblock {\em {\XPEH Phys.Rev.}}, {\bbf D66}, 041701, 2002.

\bibitem{Ward:2004gr}
R.S. Ward.
\newblock {\ssl Skyrmions and Faddeev-Hopf solitons}.
\newblock {\em {\XPEH Phys.Rev.}}, {\bbf D70}, 061701, 2004.

\bibitem{Ward}
R.S. Ward.
\newblock {\ssl Hopf solitons on the lattice}.
\newblock {\em {\XPEH J.Phys.}}, {\bbf A39}, L105--L109, 2006.

\bibitem{Witten:1984eb}
E.~Witten.
\newblock {\ssl Superconducting strings}.
\newblock {\em {\XPEH Nucl.Phys.}}, {\bbf B249}, 557--592, 1985.

\bibitem{Zakharov}
V.A. Zakharov and E.A. Kuznetsov.
\newblock {\ssl Hamiltonian formalism for nonlinear waves}.
\newblock {\em {\XPEH Phys.Usp.}}, {\bbf 40}, 1087--1116, 1997.

\end{thebibliography}

\end{document}